\def\msol{{\cal M}_\odot}
\def\lsol{{\cal L}_{\odot,B}}
\def\lso{{\cal L}_{\odot}}
\def\lesssim{\,\lower 1mm \hbox{\ueber{\sim}{<}}\,}
\def\grsim{\,\lower 1mm \hbox{\ueber{\sim}{>}}\,}
% This is cp-aa.ini, the INITEX file to create a format file for
% the Springer journal Astronomy & Astrophysics, based on plain TeX,
% using Computer Modern fonts                           version 3.0
%                                (earlier versions were called aa.cmm)
%%%%%%%%%%%%%%%%%%%%%%%%%%%%%%%%%%%%%%%%%%%%%%%%%%%%%%%%%%%%%%%%%%%%%%
\catcode`\{=1 % left brace is begin-group character
\catcode`\}=2 % right brace is end-group character
\catcode`\#=6 % hash mark is macro parameter character
%%% removed from cp-aa.ini:
%\let\x=\input \def\input#1 {\let\input=\x \let\x=\undefined}
%
\let\qayb=\font
\def\font#1=#2 {\def\qayba{#1}\def\qaybb{#2}
                \def\qaybf{\preloaded}
                \def\qaybg{\tenrm}\def\qaybh{\tenex}
                \def\qaybu{cmr7}\def\qaybv{cmtt10}\def\qaybw{cmssbx10}
                \ifx\qayba\qaybg \qayb#1=#2
                \else\ifx\qayba\qaybh \qayb#1=#2
                \else\ifx\qayba\qaybf
                  \ifx\qaybb\qaybu \qayb#1=#2
                  \else\ifx\qaybb\qaybv \qayb#1=#2
                  \else\ifx\qaybb\qaybw \qayb#1=#2 \fi\fi\fi
                \fi\fi\fi}
\let\qayc=\skewchar \def\skewchar#1=#2 {\relax}
\let\qayd=\textfont \def\textfont#1=#2{\relax}
\let\qaye=\scriptfont \def\scriptfont#1=#2{\relax}
\let\qayf=\scriptscriptfont \def\scriptscriptfont#1=#2{\relax}
%
%\x plain                                        %%% removed from cp-aa.ini
%
\let\font=\qayb \let\qayb=\undefined
\let\skewchar=\qayc  \let\qayc=\undefined
\let\textfont=\qayd  \let\qayd=\undefined
\let\scriptfont=\qaye  \let\qaye=\undefined
\let\scriptscriptfont=\qayf  \let\qayf=\undefined
\def\fonttype{cm}              
\def\Initexing{}               
\font\sevenrm=cmr7
\font\fiverm=cmr5
\font\teni=cmmi10 % math italic
\font\seveni=cmmi7
\font\fivei=cmmi5
\font\tensy=cmsy10 % math symbols
\font\sevensy=cmsy7
\font\fivesy=cmsy5
\font\tenbf=cmbx10 % boldface extended
\font\sevenbf=cmbx7
\font\fivebf=cmbx5
\font\tentt=cmtt10 % typewriter
\font\tenit=cmti10 % text italic

\skewchar\teni='177 \skewchar\seveni='177 \skewchar\fivei='177
\skewchar\tensy='60 \skewchar\sevensy='60 \skewchar\fivesy='60

\textfont0=\tenrm \scriptfont0=\sevenrm \scriptscriptfont0=\fiverm
\textfont1=\teni \scriptfont1=\seveni \scriptscriptfont1=\fivei
\textfont2=\tensy \scriptfont2=\sevensy \scriptscriptfont2=\fivesy
\textfont3=\tenex \scriptfont3=\tenex \scriptscriptfont3=\tenex
\textfont\itfam=\tenit
\textfont\bffam=\tenbf \scriptfont\bffam=\sevenbf
\scriptscriptfont\bffam=\fivebf
\textfont\ttfam=\tentt
\rm
% This is values.aa
% i.e. Values of package AA
% it contains sizes and measures specific for this macro package
% indention of equations
\newskip\mathindent      \mathindent=0pt
% \titlea
\newskip\tabefore \tabefore=20pt plus 10pt minus 5pt      % space above
\newskip\taafter  \taafter=10pt                           % space below
\newskip\tabaselineskip
\tabaselineskip=20pt
% \titleb
\newskip\tbbeforeback    \tbbeforeback=-20pt              % corrective space to a \titlea
\newskip\tbbefore        \tbbefore=17pt plus 7pt minus3pt % spaceabove
\newskip\tbafter         \tbafter=8pt                     % space below
\newskip\tbbaselineskip
\tbbaselineskip=17pt
% \titlec
\newskip\tcbeforeback    \tcbeforeback=-3pt               % corrective space to a \titleb
\advance\tcbeforeback by -10pt                            % corrective space to a \titleb
\newskip\tcbefore        \tcbefore=10pt plus 5pt minus 1pt% space above
\newskip\tcafter         \tcafter=6pt                     % space below
% \titled
\newskip\tdbeforeback    \tdbeforeback=-3pt                  % corrective space to a \titlec
\advance\tdbeforeback by -10pt                               % corrective space to a \titlec
\newskip\tdbefore        \tdbefore=10pt plus 4pt minus 1pt   % space above
% \petit
\newskip\petitsurround
\petitsurround=6pt\relax
% \ack
\newskip\ackbefore      \ackbefore=10pt plus 5pt             % space above
\newskip\ackafter       \ackafter=6pt                        % space below
% indention of lists
\newdimen\itemindent    \newdimen\itemitemindent
\itemindent=1.5em       \itemitemindent=2\itemindent
% This is cmlayout.tex
% it sets up some measures and defines macros needed by the font
% selection
\catcode`@=11    % use @ as a normal character
\normallineskip=1pt
\normallineskiplimit=0pt
\newskip\ttglue%
\def\ifundefin@d#1#2{%
\expandafter\ifx\csname#1#2\endcsname\relax}
\def\getf@nt#1#2#3#4{%
\ifundefin@d{#1}{#2}%
\global\expandafter\font\csname#1#2\endcsname=#3#4%
\fi\relax
}
\newfam\sffam
\newfam\scfam
\def\makesize#1#2#3#4#5#6#7{%
%%% these fonts cannot be loaded dynamically, because there
%%% is no macro we can overload
 \getf@nt{rm}{#1}{cmr}{#2}%
 \getf@nt{rm}{#3}{cmr}{#4}%
 \getf@nt{rm}{#5}{cmr}{#6}%
 \getf@nt{mi}{#1}{cmmi}{#2}%
 \getf@nt{mi}{#3}{cmmi}{#4}%
 \getf@nt{mi}{#5}{cmmi}{#6}%
 \getf@nt{sy}{#1}{cmsy}{#2}%
 \getf@nt{sy}{#3}{cmsy}{#4}%
 \getf@nt{sy}{#5}{cmsy}{#6}%

%%% skewchar
 \skewchar\csname mi#1\endcsname ='177
 \skewchar\csname mi#3\endcsname ='177
 \skewchar\csname mi#5\endcsname ='177
 \skewchar\csname sy#1\endcsname ='60
 \skewchar\csname sy#3\endcsname ='60
 \skewchar\csname sy#5\endcsname ='60

\expandafter\def\csname#1size\endcsname{%
 \normalbaselineskip=#7
 \normalbaselines
 \setbox\strutbox=\hbox{\vrule height0.75\normalbaselineskip%
    depth0.25\normalbaselineskip width0pt}%
 %
%%% "roman"
 \textfont0=\csname rm#1\endcsname
 \scriptfont0=\csname rm#3\endcsname
 \scriptscriptfont0=\csname rm#5\endcsname
    \def\oldstyle{\fam1\csname mi#1\endcsname}%
%%% "mathit"
 \textfont1=\csname mi#1\endcsname
 \scriptfont1=\csname mi#3\endcsname
 \scriptscriptfont1=\csname mi#5\endcsname
%%% "mathsy"
 \textfont2=\csname sy#1\endcsname
 \scriptfont2=\csname sy#3\endcsname
 \scriptscriptfont2=\csname sy#5\endcsname
%%% "mathex"
 \textfont3=\tenex\scriptfont3=\tenex\scriptscriptfont3=\tenex
   \def\rm{%
 \fam0\csname rm#1\endcsname%
   }%
   \def\it{%
 \getf@nt{it}{#1}{cmti}{#2}%
 \textfont\itfam=\csname it#1\endcsname
 \fam\itfam\csname it#1\endcsname
   }%
   \def\sl{%
 \getf@nt{sl}{#1}{cmsl}{#2}%
 \textfont\slfam=\csname sl#1\endcsname
 \fam\slfam\csname sl#1\endcsname}%
   \def\bf{%
 \getf@nt{bf}{#1}{cmbx}{#2}%
 \getf@nt{bf}{#3}{cmbx}{#4}%
 \getf@nt{bf}{#5}{cmbx}{#6}%
 \textfont\bffam=\csname bf#1\endcsname
 \scriptfont\bffam=\csname bf#3\endcsname
 \scriptscriptfont\bffam=\csname bf#5\endcsname
 \fam\bffam\csname bf#1\endcsname}%
   \def\tt{%
 \getf@nt{tt}{#1}{cmtt}{#2}%
 \textfont\ttfam=\csname tt#1\endcsname
 \fam\ttfam\csname tt#1\endcsname
 \ttglue=.5em plus.25em minus.15em
   }%
  \def\sf{%
\getf@nt{sf}{#1}{cmss}{10 at #2pt}%
\textfont\sffam=\csname sf#1\endcsname
\fam\sffam\csname sf#1\endcsname}%
   \def\sc{%
 \getf@nt{sc}{#1}{cmcsc}{10 at #2pt}%
 \textfont\scfam=\csname sc#1\endcsname
 \fam\scfam\csname sc#1\endcsname}%
\rm }}
\makesize{Xf}{10}{VIIf}{7}{Vf}{5}{12pt}
\newfam\mibfam
\def\normalsize{\Xfsize
\def\sf{%
   \getf@nt{sf}{Xf}{cmss}{10}%
   \getf@nt{sf}{VIIf}{cmss}{10 at 7pt}%
   \getf@nt{sf}{Vf}{cmss}{10 at 5pt}%
   \textfont\sffam=\csname sfXf\endcsname
   \scriptfont\sffam=\csname sfVIIf\endcsname
   \scriptscriptfont\sffam=\csname sfVf\endcsname
   \fam\sffam\csname sfXf\endcsname}%
\def\mib{%
   \getf@nt{mib}{Xf}{cmmib}{10}%
   \getf@nt{mib}{VIIf}{cmmib}{10 at7pt}%
   \getf@nt{mib}{Vf}{cmmib}{10 at5pt}%
   \textfont\mibfam=\csname mibXf\endcsname
   \scriptfont\mibfam=\csname mibVIIf\endcsname
   \scriptscriptfont\mibfam=\csname mibVf\endcsname
   \fam\mibfam\csname mibXf\endcsname}%
\def\boldmath{\textfont1=\mibXf \scriptfont1=\mibVIIf
\scriptscriptfont1=\mibVf}%
\if Y\REFEREE \normalbaselineskip=2\normalbaselineskip
\normallineskip=0.1\normalbaselineskip\normalbaselines
\fi
\rm}
\Xfsize
\it\bf\tt\rm
\def\tenrm{\rmXf}
\def\fiverm{\rmVf}
\def\teni{\miXf}
\def\fivei{\miVf}
\def\tensy{\syXf}
\def\fivesy{\syVf}
\def\tenbf{\bfXf}
\def\fivebf{\bfVf}
\def\tentt{\ttXf}

\def\tenit{\itXf}
\let\REFEREE=N
\normalsize
\it\bf\tt\sf\mib\rm
\catcode`@=12 % reset catcode
% This is layout.aa
% it sets up some measures and defines miscellaneous macros
\newdimen\fullhsize
\newcount\verybad \verybad=1010
\let\lr=L%
\baselineskip=12pt
\vsize=56\baselineskip
\hoffset=-1true cm
\voffset=-1true cm
\fullhsize=180mm
\newdimen\halfsize
\halfsize=88mm
\hsize=\halfsize
\def\fullline{\hbox to\fullhsize}
\def\makefootline{\baselineskip=12pt \fullline{\the\footline}}
\def\makeheadline{\vbox to 0pt{\vskip-22.5pt
            \fullline{\vbox to 8.5pt{}\the\headline}\vss}\nointerlineskip}
\hfuzz=2pt
\vfuzz=2pt
\tolerance=1000
\abovedisplayskip=3 mm plus6pt minus 4pt
\belowdisplayskip=3 mm plus6pt minus 4pt
\abovedisplayshortskip=0mm plus6pt
\belowdisplayshortskip=2 mm plus4pt minus 4pt
\parindent=1.5em
\newdimen\stdparindent\stdparindent\parindent
\frenchspacing
% Paginierung
\nopagenumbers
\predisplaypenalty=600        % Make a page break before a display harder
\displaywidowpenalty=2000     % and even harder for a widow display.
\def\widowsandclubs#1{\global\verybad=#1
   \global\widowpenalty=\the\verybad1      % default: 10101
   \global\clubpenalty=\the\verybad2  }    % default: 10102
\widowsandclubs{1010}
% This is runhead.aa
% it sets up the page layout for the first page of an article
% and enables the running head and its changing mechanism
%
\catcode`@=11 % use @ as a normal character
\let\firstpage=Y%
\def\paglay{\headline={\normalsize
\ifx Y\firstpage
 \global\let\firstpage=N%
 \let\REFEREE=N%
 \vbox{\hsize=.75\fullhsize
       \hrule
       \line{\vrule\kern3pt
             \vbox{\kern3pt
                   \hbox{\bf A\&A manuscript no.}
                   \hbox{(will be inserted by hand later)}
                   \kern3pt
                   \hrule
                   \kern3pt
                   \hbox{\bf Your thesaurus codes are:}
                   \hbox{\rightskip=0pt plus3em\advance\hsize by-7pt
                         \vbox{\bf\noindent\ignorespaces\the\THESAURUS}}%
                   \kern3pt}%
             \hfil
             \kern3pt
             \vrule}%
       \hrule}%
 \hfil\llap{\AALogo}%
\else
   \koltitle
\fi}}
\def\koltest{\setbox0=\hbox{\qquad\koltext}%
   \ifdim\wd0>\fullhsize
      \infuser{^^JThe running head built automatically from
               \string\AUTHOR\space and \string\MAINTITLE
               ^^Jexceeds the pagewidth, supply a shorter form
               using \string\AUTHORRUNNINGHEAD\string{...\string}
               ^^Jand/or \string\MAINTITLERUNNINGHEAD\string{...\string}
               after the \string\maketitle-command.}%
      \global\AUTHOR={Please give a shorter version with}%
      \global\MAINTITLE={{\tt\string\AUTHORRUNNINGHEAD\ }and/or%
                         {\tt\ \string\MAINTITLERUNNINGHEAD}}%
   \fi
   \global\let\kolTest=\relax}
\let\kolTest=\koltest
{\catcode`@=\active
\gdef\koltitle{%
   \petit
   \catcode`\@=\active
   \def@##1{}%
   \def\FOOTNOTE##1{}%
   \kolTest
   \ifodd\pageno
      \hfil\enspace
      \koltext
      \enspace\hfil\llap{\folio}%
   \else
      \rlap{\folio}%
      \hfil\enspace
      \koltext
      \enspace\hfil
   \fi}}
\def\koltext{\ignorespaces\the\AUTHOR\unskip: \ignorespaces
\the\MAINTITLE\unskip}
\def\kolTEST{\setbox0=\hbox{\qquad\koltext}%
   \ifdim\wd0>\fullhsize
      \infuser{^^JThe running head you supplied with
               \string\AUTHORRUNNINGHEAD\space and/or
               ^^J\string\MAINTITLERUNNINGHEAD\space exceeds the
               pagewidth, give a shorter form
               ^^Jusing \string\AUTHORRUNNINGHEAD\string{...\string}
               and/or \string\MAINTITLERUNNINGHEAD\string{...\string}
               ^^Jafter the \string\maketitle-command.}%
      \global\AUTHOR={Please give a shorter running head with}%
      \global\MAINTITLE={{\tt\string\AUTHORRUNNINGHEAD\ }and/or%
                         {\tt\ \string\MAINTITLERUNNINGHEAD}}%
   \fi
   \global\let\kolTest=\relax}
\def\MAINTITLERUNNINGHEAD#1{\global\MAINTITLE={#1}%
\infuser{^^JMAINTITLE part of running head has been changed}%
\let\kolTest=\kolTEST}
\def\AUTHORRUNNINGHEAD#1{\global\AUTHOR={#1}%
\infuser{^^JAUTHOR part of running head has been changed}%
\let\kolTest=\kolTEST}
\catcode`@=12 % reset catcode
% This is aacpetit.aa
% here the small print is defined
\catcode`@=11    % use @ as a normal character
\makesize{IXf}{9}{VIf}{6}{Vf}{5}{11pt}
\IXfsize\it\bf\tt\rm
\normalsize
\def\petit{\IXfsize
   \def\sf{%
      \getf@nt{sf}{IXf}{cmss}{9}%
      \getf@nt{sf}{VIf}{cmss}{10 at 6pt}%
      \getf@nt{sf}{Vf}{cmss}{10 at 5pt}%
      \textfont\sffam=\csname sfIXf\endcsname
      \scriptfont\sffam=\csname sfVIf\endcsname
      \scriptscriptfont\sffam=\csname sfVf\endcsname
      \fam\sffam\csname sfIXf\endcsname
}%
\def\mib{%
   \getf@nt{mib}{IXf}{cmmib}{10 at 9pt}%
   \getf@nt{mib}{VIf}{cmmib}{10 at 6pt}%
   \getf@nt{mib}{Vf}{cmmib}{10 at 5pt}%
   \textfont\mibfam=\csname mibIXf\endcsname
   \scriptfont\mibfam=\csname mibVIf\endcsname
   \scriptscriptfont\mibfam=\csname mibVf\endcsname
   \fam\mibfam\csname mibIXf\endcsname}%
\def\boldmath{\textfont1=\mibIXf\scriptfont1=\mibVIf
\scriptscriptfont1=\mibVf}%
 \if Y\REFEREE \normalbaselineskip=2\normalbaselineskip
 \normallineskip=0.1\normalbaselineskip\fi
 \setbox\strutbox=\hbox{\vrule height9pt depth2pt width0pt}%
 \normalbaselines\rm}%
%--------------------------------------------------------------------
\def\begpet{\vskip\petitsurround
\bgroup\petit}%  Beginn eines Paragraphen in petit
\def\endpet{\vskip\petitsurround
\egroup}%  Ende eines Paragraphen in petit
\petit\sf\mib
\normalsize
\catcode`@=12 % reset catcode
% This is aacmfont.aa
% It defines the fonts used for the cm version of the a&a package
% and connects the names to fonts already defined
%
% Fonts for \MAINTITLE
 \font \tamt            = cmmib10 scaled \magstep3
 \font \tams            = cmmib10 scaled \magstep1
%\font \tamss           = cmmib10  = \mibXf
 \let  \tamss=\mibXf
 \font \tast            = cmsy10 scaled \magstep3
 \font \tass            = cmsy10 scaled \magstep1
%\font \tasss           = cmsy10   = \syXf
 \let  \tasss=\syXf
 \font \tatt            = cmbx10 scaled \magstep3
 \font \tats            = cmbx10 scaled \magstep1
%\font \tatss           = cmbx10   = \bfXf
 \let  \tatss=\bfXf
% Fonts for \SUBTITLE
 \font \tbmt            = cmmib10 scaled \magstep2
%\font \tbms            = cmmib10  = \mibXf
 \let  \tbms=\mibXf
%\font \tbmss           = cmmib10 at 9pt = \mibIXf
 \let  \tbmss=\mibIXf
 \font \tbst            = cmsy10 scaled \magstep2
%\font \tbss            = cmsy10   = \syXf
 \let  \tbss=\syXf
%\font \tbsss           = cmsy9    = \syIXf
 \let  \tbsss=\syIXf
 \font \tbtt            = cmbx10 scaled \magstep2
%\font \tbts            = cmbx10   = \bfXf
 \let  \tbts=\bfXf
%\font \tbtss           = cmbx9    = \bfIXf
 \let  \tbtss=\bfIXf
 \font \headnotefont    = cmti10 scaled \magstep3
% This is ucgreek.aa
% the definition of versal greek characters
\mathchardef\Gamma  ="0000
\mathchardef\Delta  ="0001
\mathchardef\Theta  ="0002
\mathchardef\Lambda ="0003
\mathchardef\Xi     ="0004
\mathchardef\Pi     ="0005
\mathchardef\Sigma  ="0006
\mathchardef\Upsilon="0007
\mathchardef\Phi    ="0008
\mathchardef\Psi    ="0009
\mathchardef\Omega  ="000A
% This is aamisc.tex
% here miscellaneous macros are defined
%
% Makros to communicate with the user
\catcode`@=11 % use @ as a normal character
\newwrite\@info
\def\infuser#1{\immediate\write\@info{#1}}
\catcode`@=12 % at signs are no longer letters
\newlinechar=`\^^J % Zeilenumbruchhilfe fuer TeX-Meldungen am Bildschirm
\def\newline{\hfill\break}% makes a new line in the text :)
%-------------------------------------------------------------------
% draws a frame of height #1 cm times \hsize
\def\rahmen#1{\vbox{\hrule\line{\vrule\vbox to#1true
cm{\vfil}\hfil\vrule}\vfil\hrule}}
%-------------------------------------------------------------------
% shortcuts
\let\ts=\thinspace
\def\,{\relax\ifmmode\mskip\thinmuskip\else\thinspace\fi}
%-------------------------------------------------------------------
\def\unvskip{%
   \ifvmode
      \ifdim\lastskip=0pt
      \else
         \vskip-\lastskip
      \fi
   \fi}
% This is fleqn.tex
% it convinces TeX not to center the displayed formulas
\catcode`@=11    % use @ as a normal character
%
% allocate token registers for the equation and its number
\newtoks\eq\newtoks\eqn
%
% the horizontal size for a displayed formula is the actual \hsize
% minus the indention of the formula depending on the package
\newdimen\mathhsize
\def\calcmathhsize{\mathhsize=\hsize
\advance\mathhsize by-\mathindent}
\calcmathhsize
%
% define \eqalign@new, \displaylines@new, \(l)eqalignno@new to use this
% size but remember the old macros to make switching possible
\let\eqalign@old=\eqalign
\let\displaylines@old=\displaylines
\let\eqalignno@old=\eqalignno
\let\leqalignno@old=\leqalignno
\def\eqalign@new#1{\null\vcenter{\openup\jot\m@th
  \ialign{\strut\hfil$\displaystyle{##}$&$\displaystyle{{}##}$\hfil
      \crcr#1\crcr}}}
\def\displaylines@new#1{{}$\displ@y
\hbox{\vbox{\halign{$\@lign\hfil\displaystyle##\hfil$\crcr
    #1\crcr}}}${}}
\def\eqalignno@new#1{{}$\displ@y
  \hbox{\vbox{\halign
to\mathhsize{\hfil$\@lign\displaystyle{##}$\tabskip\z@skip
    &$\@lign\displaystyle{{}##}$\hfil\tabskip\centering
    &\llap{$\@lign##$}\tabskip\z@skip\crcr
    #1\crcr}}}${}}
\def\leqalignno@new#1{{}$\displ@y
\hbox{\vbox{\halign
to\mathhsize{\qquad\hfil$\@lign\displaystyle{##}$\tabskip\z@skip
    &$\@lign\displaystyle{{}##}$\hfil\tabskip\centering
    &\kern-\mathhsize\rlap{$\@lign##$}\tabskip\hsize\crcr
    #1\crcr}}}${}}
% main macro borrowed from the dirty tricks of the TeXbook
\def\generaldisplay{%
\ifeqno
       \ifleqno\leftline{$\displaystyle\the\eqn\quad\the\eq$}%
       \else\noindent\kern\mathindent\hbox to\mathhsize{$\displaystyle
             \the\eq\hfill\the\eqn$}%
       \fi
\else
       \kern\mathindent
       \hbox to\mathhsize{$\displaystyle\the\eq$\hss}%
\fi
\global\eq={}\global\eqn={}}%
%
% flags and defaults
\newif\ifeqno\newif\ifleqno
% check for the "$$"s
\def\displaysetup#1$${\displaytest#1\eqno\eqno\displaytest}
% look for equation numbers
\def\displaytest#1\eqno#2\eqno#3\displaytest{%
\if!#3!\ldisplaytest#1\leqno\leqno\ldisplaytest
\else\eqnotrue\leqnofalse\eqn={#2}\eq={#1}\fi
\generaldisplay$$}
\def\ldisplaytest#1\leqno#2\leqno#3\ldisplaytest{\eq={#1}%
\if!#3!\eqnofalse\else\eqnotrue\leqnotrue\eqn={#2}\fi}
%
% establish a switching between normal and flush left equations
%
\def\flushleftequations{%
   \infuser{^^JEquations are \ifx\Initexing\undefined now \fi
            typeset flush left.}%
   \everydisplay={\displaysetup}%
   \let\eqalign=\eqalign@new
   \let\displaylines=\displaylines@new
   \let\eqalignno=\eqalignno@new
   \let\leqalignno=\leqalignno@new}
\def\centeredequations{%
   \infuser{^^JEquations are \ifx\Initexing\undefined now \fi
            centered.}%
   \everydisplay={}%
   \let\eqalign=\eqalign@old
   \let\displaylines=\displaylines@old
   \let\eqalignno=\eqalignno@old
   \let\leqalignno=\leqalignno@old}
%
% set the default to flush left equations
\flushleftequations
\catcode`@=12 % at signs are no longer letters
% This is autnum.tex
% it defines a macro that steps and typesets a counter for equations
\newcount\eqnum\eqnum=0% register
%
% use: $$<formula>\eqno\autnum$$
%  or: $$\eqalignno{a&=b&\autnum\cr}$$
\def\autnum{\global\advance\eqnum by 1\relax{\rm(\the\eqnum)}}
% This is item.tex
% is defines a new \litem(item) with a left aligned argument
% and redefines plain \item(item)
%
\catcode`@=11    % use @ as a normal character

%
% \ch@ckitem(item)mark measures the width of the \item(item)s mark
% and issues a warning if it will not fit in the space provided
\def\ch@ckitemmark#1{\setbox0=\hbox{\enspace#1}%
\ifdim\wd0>\itemindent
   \infuser{^^J\string\item: Your mark `\string#1' is too wide. }%
\fi}
\def\ch@ckitemitemmark#1{\setbox0=\hbox{\kern\itemindent\enspace#1}%
\ifdim\wd0>\itemitemindent
   \infuser{^^J\string\itemitem: Your mark `\string#1' is too wide. }%
\fi}
%
% \set@item@mark is used to produce the \item's mark
\def\set@item@mark#1{\ch@ckitemmark{#1}%
\hbox to\itemindent{#1\hss}\ignorespaces}
% \set@itemitem@mark is used to produce the \itemitem's mark
\def\set@itemitem@mark#1{\ch@ckitemitemmark{#1}%
\dimen0=\itemitemindent
\advance\dimen0 by-\itemindent
\kern\itemindent\hbox to\dimen0{#1\hss}\ignorespaces}
%
% \setitem(item)indent takes its argument as the widest mark
% of an \item(item) and changes the \item(item)indent accordingly
\def\setitemindent#1{\setbox0=\hbox{\ignorespaces#1\unskip\enspace}%
\itemindent=\wd0\relax
\ifx\quiet\undefined
\infuser{^^J\string\setitemindent: Mark width modified to hold
         ^^J`\string#1' plus an \string\enspace\space gap. }%
\fi
}
\def\setitemitemindent#1{\setbox0=\hbox{\ignorespaces#1\unskip\enspace}%
\itemitemindent=\wd0\relax
\ifx\quiet\undefined
\infuser{^^J\string\setitemitemindent: Mark width modified to hold
         ^^J`\string#1' plus an \string\enspace\space gap. }%
\fi
\advance\itemitemindent by\itemindent}
%
% \item is redefined to produce a right aligned mark with
% a fixed gap, the hanging indentation has the width \itemindent.
% \itemitem is redefined to produce a left aligned mark with indention
% \itemindent, the hanging indentation has the width \itemitemindent.
% If there are flush left equations (\mathhsize is defined)
% that size is also corrected for use inside an \item(item).
\ifx\undefined\mathhsize
   \def\item{\par\noindent
   \hangindent\itemindent\hangafter=1\relax
   \set@item@mark}
   \def\itemitem{\par\noindent
   \hangindent\itemitemindent\hangafter=1\relax
   \set@itemitem@mark}
\else
   \def\item{\par\noindent\advance\mathhsize by-\itemindent
   \hangindent\itemindent\hangafter=1\relax
   \everypar={\global\mathhsize=\hsize
   \global\advance\mathhsize by-\mathindent
   \global\everypar={}}\set@item@mark}
   \def\itemitem{\par\noindent\advance\mathhsize by-\itemitemindent
   \hangindent\itemitemindent\hangafter=1\relax
   \everypar={\global\mathhsize=\hsize
   \global\advance\mathhsize by-\mathindent
   \global\everypar={}}\set@itemitem@mark}
\fi
\catcode`@=12 % at signs are no longer letters
% This is aatypset.tex
% it defines the final actions of the macro package
% i.e. printing of the identifying message and the
% characters invented by the author
\catcode`@=11    % use @ as a normal character
\newcount\the@end \global\the@end=0
\newbox\springer@macro \setbox\springer@macro=\vbox{}
\def\typeset{\setbox\springer@macro=\vbox{\begpet\noindent
   This article was processed by the author using
   Sprin\-ger-Ver\-lag \TeX{} A\&A macro package 1992.\par
   \egroup}\global\the@end=1}
% \show@special lists the contents of the (user defined) control
% sequence "specialn" where n is the lowercase roman numeral
% representation of the counter \footcount and then steps that counter.
\def\show@special{%
   \smallskip
   \noindent special character \#\number\sterne:
   \csname special\romannumeral\sterne\endcsname
   \advance\sterne by 1\relax
   \testspeci@l}
%
% \testspeci@l checks whether the control sequence "specialn" is
% defined, causing to print it and recursing to n+1 if yes,
% and aborting otherwise.
\def\testspeci@l{%
   \expandafter
   \ifx
      \csname special\romannumeral\sterne\endcsname
      \relax
      \let\next\relax
   \else
      \let\next\show@special
   \fi
   \next}
\outer\def\bye{\bigskip\typeset
\sterne=1 \testspeci@l
\if R\lr\null\fi\vfill\supereject\end}
\catcode`@=12    % reset catcode
% This is aalogo.tex
% it defines the A&A logo for the first page of an article
\def\AALogo{\setbox254=\hbox{ ASTROPHYSICS }%
\vbox{\baselineskip=10pt\hrule\hbox{\vrule\vbox{\kern3pt
\hbox to\wd254{\hfil ASTRONOMY\hfil}
\hbox to\wd254{\hfil AND\hfil}\copy254
\hbox to\wd254{\hfil\number\day.\number\month.\number\year\hfil}
\kern3pt}\vrule}\hrule}}
% This is aalegend.tex
% it defines the macro to produce figure captions
\def\figure#1#2{\medskip\noindent{\petit{\bf Fig.\ts#1.\
}\ignorespaces#2\par}}
% This is tabcap.tex
% it defines the macro to produce the table heads
% Tabellenkoepfe
\def\tabcap#1#2{\smallskip\noindent{\bf Table\ts\ignorespaces
#1\unskip.\ }\ignorespaces #2\vskip3mm}
% This is counters.tex
% it defines and initializes the counters needed
\catcode`@=11    % use @ as a normal character
\expandafter \newcount \csname c@Tl\endcsname
    \csname c@Tl\endcsname=0
\expandafter \newcount \csname c@Tm\endcsname
    \csname c@Tm\endcsname=0
\expandafter \newcount \csname c@Tn\endcsname
    \csname c@Tn\endcsname=0
\expandafter \newcount \csname c@To\endcsname
    \csname c@To\endcsname=0
\expandafter \newcount \csname c@Tp\endcsname
    \csname c@Tp\endcsname=0
\expandafter \newcount \csname c@fn\endcsname
    \csname c@fn\endcsname=0
\def \stepc#1    {\global
    \expandafter
    \advance
    \csname c@#1\endcsname by 1}
\def \resetcount#1    {\global
    \csname c@#1\endcsname=0}
\def\@nameuse#1{\csname #1\endcsname}
\def\arabic#1{\@arabic{\@nameuse{c@#1}}}
\def\@arabic#1{\ifnum #1>0 \number #1\fi}
\catcode`@=12    % reset catcode
% This is maintitl.aa
% it defines the macros to start and end the main title
\catcode`@=11 % use @ as a normal character
\def \aTa  { \goodbreak
\bgroup
\par
\let\bf=\relax\let\boldmath=\relax\let\bffam=\z@
\textfont0=\tatt \scriptfont0=\tats \scriptscriptfont0=\tatss
\textfont1=\tamt \scriptfont1=\tams \scriptscriptfont1=\tamss
\textfont2=\tast \scriptfont2=\tass \scriptscriptfont2=\tasss
\ifx\grfam\undefined\else
   \textfont\grfam=\tagrt \scriptfont\grfam=\tagrs
   \scriptscriptfont\grfam=\tagrss
\fi
\baselineskip=\tabaselineskip
\lineskiplimit=0pt\lineskip=0pt
\rightskip=0pt plus4cm
\pretolerance=10000
\noindent
\tatt}
% --------------------------------------------------------------
% Ende Ueberschrift
%
\def\eTa{\vskip10pt\egroup
         \noindent
         \ignorespaces}
\catcode`@=12 % at signs are no longer letters
% This is subtitl.tex
% it defines the macros to start and end the subtitle
\catcode`@=11 % use @ as a normal character
\def\aTb{\goodbreak
\bgroup
\par
\let\bf=\relax\let\boldmath=\relax\let\bffam=\z@
\textfont0=\tbtt \scriptfont0=\tbts \scriptscriptfont0=\tbtss
\textfont1=\tbmt \scriptfont1=\tbms \scriptscriptfont1=\tbmss
\textfont2=\tbst \scriptfont2=\tbss \scriptscriptfont2=\tbsss
\ifx\grfam\undefined\else
   \textfont\grfam=\tbgrt \scriptfont\grfam=\tbgrs
   \scriptscriptfont\grfam=\tbgrss
\fi
\baselineskip=\tbbaselineskip
\lineskip=0pt\lineskiplimit=0pt
\rightskip=0pt plus4cm
\pretolerance=10000
\noindent
\tbtt}
% ---------------------------------------------------------------
% Ende Ueberschrift 2. Ordnung
%
\def\eTb{\vskip10pt
    \egroup
    \noindent
    \ignorespaces}
\catcode`@=12 % at signs are no longer letters
% This is aatita.tex
% it defines the heading of first order that is numbered automatically
\catcode`@=11    % use @ as a normal character
\newcount\section@penalty  \section@penalty=0
\newcount\subsection@penalty  \subsection@penalty=0
\newcount\subsubsection@penalty  \subsubsection@penalty=0
\def\titlea#1{\par\stepc{Tl}
    \resetcount{Tm}
    \bgroup
       \normalsize
       \bf \rightskip 0pt plus4em
       \pretolerance=20000
       \boldmath
       \setbox0=\vbox{\vskip\tabefore
          \noindent
          \arabic{Tl}.\
          \ignorespaces#1
          \vskip\taafter}
       \dimen0=\ht0\advance\dimen0 by\dp0
       \advance\dimen0 by 2\baselineskip
       \advance\dimen0 by\pagetotal
       \ifdim\dimen0>\pagegoal
          \ifdim\pagetotal>\pagegoal
          \else\eject\fi\fi
       \vskip\tabefore
       \penalty\section@penalty \global\section@penalty=-200
       \global\subsection@penalty=10007
       \noindent
       \arabic{Tl}.\
       \ignorespaces#1\par
    \egroup
    \nobreak
    \vskip\taafter
    \parindent=0pt
    \let\lasttitle=A%
\everypar={\parindent=\stdparindent
    \penalty\z@\let\lasttitle=N\everypar={}}%
    \ignorespaces}
\catcode`@=12    % reset catcode
% This is aatitb.tex
% it defines the heading of second order that is numbered automatically
\catcode`@=11    % use @ as a normal character
\def\titleb#1{\par\stepc{Tm}
    \resetcount{Tn}
    \if N\lasttitle\else\vskip\tbbeforeback\fi
    \bgroup
       \normalsize
       \raggedright
       \pretolerance=10000
       \it
       \setbox0=\vbox{\vskip\tbbefore
          \normalsize
          \raggedright
          \pretolerance=10000
          \noindent \it \arabic{Tl}.\arabic{Tm}.\ \ignorespaces#1
          \vskip\tbafter}
       \dimen0=\ht0\advance\dimen0 by\dp0\advance\dimen0 by 2\baselineskip
       \advance\dimen0 by\pagetotal
       \ifdim\dimen0>\pagegoal
          \ifdim\pagetotal>\pagegoal
          \else \if N\lasttitle\eject\fi \fi\fi
       \vskip\tbbefore
       \if N\lasttitle \penalty\subsection@penalty \fi
       \global\subsection@penalty=-100
       \global\subsubsection@penalty=10007
       \noindent \arabic{Tl}.\arabic{Tm}.\ \ignorespaces#1\par
    \egroup
    \nobreak
    \vskip\tbafter
    \let\lasttitle=B%
    \parindent=0pt
    \everypar={\parindent=\stdparindent
       \penalty\z@\let\lasttitle=N\everypar={}}%
       \ignorespaces}
\catcode`@=12    % reset catcode
% This is aatitc.tex
% it defines the heading of third order that is numbered automatically
\catcode`@=11    % use @ as a normal character
\def\titlec#1{\par\stepc{Tn}
    \resetcount{To}
    \if N\lasttitle\else\vskip\tcbeforeback\fi
    \bgroup
       \normalsize
       \raggedright
       \pretolerance=10000
       \setbox0=\vbox{\vskip\tcbefore
          \noindent
          \arabic{Tl}.\arabic{Tm}.\arabic{Tn}.\
          \ignorespaces#1\vskip\tcafter}
       \dimen0=\ht0\advance\dimen0 by\dp0\advance\dimen0 by 2\baselineskip
       \advance\dimen0 by\pagetotal
       \ifdim\dimen0>\pagegoal
           \ifdim\pagetotal>\pagegoal
           \else \if N\lasttitle\eject\fi \fi\fi
       \vskip\tcbefore
       \if N\lasttitle \penalty\subsubsection@penalty \fi
       \global\subsubsection@penalty=-50
       \noindent
       \arabic{Tl}.\arabic{Tm}.\arabic{Tn}.\
       \ignorespaces#1\par
    \egroup
    \nobreak
    \vskip\tcafter
    \let\lasttitle=C%
    \parindent=0pt
    \everypar={\parindent=\stdparindent
       \penalty\z@\let\lasttitle=N\everypar={}}%
       \ignorespaces}
\catcode`@=12    % reset catcode
% This is aatitd.tex
% it defines the heading of fourth order that is numbered automatically
\def\titled#1{\par\stepc{To}
    \resetcount{Tp}
    \if N\lasttitle\else\vskip\tdbeforeback\fi
    \vskip\tdbefore
    \bgroup
       \normalsize
       \if N\lasttitle \penalty-50 \fi
       \it \noindent \ignorespaces#1\unskip\
    \egroup\ignorespaces}
% This is referenc.tex
% it defines the macros needed to produce the references
%
\def\begref#1{\par
   \unvskip
   \goodbreak\vskip\tabefore
   {\noindent\bf\ignorespaces#1%
   \par\vskip\taafter}\nobreak\let\INS=N}
\def\ref{\if N\INS\let\INS=Y\else\goodbreak\fi
   \hangindent\stdparindent\hangafter=1\noindent\ignorespaces}
\def\endref{\goodbreak}% Ende der Referenzen
% This is acknow.tex
% it prints the acknowledgement text
\def\acknow#1{\par
   \unvskip
   \vskip\tcbefore
   \noindent{\it Acknowledgements\/}. %
   \ignorespaces#1\par
   \vskip\tcafter}
% This is appendix.aa
% it starts an appendix section without automatic numeration
\def\appendix#1{\vskip\tabefore
    \vbox{\noindent{\bf Appendix #1}\vskip\taafter}%
    \global\eqnum=0\relax
    \nobreak\noindent\ignorespaces}
% This is referee.tex
% it allows the author to produce a referees copy
% single column wider line spacing
\newbox\refereebox
\setbox\refereebox=\vbox
to0pt{\vskip0.5cm\fullline{\hrulefill\tentt\lower0.5ex
\hbox{\kern5pt referee's copy\kern5pt}\hrulefill}\vss}%
\def\refereelayout{\let\REFEREE=Y\footline={\copy\refereebox}%
    \infuser{^^JA referee's copy will be produced}\par
    \if N\lr\else\onecolumn\topskip=10pt\fi
    \normalsize
    \calcmathhsize}
% This is symbols.tex
% the symbols not available in plain TeX are constructed
% by overprinting some characters

\def\sq{\hbox{\rlap{$\sqcap$}$\sqcup$}}
\def\degr{\hbox{$^\circ$}}
\def\arcmin{\hbox{$^\prime$}}
\def\arcsec{\hbox{$^{\prime\prime}$}}
\def\utw{\smash{\rlap{\lower5pt\hbox{$\sim$}}}}
\def\udtw{\smash{\rlap{\lower6pt\hbox{$\approx$}}}}

\def\farcm{\hbox{$.\mkern-4mu^\prime$}}

 %reelle Zahlen
 %natuerliche Zahlen

\def\bbbc{{\mathchoice {\setbox0=\hbox{$\displaystyle\rm C$}\hbox{\hbox
to0pt{\kern0.4\wd0\vrule height0.9\ht0\hss}\box0}}
{\setbox0=\hbox{$\textstyle\rm C$}\hbox{\hbox
to0pt{\kern0.4\wd0\vrule height0.9\ht0\hss}\box0}}
{\setbox0=\hbox{$\scriptstyle\rm C$}\hbox{\hbox
to0pt{\kern0.4\wd0\vrule height0.9\ht0\hss}\box0}}
{\setbox0=\hbox{$\scriptscriptstyle\rm C$}\hbox{\hbox
to0pt{\kern0.4\wd0\vrule height0.9\ht0\hss}\box0}}}}
\def\bbbq{{\mathchoice {\setbox0=\hbox{$\displaystyle\rm Q$}\hbox{\raise
0.15\ht0\hbox to0pt{\kern0.4\wd0\vrule height0.8\ht0\hss}\box0}}
{\setbox0=\hbox{$\textstyle\rm Q$}\hbox{\raise
0.15\ht0\hbox to0pt{\kern0.4\wd0\vrule height0.8\ht0\hss}\box0}}
{\setbox0=\hbox{$\scriptstyle\rm Q$}\hbox{\raise
0.15\ht0\hbox to0pt{\kern0.4\wd0\vrule height0.7\ht0\hss}\box0}}
{\setbox0=\hbox{$\scriptscriptstyle\rm Q$}\hbox{\raise
0.15\ht0\hbox to0pt{\kern0.4\wd0\vrule height0.7\ht0\hss}\box0}}}}
\def\bbbt{{\mathchoice {\setbox0=\hbox{$\displaystyle\rm
T$}\hbox{\hbox to0pt{\kern0.3\wd0\vrule height0.9\ht0\hss}\box0}}
{\setbox0=\hbox{$\textstyle\rm T$}\hbox{\hbox
to0pt{\kern0.3\wd0\vrule height0.9\ht0\hss}\box0}}
{\setbox0=\hbox{$\scriptstyle\rm T$}\hbox{\hbox
to0pt{\kern0.3\wd0\vrule height0.9\ht0\hss}\box0}}
{\setbox0=\hbox{$\scriptscriptstyle\rm T$}\hbox{\hbox
to0pt{\kern0.3\wd0\vrule height0.9\ht0\hss}\box0}}}}
\def\bbbs{{\mathchoice
{\setbox0=\hbox{$\displaystyle     \rm S$}\hbox{\raise0.5\ht0\hbox
to0pt{\kern0.35\wd0\vrule height0.45\ht0\hss}\hbox
to0pt{\kern0.55\wd0\vrule height0.5\ht0\hss}\box0}}
{\setbox0=\hbox{$\textstyle        \rm S$}\hbox{\raise0.5\ht0\hbox
to0pt{\kern0.35\wd0\vrule height0.45\ht0\hss}\hbox
to0pt{\kern0.55\wd0\vrule height0.5\ht0\hss}\box0}}
{\setbox0=\hbox{$\scriptstyle      \rm S$}\hbox{\raise0.5\ht0\hbox
to0pt{\kern0.35\wd0\vrule height0.45\ht0\hss}\raise0.05\ht0\hbox
to0pt{\kern0.5\wd0\vrule height0.45\ht0\hss}\box0}}
{\setbox0=\hbox{$\scriptscriptstyle\rm S$}\hbox{\raise0.5\ht0\hbox
to0pt{\kern0.4\wd0\vrule height0.45\ht0\hss}\raise0.05\ht0\hbox
to0pt{\kern0.55\wd0\vrule height0.45\ht0\hss}\box0}}}}
\def\bbbz{{\mathchoice {\hbox{$\sf\textstyle Z\kern-0.4em Z$}}
{\hbox{$\sf\textstyle Z\kern-0.4em Z$}}
{\hbox{$\sf\scriptstyle Z\kern-0.3em Z$}}
{\hbox{$\sf\scriptscriptstyle Z\kern-0.2em Z$}}}}
\def\diameter{{\ifmmode\oslash\else$\oslash$\fi}}
%
% redefine \Re and \Im

%

% This is vec.tex
% it defines vectors in boldface math and tensors in sans serif style
%
\catcode`@=11 % use @ as a normal character
\def\vec#1{{\boldmath\bf
\textfont\z@=\textfont\bffam\scriptfont\z@=\scriptfont\bffam
\scriptscriptfont\z@=\scriptscriptfont\bffam
\ifmmode
\mathchoice{\hbox{$\displaystyle#1$}}{\hbox{$\textstyle#1$}}
{\hbox{$\scriptstyle#1$}}{\hbox{$\scriptscriptstyle#1$}}\else
$#1$\fi}}
\def\tens#1{\relax\ifmmode
\mathchoice{\hbox{$\displaystyle\sf#1$}}{\hbox{$\textstyle\sf#1$}}
{\hbox{$\scriptstyle\sf#1$}}{\hbox{$\scriptscriptstyle\sf#1$}}\else
$\sf#1$\fi}
\catcode`@=12 % at signs are no longer letters
% This is titsetup.tex
% it defines variables and macros for the title page
% Variablenvereinbarung fuer Titelseitenautomatik
\newcount\sterne \sterne=0
\newdimen\fullhead
{\catcode`@=11    % use @ as a normal character
\def\newtoks{\alloc@5\toks\toksdef\@cclvi}
\outer\gdef\makenewtoks#1{\newtoks#1#1={ ????? }}}
\makenewtoks\DATE
\makenewtoks\MAINTITLE
\makenewtoks\SUBTITLE
\makenewtoks\AUTHOR
\makenewtoks\INSTITUTE
\makenewtoks\ABSTRACT
\makenewtoks\KEYWORDS
\makenewtoks\THESAURUS
\makenewtoks\OFFPRINTS
\makenewtoks\HEADNOTE
% Klammeraffe(@-Zeichen) wird Schluesselbuchstabe fuer affiliations
% solange Titel-page aktiv ist (Benutzung in AUTHOR und INSTITUTE)
\let\INS=N%
{\catcode`\@=\active
% Aktionen, die bei Antreffen des @-Zeichens zu machen sind;
% zwei Faelle a) @ bei AUTHOR, b) @ bei INSTITUTE
\gdef@#1{%
\if N\INS
   $^{#1}$%
\else
   \let\INS=Y%
   \par
   \noindent\hbox to0.5\stdparindent{$^{#1}$\hfil}\ignorespaces
\fi}%
}% @ wieder normal
% Automatische Fussnotennumerierung mit wachsender Anzahl Sterne
\def\mehrsterne{\global\advance\sterne by1\relax}%
% This is footnote.tex
% it defines all footnote macros
\def\footnoterule{\kern-3pt\hrule width 2true cm\kern2.6pt}% Trennlinie
%-------------------------------------------------------------------------------
\def\makeOFFPRINTS#1{\bgroup\let\REFEREE=N \normalsize
       \if N\lr\hsize=\fullhsize\else\hsize=\halfsize\fi
       \lineskiplimit=0pt\lineskip=0pt
       \def\textindent##1{\noindent{\it Send offprint
          requests to\/}: }\relax
       \vfootnote{nix}{\ignorespaces#1}\egroup}
%-------------------------------------------------------------------------------
% Fussnote mit stern(*) gekennzeichnet
\def\makesterne{\count254=0\loop\ifnum\count254<\sterne
\advance\count254 by1\star\repeat}
\def\FOOTNOTE#1{\bgroup\let\REFEREE=N
       \ifhmode\unskip\fi
       \mehrsterne$^{\makesterne}$\relax
       \normalsize
       \if N\lr\hsize=\fullhsize\else\hsize=\halfsize\fi
       \lineskiplimit=0pt\lineskip=0pt
       \def\textindent##1{\noindent\hbox
       to\stdparindent{##1\hss}}\relax
       \vfootnote{$^{\makesterne}$}{\ignorespaces#1}\egroup}
%

%
% Automatisch numerierte Fussnote
\def\fonote#1{\relax\ifhmode\unskip\fi
       \mehrsterne$^{\the\sterne}$\bgroup
       \normalsize
       \if N\lr\hsize=\fullhsize\else\hsize=\halfsize\fi
       \def\textindent##1{\noindent\hbox
       to\stdparindent{##1\hss}}\relax
       \vfootnote{$^{\the\sterne}$}{\ignorespaces#1}\egroup}
% This is maketitl.tex
% it defines all macros to construct the title page
%
\def\missmsg#1{\infuser{^^JMissing #1 }}
%% test if missing:
%
\def\tstmiss#1#2#3#4#5{%
\edef\test{\the #1}%
\ifx\test\missing%
  #2\relax%  infuser
  #3%   action if missing
\else
  \ifx\test\missingi%
    #2\relax%  infuser
    #3%   action if missing
  \else #4%  action if existing
  \fi
\fi
#5%   action at any rate
}%
\def\strich{\par
\vbox to0pt{\hrule width\hsize\vss}\vskip-1.2\baselineskip
\vskip0pt plus3\baselineskip\relax}%
%

%
% Hauptmacro fuer automatische Titelseite
\def\maketitle{\paglay%
\def\missing{ ????? }% Schablone zur Erkennung von nicht ausgefuellten
\def\missingi{ }% Schablone zur Erkennung von nicht ausgefuellten
% Feldern
%
% Methode zum Test auf Leerfelder: Variable wird mit \missing=" ????? "
% verglichen; Fallunterscheidung.
% \edef\test{\the\VARIABLE} macht die Variable dem Vergleich zugaenglich
% \ifx\test = \missing    d.h. steht in Variable der Text: " ????? "?
% !!! Aktion fuer Fall ja, hier wurde also nichts eingetragen !!!
% \else !!! Fall fuer ausgefuellte Variable!!! \fi
{\parskip=0pt\relax
\setbox0=\vbox{\hsize=\fullhsize\null\vskip2truecm
\tstmiss%
  {\HEADNOTE}%
  {}%
  {}%
  {%   write HEADNOTE:
   \noindent{\headnotefont\ignorespaces\the\HEADNOTE}\vskip8mm
   }%
  {}%
\tstmiss%
  {\MAINTITLE}%
  {}%
  {\global\MAINTITLE={MAINTITLE should be given}}%
  {}%
  {%   write MAINTITLE:
   \aTa\ignorespaces\the\MAINTITLE\eTa}%
\tstmiss%
  {\SUBTITLE}%
  {}%
  {}%
  {%   write SUBTITLE:
   \aTb\ignorespaces\the\SUBTITLE\eTb}%
  {}%
\tstmiss%
  {\AUTHOR}%
  {}%
  {\AUTHOR={Name(s) and initial(s) of author(s) should be given}}
  {}%
  {%   write AUTHOR:
\noindent{\bf\ignorespaces\the\AUTHOR\vskip4pt}}%
\tstmiss%
  {\INSTITUTE}%
  {}%
  {\INSTITUTE={Address(es) of author(s) should be given.}}%
  {}%
  {%   write INSTITUTE:
   \let\INS=E
\noindent\ignorespaces\the\INSTITUTE\vskip10pt}%
\tstmiss%
  {\DATE}%
  {}%
  {\DATE={$[$the date of receipt and acceptance should be inserted
later$]$}}%
  {}%
  {%   write DATE:
{\noindent\ignorespaces\the\DATE\vskip21pt}\bf A}%
}%
\global\fullhead=\ht0\global\advance\fullhead by\dp0
\global\advance\fullhead by10pt\global\sterne=0
{\hsize=\halfsize\null\vskip2truecm
\tstmiss%
  {\OFFPRINTS}%
  {}%
  {}%
  {\makeOFFPRINTS{\the\OFFPRINTS}}%
  {}%
\hsize=\fullhsize
\tstmiss%
  {\HEADNOTE}%
  {}%
  {}%
  {%   write HEADNOTE:
   \noindent{\headnotefont\ignorespaces\the\HEADNOTE}\vskip8mm
   }%
  {}%
\tstmiss%
  {\MAINTITLE}%
  {\missmsg{MAINTITLE}}%
  {\global\MAINTITLE={MAINTITLE should be given}}%
  {}%
  {%   write MAINTITLE:
   \aTa\ignorespaces\the\MAINTITLE\eTa}%
\tstmiss%
  {\SUBTITLE}%
  {}%
  {}%
  {%   write SUBTITLE:
   \aTb\ignorespaces\the\SUBTITLE\eTb}%
  {}%
\tstmiss%
  {\AUTHOR}%
  {\missmsg{name(s) and initial(s) of author(s)}}%
  {\AUTHOR={Name(s) and initial(s) of author(s) should be given}}
  {}%
  {%   write AUTHOR:
\noindent{\bf\ignorespaces\the\AUTHOR\vskip4pt}}%
\tstmiss%
  {\INSTITUTE}%
  {\missmsg{address(es) of author(s)}}%
  {\INSTITUTE={Address(es) of author(s) should be given.}}%
  {}%
  {%   write INSTITUTE:
   \let\INS=E
\noindent\ignorespaces\the\INSTITUTE\vskip10pt}%
\tstmiss%
  {\DATE}%
  {\infuser{^^JThe date of receipt and acceptance should be inserted
later.}}%
  {\DATE={$[$the date of receipt and acceptance should be inserted
later$]$}}%
  {}%
  {%   write DATE:
{\noindent\ignorespaces\the\DATE\vskip21pt}}%
}%
\tstmiss%
  {\THESAURUS}%
  {\infuser{^^JThesaurus codes are not given.}}%
  {\global\THESAURUS={missing; you have not inserted them}}%
  {}%
  {}%
\normalsize
\tstmiss%
  {\ABSTRACT}%
  {\missmsg{ABSTRACT}}%
  {\ABSTRACT={Not yet given.}}%
  {}%
  {\noindent{\bf Abstract. }\ignorespaces\the\ABSTRACT\vskip0.5true cm}%
\tstmiss%
  {\KEYWORDS}%
  {\missmsg{KEYWORDS}}%
  {\KEYWORDS={Not yet given.}}%
  {}%
  {\noindent{\bf Key words: }\ignorespaces\the\KEYWORDS
  \strich}%
\global\sterne=0
\global\catcode`\@=12
}\normalsize}%Ende von maketitle
\catcode`@=11    % use @ as a normal character
\newdimen\@txtwd  \@txtwd=\hsize
\newdimen\@txtht  \@txtht=\vsize
\newdimen\@colht  \@colht=\vsize
\newdimen\@colwd  \@colwd=-1pt
\newdimen\@colsavwd
%%%
%%% =========== Macros for initializing the whole thing ==========
%%%
\newcount\in@t \in@t=0
\def\initlr{\if N\lr \ifdim\@colwd<0pt \global\@colwd=\hsize \fi
   \else\global\let\lr=L\ifdim\@colwd<0pt \global\@colwd=\hsize
      \global\divide\@colwd\tw@ \global\advance\@colwd by -10pt
   \fi\fi\global\advance\in@t by 1}
\def\setuplr#1#2#3{\let\lr=O \ifx#1\lr\global\let\lr=N
      \else\global\let\lr=L\fi
   \@txtht=\vsize \@colht=\vsize \@txtwd=#2 \@colwd=#3
   \if N\lr \else\multiply\@colwd\tw@ \fi
   \ifdim\@colwd>\@txtwd\if N\lr
        \errmessage{The text width is less than the column width}%
      \else
        \errmessage{The text width is less the two times the column width}%
      \fi \global\@colwd=\@txtwd
      \if N\lr\divide\@colwd by 2\fi
   \else \global\@colwd=#3 \fi \initlr \@colsavwd=#3
   \global\@insmx=\@txtht
   \global\hsize=\@colwd}
%% ----------- switching between one and two column output ------
\def\twocolumns{\@fillpage\eject\global\let\lr=L \@makecolht
   \global\@colwd=\@colsavwd \global\hsize=\@colwd}
\def\onecolumn{\@fillpage\eject\global\let\lr=N \@makecolht
   \global\@colwd=\@txtwd \global\hsize=\@colwd}
\def\newpage{\@fillpage\eject}
\def\@fillpage{\vfill\supereject\if R\lr \null\vfill\eject\fi}

%% ----------------------------------------------------------------
%%%
%%% =============== Macros used by the output routine ===============
%%%
\newbox\@leftcolumn
\newbox\@rightcolumn
\newbox\@outputbox
\newbox\@tempboxa
\newbox\@keepboxa
\newbox\@keepboxb
\newbox\@bothcolumns
\newbox\@savetopins
\newbox\@savetopright
\newcount\verybad \verybad=1010
%% -------- \@makecolumn puts the current column in a box ---------
\def\@makecolumn{\ifnum \in@t<1\initlr\fi
   \ifnum\outputpenalty=\the\verybad1  %%% i.e. 10101 if \verybad=1010
      \if L\lr\else\advance\pageno by1\fi
      \infuser{Warning: There is a 'widow' line
      at the top of page \the\pageno\if R\lr (left)\fi.
      This is unacceptable.} \if L\lr\else\advance\pageno by-1\fi \fi
   \ifnum\outputpenalty=\the\verybad2
      \infuser{Warning: There is a 'club' line
      at the bottom of page \the\pageno\if L\lr(left)\fi.
      This is unacceptable.} \fi
   \if L\lr \ifvoid\@savetopins\else\@colht=\@txtht\fi \fi
   \if R\lr \ifvoid\@bothcolumns \ifvoid\@savetopright
       \else\@colht=\@txtht\fi\fi\fi
   \global\setbox\@outputbox
   \vbox to\@colht{\boxmaxdepth\maxdepth
   % One-column top insertions are held back if there is already a
   % two-column floating insertion and the one-column top insertion
   % doesn't fit entirely in the column.
   \if L\lr \ifvoid\@savetopins\else\unvbox\@savetopins\fi \fi
   \if R\lr \ifvoid\@bothcolumns \ifvoid\@savetopright\else
       \unvbox\@savetopright\fi\fi\fi
   \ifvoid\topins\else\ifnum\count\topins>0
         \ifdim\ht\topins>\@colht
            \infuser{^^JError: Too many or too large single column
            box(es) on this page.}\fi
         \unvbox\topins
      \else
         \global\setbox\@savetopins=\vbox{\ifvoid\@savetopins\else
         \unvbox\@savetopins\penalty-500\fi \unvbox\topins} \fi\fi
   \dimen@=\dp\@cclv \unvbox\@cclv % open up \box255
   \ifvoid\bottomins\else\unvbox\bottomins\fi
   \ifvoid\footins\else % footnote info is present
     \vskip\skip\footins
     \footnoterule
     \unvbox\footins\fi
   \ifr@ggedbottom \kern-\dimen@ \vfil \fi}%
}
%% --------- \@outputpage puts the columns and the top insertions
%% --------- together and puts them out
\def\@outputpage{\@dooutput{\lr}}
\def\@colbox#1{\hbox to\@colwd{\box#1\hss}}
\def\@dooutput#1{\global\topskip=10pt
  \ifdim\ht\@bothcolumns>\@txtht
    \if #1N
       \unvbox\@outputbox
    \else
       \unvbox\@leftcolumn\unvbox\@outputbox
    \fi
    \global\setbox\@tempboxa\vbox{\hsize=\@txtwd\makeheadline
       \vsplit\@bothcolumns to\@txtht
       \makefootline\hsize=\@colwd}%
    \infuser{^^JError: Too many double column boxes on this page.}%
    \shipout\box\@tempboxa\advancepageno
    \unvbox255 \penalty\outputpenalty
  \else
    \global\setbox\@tempboxa\vbox{\hsize=\@txtwd\makeheadline
       \ifvoid\@bothcolumns\else\unvbox\@bothcolumns\fi
       \hsize=\@colwd
       \if #1N
          \hbox to\@txtwd{\@colbox{\@outputbox}\hfil}%
       \else
          \hbox to\@txtwd{\@colbox{\@leftcolumn}\hfil\@colbox{\@outputbox}}%
       \fi
       \hsize=\@txtwd\makefootline\hsize=\@colwd}%
    \shipout\box\@tempboxa\advancepageno
  \fi
  \ifnum \special@pages>0 \s@count=100 \page@command
      \xdef\page@command{}\global\special@pages=0 \fi
  }
%% -------- \balance@right@left balances the columns on the last
%% -------- page of text.
\def\balance@right@left{\dimen@=\ht\@leftcolumn
    \advance\dimen@ by\ht\@outputbox
    \advance\dimen@ by\ht\springer@macro
    \dimen2=\z@ \global\the@end=0
    % put both columns together and compensate \vfill at the end
    \ifdim\dimen@>70pt\setbox\z@=\vbox{\unvbox\@leftcolumn
          \unvbox\@outputbox}%
       \loop
          \dimen@=\ht\z@
          \advance\dimen@ by0.5\topskip
          \advance\dimen@ by\baselineskip
          \advance\dimen@ by\ht\springer@macro
          \advance\dimen@ by\dimen2
          \divide\dimen@ by2
          \splittopskip=\topskip
          % Now split it to two parts of about the same height
          {\vbadness=10000
             \global\setbox3=\copy\z@
             \global\setbox1=\vsplit3 to \dimen@}%
          \dimen1=\ht3 \advance\dimen1 by\ht\springer@macro
       \ifdim\dimen1>\ht1 \advance\dimen2 by\baselineskip\repeat
       \dimen@=\ht1
       % Restore the column boxes and adjust
       \global\setbox\@leftcolumn
          \hbox to\@colwd{\vbox to\@colht{\vbox to\dimen@{\unvbox1}\vfil}}%
       \global\setbox\@outputbox
          \hbox to\@colwd{\vbox to\@colht{\vbox to\dimen@{\unvbox3
             \vfill\box\springer@macro}\vfil}}%
    \else
       \setbox\@leftcolumn=\vbox{unvbox\@leftcolumn\bigskip
          \box\springer@macro}%
    \fi}
%
%%%
%%% ================== Insertion routines ======================
%%%
%%% This follows loosely the definition of \topins by Knuth but without
%%% the need to distinguish between 3 different kinds of insertions.
%%% See the TeXBook p.363.
%%% Insertions in right columns are first saved in a box (\rightins)
%%% and inserted to \bothins after this column has been shipped out.
%
\newinsert\bothins
\newbox\rightins
\skip\bothins=\z@skip
\count\bothins=1000
\dimen\bothins=\@txtht \advance\dimen\bothins by -\bigskipamount
\def\bothtopinsert{\par\begingroup\setbox\z@\vbox\bgroup
    \hsize=\@txtwd\parskip=0pt\par\noindent\bgroup}
\def\endbothinsert{\egroup\egroup
  \if R\lr
    \right@nsert
  \else    % L\lr or N\lr
    % If double column insertions don't fit into the current columm
    % keep them until the next page starts.
    \dimen@=\ht\z@ \advance\dimen@ by\dp\z@ \advance\dimen@ by\pagetotal
    \advance\dimen@ by \bigskipamount \advance\dimen@ by \topskip
    \advance\dimen@ by\ht\topins \advance\dimen@ by\dp\topins
    \advance\dimen@ by\ht\bottomins \advance\dimen@ by\dp\bottomins
    \advance\dimen@ by\ht\@savetopins \advance\dimen@ by\dp\@savetopins
    \ifdim\dimen@>\@colht\right@nsert\else\left@nsert\fi
  \fi  \endgroup}
\def\right@nsert{\global\setbox\rightins\vbox{\ifvoid\rightins
    \else\unvbox\rightins\fi\penalty100
    \splittopskip=\topskip
    \splitmaxdepth\maxdimen \floatingpenalty200
    \dimen@\ht\z@ \advance\dimen@\dp\z@
    \box\z@\nobreak\bigskip}}
\def\left@nsert{\insert\bothins{\penalty100
    \splittopskip=\topskip
    \splitmaxdepth\maxdimen \floatingpenalty200
    \box\z@\nobreak\bigskip}
    \@makecolht}
\newdimen\@insht    \@insht=\z@
\newdimen\@insmx    \@insmx=\vsize
%% ------ \@makecolht computes the available height of the current column.
\def\@makecolht{\global\@colht=\@txtht \@compinsht
    \global\advance\@colht by -\@insht \global\vsize=\@colht
    \global\dimen\topins=\@colht}
\def\@compinsht{\if R\lr
       \dimen@=\ht\@bothcolumns \advance\dimen@ by\dp\@bothcolumns
       \ifvoid\@bothcolumns \advance\dimen@ by\ht\@savetopright
          \advance\dimen@ by\dp\@savetopright \fi
    \else
       \dimen@=\ht\bothins \advance\dimen@ by\dp\bothins
       \advance\dimen@ by\ht\@savetopins \advance\dimen@ by\dp\@savetopins
    \fi
    \ifdim\dimen@>\@insmx
       \global\@insht=\dimen@
    \else\global\@insht=\dimen@
    \fi}
\newinsert\bottomins
\skip\bottomins=\z@skip
\count\bottomins=1000
%%%
%%% Special pages to be inserted
%%%
\xdef\page@command{}
\newcount\s@count
\newcount\special@pages \special@pages=0
\def\specialpage#1{\global\advance\special@pages by1
    \global\s@count=\special@pages
    \global\advance\s@count by 100
    \global\setbox\s@count
    \vbox to\@txtht{\hsize=\@txtwd\parskip=0pt
    \par\noindent\noexpand#1\vfil}%
    \def\protect{\noexpand\protect\noexpand}%
    \xdef\page@command{\page@command
         \protect\global\advance\s@count by1
         \protect\begingroup
         \protect\setbox\z@\vbox{\protect\makeheadline
                                    \protect\box\s@count
            \protect\makefootline}%
         \protect{\shipout\box\z@}%
         \protect\endgroup\protect\advancepageno}%
    \let\protect=\relax
   }
%%%
%%%
%%% This little macro adjusts the top of figure boxes with the
%%% the top of the column. Previously they were adjusted with the
%%% the baseline of the first row in a column.
\def\@startins{\vskip \topskip\hrule height\z@
   \nobreak\vskip -\topskip\vskip3.7pt}
%%%
%%%  ==============  The main output routine ===============
%%%
%%%           The output routine was adapted to A&A
%%%
\let\retry=N
\output={\@makecolht \global\topskip=10pt \let\retry=N%
   \ifnum\count\topins>0 \ifdim\ht\topins>\@colht
       \global\count\topins=0 \global\let\retry=Y%
       \unvbox\@cclv \penalty\outputpenalty \fi\fi
   \if N\retry
    \if N\lr     % this is for single column output
       \@makecolumn
       \ifnum\the@end>0
          \setbox\z@=\vbox{\unvcopy\@outputbox}%
          \dimen@=\ht\z@ \advance\dimen@ by\ht\springer@macro
          \ifdim\dimen@<\@colht
             \setbox\@outputbox=\vbox to\@colht{\box\z@
             \unskip\vskip12pt plus0pt minus12pt
             \box\springer@macro\vfil}%
          \else \box\springer@macro \fi
          \global\the@end=0
       \fi
       \ifvoid\bothins\else\global\setbox\@bothcolumns\box\bothins\fi
       \@outputpage
       \ifvoid\rightins\else
       %  Hold \rightins back if there is already a \@savetopins
       \ifvoid\@savetopins\insert\bothins{\unvbox\rightins}\fi
       \fi
    \else
       \if L\lr    % this is the left of two columns
          \@makecolumn
          \global\setbox\@leftcolumn\box\@outputbox \global\let\lr=R%
          \ifx Y\firstpage %\ifnum\pageno=1
             \infuser{^^J[left\the\pageno]}%
          \else
             \infuser{[left\the\pageno]}\fi
          \ifvoid\bothins\else\global\setbox\@bothcolumns\box\bothins\fi
          \global\dimen\bothins=\z@
          \global\count\bothins=0
          \ifx Y\firstpage %\ifnum\pageno=1
             \global\topskip=\fullhead\fi
       \else    % the right column
          \@makecolumn
          \ifnum\the@end>0\ifnum\pageno>1\balance@right@left\fi\fi
          \@outputpage \global\let\lr=L%
          \global\dimen\bothins=\maxdimen
          \global\count\bothins=1000
          \ifvoid\rightins\else
          %  Hold \rightins back if there is already a \@savetopins
             \ifvoid\@savetopins \insert\bothins{\unvbox\rightins}\fi
          \fi
       \fi
    \fi
    \global\let\last@insert=N \put@default
    \ifnum\outputpenalty>-\@MM\else\dosupereject\fi
    \ifvoid\@savetopins\else
      \ifdim\ht\@savetopins>\@txtht
        \global\setbox\@tempboxa=\box\@savetopins
        \global\setbox\@savetopins=\vsplit\@tempboxa to\@txtht
        \global\setbox\@savetopins=\vbox{\unvbox\@savetopins}%
        \global\setbox\@savetopright=\box\@tempboxa \fi
    \fi
    \@makecolht
    \global\count\topins=1000
   \fi
   }
%
%%% ----------  Start one- or two-column output  ---------
%
\if N\lr
   \setuplr{O}{\fullhsize}{\hsize}% O = one column
\else
   \setuplr{T}{\fullhsize}{\hsize}% T = two columns
\fi
% This is aafigure.tex
%
% The macros cover the case of single column format as well
% as double column format. In single column format all 'double
% column' insertions are reduced to single column insertions.
% While real double column insertions will always appear at
% the top of page, the placement of single column insertions
% can be influenced by the definition of \put@default.
% Possible positions are at the current position (only if
% \insert@here expands to Y, there is no top and no bottom insertion
% so far, and there is enough space), at the bottom of the column
% (only if \insert@at@the@bottom expands to Y and there is enough
% space in this column) and at the top of the column but below
% any double column insertion.
% The default placement of single column figures is usually at the top.
% If the figure almost fills the page if inserted at the bottom,
% it may go there.
%
\catcode`@=11
\def\put@default{\global\let\insert@here=Y
   \global\let\insert@at@the@bottom=N}%
% Allow figures to be inserted a the current position (if possible):
\def\puthere{\global\let\insert@here=Y%
    \global\let\insert@at@the@bottom=N}
% All figures are inserted at the top:
\def\putattop{\global\let\insert@here=N%
    \global\let\insert@at@the@bottom=N}
% Figures are inserted at the bottom (if possible):
\def\putatbottom{\global\let\insert@here=N%
    \global\let\insert@at@the@bottom=X}
%--------------------------------------------------------------------
\put@default
\let\last@insert=N        % Always reset to 'N' when a column is finished
\def\end@skip{\smallskip} % This space is added except after bottom insertions
\newdimen\min@top
\newdimen\min@here
\newdimen\min@bot
\min@top=10cm
\min@here=4cm   % do not insert figures after a few lines of text only
\min@bot=\topskip % figures may be at the bottom but there is a \@startins
\def\figfuzz{\vskip 0pt plus 6pt minus 3pt}  % more flexible spacing
%--------------------------------------------------------------------
\def\check@here@and@bottom#1{\relax
   %%% Several conditions have to be true if a figure or table can be
   %%% inserted at the current position or at the bottom of the page.
   %%% These conditions should preserve the
   %%% order of single column figures and put floating figures
   %%% always to the top of a column. However, exceptions are
   %%% possible with a \puthere\begfig{...}\endfig appearing at
   %%% the current position and a later \begfig{...}\endfig
   %%% appearing at the top of the same column.
   %
   \ifvoid\topins\else       \global\let\insert@here=N\fi
   \if B\last@insert         \global\let\insert@here=N\fi
   \if T\last@insert         \global\let\insert@here=N\fi
   \ifdim #1<\min@bot        \global\let\insert@here=N\fi
   \ifdim\pagetotal>\@colht  \global\let\insert@here=N\fi
   \ifdim\pagetotal<\min@here\global\let\insert@here=N\fi
   \if X\insert@at@the@bottom\global\let\insert@at@the@bottom=Y
     \else\if T\last@insert  \global\let\insert@at@the@bottom=N\fi
          \if H\last@insert  \global\let\insert@at@the@bottom=N\fi
          \ifvoid\topins\else\global\let\insert@at@the@bottom=N\fi\fi
   \ifdim #1<\min@bot        \global\let\insert@at@the@bottom=N\fi
   \ifdim\pagetotal>\@colht  \global\let\insert@at@the@bottom=N\fi
   \ifdim\pagetotal<\min@top \global\let\insert@at@the@bottom=N\fi
   \ifvoid\bottomins\else    \global\let\insert@at@the@bottom=Y\fi
   \if Y\insert@at@the@bottom\global\let\insert@here=N\fi }
\def\single@column@insert#1{\relax
   \setbox\@tempboxa=\vbox{#1}%
   \dimen@=\@colht \advance\dimen@ by -\pagetotal
   \advance\dimen@ by-\ht\@tempboxa \advance\dimen0 by-\dp\@tempboxa
   \advance\dimen@ by-\ht\topins \advance\dimen0 by-\dp\topins
   \check@here@and@bottom{\dimen@}%
   \if Y\insert@here
      \par  % The insertion forces a new paragraph in this case.
      \midinsert\figfuzz\relax     %%%%%%%%%\bigskip
      \box\@tempboxa\end@skip\figfuzz\endinsert
      \global\let\last@insert=H
   \else \if Y\insert@at@the@bottom
      \begingroup\insert\bottomins\bgroup\if B\last@insert\end@skip\fi
      \floatingpenalty=20000\figfuzz\bigskip\box\@tempboxa\egroup\endgroup
      \global\let\last@insert=B
   \else
      \topinsert\box\@tempboxa\end@skip\figfuzz\endinsert
      \global\let\last@insert=T
   \fi\fi\put@default\ignorespaces}
%
% ---------------- The insertion macros for the user -------------------
%
\def\begfig#1cm#2\endfig{\par\the\everypar
\single@column@insert{\@startins\rahmen{#1}#2}\ignorespaces}
\def\begfigwid#1cm#2\endfig{\par\the\everypar
   \if N\lr  % Here the only difference to \begfig is the larger \hsize
      {\hsize=\fullhsize \begfig#1cm#2\endfig}%
   \else
      \setbox0=\vbox{\hsize=\fullhsize\bigskip#2\smallskip}%
      \dimen0=\ht0\advance\dimen0 by\dp0
      \advance\dimen0 by#1cm
      \advance\dimen0by7\normalbaselineskip\relax
      \ifdim\dimen0>\@txtht
         \infuser{^^JFigure plus legend too high, will try to put it on a
                  separate page. }%
         \begfigpage#1cm#2\endfig
      \else
         \bothtopinsert\line{\vbox{\hsize=\fullhsize
         \@startins\rahmen{#1}#2\smallskip}\hss}\figfuzz\endbothinsert
      \fi
   \fi}
\def\begfigside#1cm#2cm#3\endfig{\par\the\everypar
   \if N\lr  % Here the only difference to \begfig is the larger \hsize
      {\hsize=\fullhsize \begfig#1cm#3\endfig}%
   \else
      \dimen0=#2true cm\relax
      \ifdim\dimen0<\hsize
         \infuser{^^JYour figure fits in a single column; why don't
                  ^^Jyou use \string\begfig\space instead of
                  \string\begfigside? }%
      \fi
      \dimen0=\fullhsize
      \advance\dimen0 by-#2true cm
      \advance\dimen0 by-1true cc\relax
      \bgroup
         \ifdim\dimen0<8true cc\relax
            \infuser{^^JNo sufficient room for the legend;
                     using \string\begfigwid. }%
            \begfigwid #1cm#3\endfig
         \else
            \ifdim\dimen0<10true cc\relax
               \infuser{^^JRoom for legend to narrow;
                        legend will be set raggedright. }%
               \rightskip=0pt plus 2cm\relax
            \fi
            \setbox0=\vbox{\def\figure##1##2{\vbox{\hsize=\dimen0\relax
                           \@startins\noindent\petit{\bf
                           Fig.\ts##1\unskip.\ }\ignorespaces##2\par}}%
                           #3\unskip}%
            \ifdim#1true cm<\ht0\relax
               \infuser{^^JText of legend higher than figure; using
                        \string\begfig. }%
               \begfigwid #1cm#3\endfig
            \else
               \def\figure##1##2{\vbox{\hsize=\dimen0\relax
                                       \@startins\noindent\petit{\bf
                                       Fig.\ts##1\unskip.\
                                       }\ignorespaces##2\par}}%
               \bothtopinsert\line{\vbox{\hsize=#2true cm\relax
               \@startins\rahmen{#1}}\hss#3\unskip}\figfuzz\endbothinsert
            \fi
         \fi
      \egroup
   \fi\ignorespaces}
\def\begfigpage#1cm#2\endfig{\par\the\everypar\specialpage{\@startins
   \vskip3.7pt\rahmen{#1}#2}\ignorespaces}%
\def\begtab#1cm#2\endtab{\par\the\everypar
\single@column@insert{#2\rahmen{#1}}\ignorespaces}
\let\begtabempty=\begtab
\def\begtabfull#1\endtab{\par\the\everypar
\single@column@insert{#1}\ignorespaces}
\def\begtabemptywid#1cm#2\endtab{\par\the\everypar
   \if N\lr
      {\hsize=\fullhsize \begtabempty#1cm#2\endtab}%
   \else
      \bothtopinsert\line{\vbox{\hsize=\fullhsize
      #2\rahmen{#1}}\hss}\medskip\endbothinsert
   \fi\ignorespaces}
\def\begtabfullwid#1\endtab{\par\the\everypar
   \if N\lr
      {\hsize=\fullhsize \begtabfull#1\endtab}%
   \else
      \bothtopinsert\line{\vbox{\hsize=\fullhsize
      \noindent#1}\hss}\medskip\endbothinsert
   \fi\ignorespaces}
\def\begtabpage#1\endtab{\par\the\everypar\specialpage{#1}\ignorespaces}
\catcode`@=12
% This is startup.aa
% It defines additional macros and settings that are special for
% the AA package or that have to be changed in the macros
% already loaded. A control sequence \firsttodo is defined that is
% called first after TeX starts up and is executed by \everyjob for
% a format file use of the package.
%
\catcode`@=11 % use @ as a normal character
\def\SpringerMacroPackageNameA{AA}
\edef\firsttodo{%
\infuser{^^JThis is \SpringerMacroPackageNameA, the plain TeX macro
package from Springer-Verlag,^^Jfor the Astronomy and Astrophysics
Main Journal
\ifx\Initexing\undefined\else format file \fi
\fonttype\space version 3.0^^J}%
\catcode`@=\active   % This is reset by the \maketitle macro
}
\everyjob={\firsttodo}% set the startup variable of TeX
\firsttodo % (do things that are first to do :-)
\catcode`@=12 % at signs are no longer letters
%%% rest removed from cp-aa.ini
%% This is dodump.aa
%% it clears the first page possibly filled with junk material
%% adjusts the page counter and calls INITEX's \dump
%\pageno=0
%\let\header=N%
%\leavevmode
%\vfil
%You just produced a \TeX{} format file for the \fonttype\space
%version of the {\tt\SpringerMacroPackageNameA} macro package from
%Springer-Verlag.
%\bigskip
%\vfil
%\eject
%\let\Initexing=\undefined
%\dump
%.

%\input pp-aa.cls
%\input p-aa.cls
%\input pp-aa.ini
%\input p-aa-fix.tex
%format MTPLAINE
%\input pp-aa.mtm    % use MT-fonts incombination with format MTPLAIN
\input psfig.tex
%
% \refereelayout           % a referee layout
%
  \MAINTITLE={Morphology of the Virgo Cluster: Gas versus Galaxies}
  \AUTHOR={S. Schindler$^{1,2,3,*}$,
           B. Binggeli$^{4}$,
           H. B\"ohringer$^{2}$}
  \INSTITUTE={ 
$^1$ Astrophysics Research Institute, Liverpool John Moores University,
   Twelve Quays House, Egerton Wharf, Birkenhead L41 1LD, UK
\newline
$^2$ Max-Planck-Institut f\"ur extraterrestrische Physik,
   Giessenbachstra\ss e, D-85478 Garching, Germany
\newline
$^3$ Max-Planck-Institut f\"ur Astrophysik, Karl-Schwarzschild-Stra\ss e 1,
   D-85478 Garching, Germany
\newline
$^4$ Astronomisches Institut der Universit\"at Basel, Venusstrasse 7,
   CH-4102 Binningen, Switzerland}
  \catcode`\@=12
  \FOOTNOTE{e-mail: sas@staru1.livjm.ac.uk}
  \catcode`\@=11
  \ABSTRACT={
  We draw a
  quantitative comparison of the distribution of the galaxies and the
  intra-cluster gas in the Virgo cluster by extending the
  morphological analysis 
  by Binggeli et al. (1987) to the intra-cluster gas. 
  We use the Virgo Cluster Catalog in combination with data from the
  ROSAT All-Sky Survey. The
  galaxy distribution and the gas distribution are relatively similar.
  The steep density gradient southwest of M87 previously observed in the
  optical is also seen in the X-ray emission. 
  In both wavebands the irregular structure of Virgo can be decomposed
  into three major subclusters centred on M87, M49, and M86. 
  A new statistical method of subcluster decomposition is applied.
  Radial galaxy and X-ray density profiles of the three subclusters
  are fitted with $\beta$-models, allowing analytic deprojection.
  Comparison of these profiles reveals no significant difference between the 
  galaxy number density and the smoothed galaxy luminosity 
  density if the inner part is excluded ($r <$ 60\arcmin, 
  dominated by the central giant galaxy),
  i.e.~there is no general luminosity segregation in the cluster.
  The gas density profile is steeper than the galaxy density
  profile in the inner part (again excluding the central galaxy),
  while this trend is reversed in the outer part. The turning point
  is around 300 kpc.
  A comparison among the subclusters shows the poorer the subcluster 
  the steeper its radial profile, i.e.~the more compact it is, both in 
  the optical and the X-rays. This is in general agreement with recent 
  N-body simulations by Navarro et al.~(1997). The subcluster profiles for
  different Hubble types confirm the well-known result that the
  distribution of the late-type
  galaxies is more extended than the early types. 
  Differential and integrated mass density profiles of the different 
  components in the M87 and M49 subclusters are presented. The total,
  gravitating mass (dominated by dark matter) is inferred from the 
  distribution of the X-ray gas assuming hydrostatic equilibrium.
  In the M87 subcluster the gas mass is about three times the mass in
  galaxies (assuming a constant $M/L$ = 20 for the single galaxies),
  while it accounts only for 8\% to 14\% of the total mass at 0.4 and 1 Mpc,
  respectively. In the M49 subcluster there is more mass in the galaxies
  than in the gas and the gas-to-total mass fraction is only 1\%, which is
  unusually low for a cluster. 
  The profiles of the projected mass-to-light ratio show relatively constant
  values around $500\msol/\lsol$.
  }
  \KEYWORDS={Galaxies: clusters: individual: Virgo cluster 
  - inter-galactic medium 
  - dark matter
  - X-rays: galaxies } 
  \THESAURUS={03 {11.03.4; 11.09.3; 12.04.1; 13.25.2} }
%
%  \OFFPRINTS={S. Schindler, address 3}      % is optional

  \DATE={ ????? }           % must be given at the last moment
%
% Please don't overrule this command, it will format your titlepage.
\maketitle
%\AUTHORRUNNINGHEAD{}
%

\titlea{Introduction}

Clusters of galaxies are the largest gravitationally bound aggregates of
matter in the universe, being a mixture of dark matter (up to 90\% of
the total gravitating mass), hot gas (up to 30\%), and galaxies
(a few \% of the total; for a recent review on clusters, see B\"ohringer 1995).
Although clusters represent the highest peaks in the large-scale density field,
most of them (even very rich ones) show significant substructure and 
therefore cannot be dynamically relaxed. This is usually taken as evidence
that cluster formation is a prolonged process, or has set in only recently
(say, around $z$ = 1). Clearly, the study of substructure in clusters
of galaxies should provide important constraints on structure formation in
the universe in general.

Traditional studies of substructure rely on various statistical tests
applied to the (projected) spatial and velocity distribution of (suspected) 
member galaxies (e.g., Fitchett 1988; West 1994; Girardi et al. 1997 and
references therein). Since the advent of X-ray satellites, and in particular 
ROSAT (Tr\"umper 1993), cluster substructure is also observed and mapped in 
the intra-cluster gas (e.g., McMillan et al.~1989; Briel et al.~1991, 1992;
White et al.~1993). 

As both the galaxies and the hot gas are sitting in a common
dark-matter potential, it is expected that both components show, to first
order, the same distribution -- which indeed is observed. However, there is 
also an important difference: the gas is dissipative, while the galaxies are 
(essentially) collisionless. In the case of subcluster merging, this can lead
to significant deviations in the X-ray versus optical structure of a cluster
(Schindler \& M\"uller 1993), evidence for which is reported, e.g., for
Abell\,754 (Zabludoff \& Zaritsky 1995) and Abell\,2255 (Burns et al.~1995).
The systematic comparison of the distribution of the X-ray gas
and the cluster galaxies can therefore be regarded as a 
powerful tool to assess the dynamical state of a cluster. 

In this paper we present such a comparison of gas versus galaxies for the
Virgo cluster of galaxies. The Virgo cluster is the nearest, fairly rich
galaxy cluster where such a study can also be carried out with greatest
detail. In spite (or rather because) of its proximity, this cluster has been
difficult to map, simply because of its large angular extent in the sky
(roughly 100 deg$^2$). For a complete, deep, high-resolution optical study
it had to await the Las Campanas Survey, which resulted in the Virgo Cluster
Catalog (VCC, Binggeli et al.~1985). Likewise, following a first large-scale
scan with GINGA (Takano et al.~1989), it took the ROSAT 
All-Sky Survey (Voges et al.~1996) to cover the cluster completely with 
good resolution in the X-ray waveband.

Based on the VCC and supplementary kinematic data, 
Binggeli et al.~(1987, hereafter BTS87; 1993) studied the spatial and 
velocity distribution of different morphological types in the Virgo cluster. 
Confirming in part earlier results by de\,Vaucouleurs and others (cf.~BTS87), 
three major subclusters centred on the giant ellipticals
M87, M86, and M49 were identified. In a subsequent study of the X-ray structure
of Virgo based on ROSAT All-Sky Survey data by B\"ohringer et 
al.~(1994) these three subclusters became evident by their extended
X-ray halos. B\"ohringer et al.~(1994) also provided a first global
comparison of the gas versus galaxy distribution and found additional evidence
for the imminent or ongoing merging of the M87 and M86 subclusters.
The Virgo cluster is a typical irregular cluster that appears to be still in
the making (see also Binggeli, 1998, for a general review).

Here we combine the VCC and ROSAT All-Sky Survey data bases to construct
and intercompare the 3D-radial profiles in galaxy number density, luminosity
density, gas density, and mass density for all three major
subclumps of the Virgo cluster.

The rest of the 
paper is organized as follows. After a brief description of the data
sources (Sect.~2), we present a purely qualitative and global comparison
of the optical versus X-ray structure of the Virgo cluster, addressing also
the differences with respect to morphological type (Sect.~3). Starting with 
Sect.~4 we concentrate on the three subclusters around M87, M86, and M49 
and describe a new method of how to decompose the cluster into subclusters.
A detailed, quantitative comparison for the radial profiles of the three
subclusters, based in large part on $\beta$-profile fitting, is then given 
in Sect.~5. Mass density and mass-to-light ratio profiles are presented in
Sect.~6. Finally, Sect.~7 gives our summary and conclusions. 

A mean Virgo cluster distance of 20 Mpc, corresponding to
a true distance modulus of
$(m - M)$ = 31.5 (cf., e.g., Tammann \& Federspiel 1996), is assumed 
throughout this paper. 

\titlea{Data}

\begfigwid 0.0cm
\psfig{figure=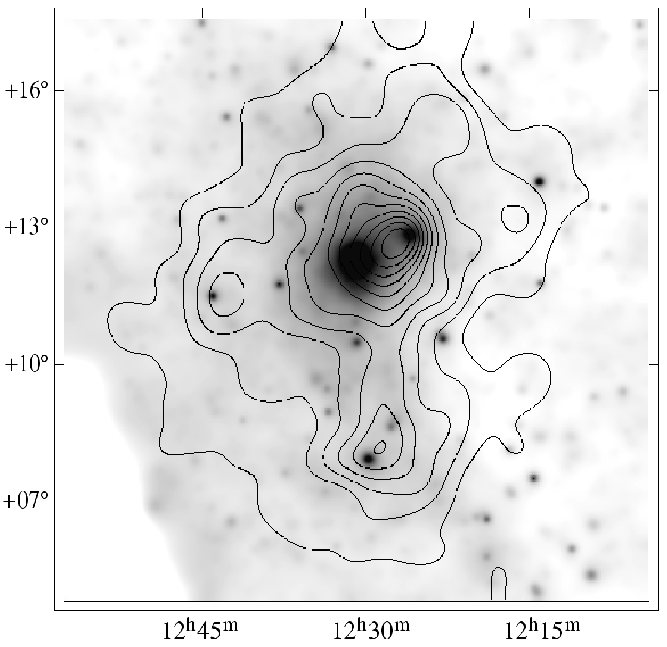,width=18.cm,clip=}
\figure{1}{Comparison of the X-ray and the optical appearance of the
Virgo cluster. The greyscale image shows, in a logarithmic scale, 
the X-ray emission as observed
by the ROSAT All-Sky Survey in the hard band (0.5 - 2.0
keV)(B\"ohringer et al. 1994, where also most of the point sources are
identified). It is smoothed with a Gaussian filter with
$\sigma = 24\arcmin$ on the faintest levels and decreasing filter size
with increasing surface brightness. The contours show the number
density of the 1292 
member galaxies of the VCC smoothed again with a Gaussian of
$\sigma$ = 24 \arcmin. The spacing of the contours is linear. The
lowest contour line and the contour spacing is $1.5\times10^{-3}$
galaxies per \sq\arcmin. The image has a size of
12.8$^{\circ}\times$12.8$^{\circ}$. North is up and West to the right.
The dominant dark blob is centred on
M87 ($\alpha_{2000} \approx$ 12$^{\rm h} 30^{\rm m}$, $\delta_{2000} \approx$
12\degr23\arcmin). For the identification of further galaxies, cf.~Fig.~2.  
}
\endfig

For the optical part of this study we use the 
positions, types and total blue ($B$) magnitudes of the 1292 member galaxies 
listed in the VCC (Binggeli et al. 1985, membership as updated in 
Binggeli et al.~1993). This catalog was compiled from
the Las Campanas Survey of the Virgo cluster, covering an area of
$\approx 140$ deg$^2$. The VCC is complete to an apparent magnitude $B$ = 18 
(corresponding to $M_{B_{\rm T}} = - 13.5$ with the adopted distance of
20 Mpc) and contains
in addition many fainter dwarf galaxy members down to $B$ = 20.
For 403 of the galaxies radial velocity information is also
available. These velocities were crucial for the
identification of the M86 subclump (Binggeli et al.~1993).
The cluster membership of a galaxy was reckoned essentially from its 
morphological appearance, guided by the available velocity data.
It has recently been confirmed by
Drinkwater et al. (1996) that the morphological membership selections
are generally correct.

For transforming observed galaxy magnitudes into true luminosities 
we assume an extinction value for galactic absorption of 0.15 mag, 
which is an average value for all Virgo galaxies from 
the RC3 (Schr\"oder 1995). 
To get a rough estimate of the mass of a galaxy (Sect.~6) we employ a
B-band mass-to-light-ratio of 20, independent of the Hubble type. The
faintest galaxies in the VCC have apparent $B$ magnitudes of 20,  
while the most luminous Virgo galaxy, M49, has $B$ = 
9.31 mag. This gives almost a factor of 20\,000 in luminosity or mass
between these extreme galaxies.

The distribution of the hot intra-cluster gas is inferred from the X-ray
brightness distribution as mapped by the ROSAT All-Sky-Survey (Voges et
al. 1996). The Virgo area was scanned from November 1990 to January 1991 
with an average exposure time of about 450 seconds and a spatial resolution
of 20 - 60\arcsec .
Here we use the X-ray emission detected in the ROSAT hard band
(channel 52 to 201 corresponding roughly to the energy range 0.5-2.0
keV). 

For constructing the X-ray radial profiles of the three subclusters (Sect.~5),
13 point sources had to be excluded. Most of these background 
(QSO or distant cluster) or foreground (stellar) sources are identified
in Fig.~1 of B\"ohringer et al.~(1994).
For the analysis of the M87 subcluster, 
the regions around M49 and M86 had additionally to be 
excluded. For the analysis of the M49
subcluster, two different approaches were used which give
insignificantly different results: (1) Excluding a rectangular region 
north of the M49 halo and (2) excluding a sector in the north with an
angle of $120^{\circ}$.

\titlea{A qualitative optical and X-ray comparison of the Virgo
cluster}

Fig.~1 is a superposition of the X-ray emission
in the Virgo cluster region measured by ROSAT (greyscale image) 
and the optical distribution of VCC member galaxies
(number density contours). 
Both distributions are smoothed with
a Gaussian filter with $\sigma$ = 24\arcmin. 
For a direct plot of all VCC
members, giving an impression of how the Virgo cluster appears in the sky, 
the reader is referred to BTS87. For easy identification of the major cluster
members and the substructures associated with them, we have repeated the
contour image for all member galaxies in Fig.~2a, where these galaxies
are plotted with different symbols. The X-ray image is repeated in Fig.~2 as 
well (Fig.~2e), this time in logarithmic contours, to guide the optical
versus X-ray comparison. The smoothing scale of $24\arcmin$ is arbitrary but
has been optimized to avoid the noise of single galaxies 
by too little smoothing on the one hand, and the smearing out of 
substructure by too much smoothing on the other.

Already at a glance the Virgo cluster appears as a complex system. There
is a pronounced double structure in the direction N-S. The Northern clump,
called ``cluster A'' in BTS87, is dominated by M87, which coincides
with the maximum of the X-ray emission in the whole area (the large dark
blob in the centre of Fig.~1). The Southern, much less pronounced clump,
called ``cluster B'' in BTS87, coincides with M49, another supergiant 
elliptical that is even slightly brighter than M87. 
However, as emphasized
in BTS87, the galaxy density contours do not peak on M87 but almost 
$1\degr$ WNW of it -- more than halfway in the direction to M86, which is 
yet another giant E galaxy
(see Fig.~2a; in Fig.~1 M86 is the second-ranked dark spot).
The reason of this ``mispointing'' of M87 is now clear: the Northern clump
itself is a double system comprising the dominating, massive subcluster
centred on M87 and a smaller subcluster centred on M86. This view is
supported by the existence of a swarm of low-velocity dwarf galaxies around
M86 (which itself has a negative velocity, Binggeli et al.~1993) and the 
extended X-ray halo of M86 (B\"ohringer
et al.~1994). The M86 subcluster seems to fall into the M87 subcluster from
behind (see also Binggeli 1998), bringing the previous picture of a single 
galaxy being stripped by
the intra-cluster medium of the Virgo cluster (Forman et al. 1979;
Rangarajan et al. 1995) in disfavour. 

We thus deal essentially with three major subclumps, or subclusters (as they
will henceforth be called): centred on the giant ellipticals M87, M86, and
M49. They are readily visible as dark spots in Fig.~1. Note that these
X-ray sources are not point sources but are extended: the hot gas is sitting
in the potential of the (extended) subclusters. 

Clearly there is more, but rather spurious substructure in the Virgo cluster:
in particular there is a lump in the East around the S0 galaxy M60, 
another one around the well-known spiral M100 in the North and a third
one WNW of M86
(see Fig.~2a). Overall, the irregular, asymmetric structure of Virgo 
can be described by two distinct axes: one stretching N-S from 
M100-M87/86-M49, and one E-NW from M60-M87-M86. Remarkably, the latter axis
is perfectly aligned with the jet of M87 (cf.~Binggeli 1998).

%%%%%%%
We emphasize that no depth effects are considered here.
There is weak evidence that the southern M49 subcluster is
slightly more distant than the northern M87 subcluster 
(Federspiel et al. 1998), but the effect is not significant.
On the other hand, Tonry et al.~(1990) and Morris \& Shanks (1998) claim
that M49 itself is lying considerably in the foreground. However, this claim is
based on the ``surface brightness fluctuations'' method for elliptical
galaxies whose reliability is still debated (for a detailed
discussion of the distance and the depth of the Virgo cluster see
Binggeli 1998). Hence it seems safe at present to assume that, to first
approximation, all Virgo subclusters and there central beasts are 
at the same distance.
%%%%%%%

\titleb{Hubble type dependence}

The number density distribution for the various galaxy types in the Virgo
cluster is shown in Fig.~2b (early-type dwarfs, i.e.~dE and dS0), 2c 
(early-type giants, i.e.~E and S0), and 2d (late types, i.e.~spirals and
irregulars). Confirming the results of BTS87
we find very different distributions for the different types.
The dwarf ellipticals (Fig.~2b) have a distribution very
similar to that of all galaxies (Fig.~2a), simply
because they make up 3/4 of the total cluster population. The
ellipticals and S0s (Fig.~2c) show a less extended distribution and are more
concentrated to the subcluster centres, i.e. the
X-ray maxima (compare with Fig.~2e). In sharp contrast, 
the distribution of the spirals and irregulars (Fig.~2d) is
very extended and shows no correlation with the X-ray emission.
Note also the difference in the type-mix between the various clumps.
The M49 subcluster is obviously spiral-rich. 

All of these trends, which of course reflect the general 
morphology-density relation (Dressler 1980), appear even more pronounced in 
the luminosity-weighted distributions shown in Fig.~3. The
luminosity distributions are obviously dominated by the light peaks 
of single bright galaxies, especially (not surprisingly) of M87, M86, M49,
and M60 (most pronounced in Fig.~3c). The compact, spherical contours
around these galaxies, also around bright spirals in Fig.~3d, merely
reflect the smoothing scale (Gaussian $\sigma$) of 24\arcmin. The galaxies
themselves are much smaller.
When M87, M86, M49, and M60 are subtracted, 
the luminosity distribution changes drastically 
(Fig.~3e, to be compared with 3a). 
There remains an extended region of enhanced galaxy density 
in the M87/M86 region, however, while 
the luminosity density around the position
of M49 becomes very low, i.e. the optical luminosity of the
M49 subcluster is very strongly dominated by M49 itself.

There is no way to obtain a smooth luminosity distribution 
when the galaxies span a factor of 20\,000 in luminosity. This problem will
be encountered again below.
\begfigwid 0.0cm
\psfig{figure=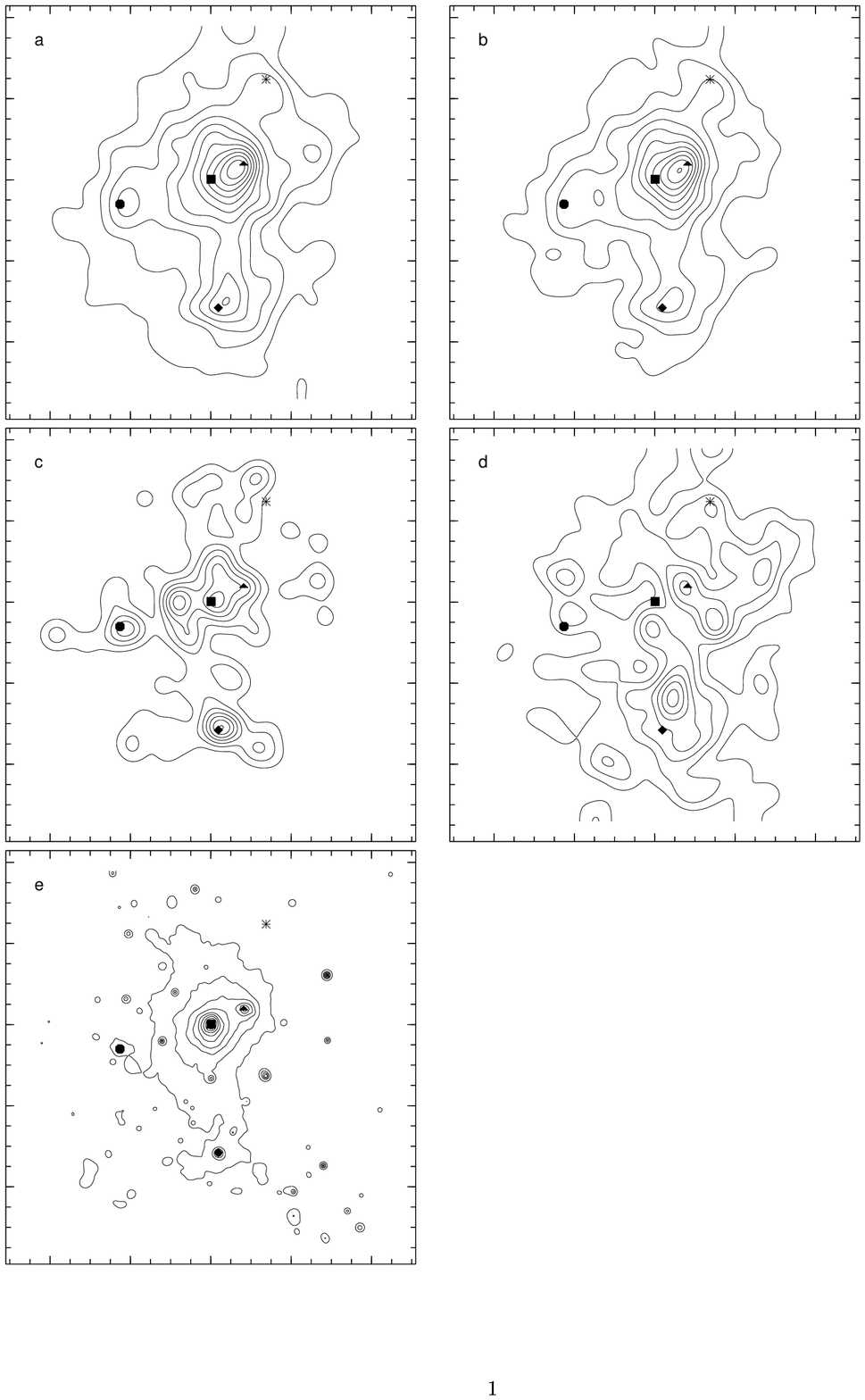,width=14.0cm,clip=} 
\figure{2}{
Distributions of the different galaxy types. Number densities
smoothed with a Gaussian of $\sigma = 24\arcmin$ are shown. (a) all
galaxy types, (b) dwarf elliptical and dwarf S0 galaxies, (c)
elliptical and S0 galaxies, (d) spiral and irregular galaxies.
The contours are linear with spacings of (a) $1.5\times10^{-3}$
galaxies/\sq\arcmin, (b) $1.3\times10^{-3}$ galaxies/\sq\arcmin, 
(c) $2.2\times10^{-4}$ galaxies/\sq\arcmin, 
(d) $3.6\times10^{-4}$ galaxies/\sq\arcmin. 
The level of the first contour line is
equal to the spacing. For better
comparison with Fig.~1 the positions of five galaxies are marked: M87
(square), M49 (diamond), M86 (triangle), M60 (octagon), M100
(star). (a) is the same as the contours in Fig.~1. For comparison the
X-ray image is shown again in (e) with logarithmically spaced contours.
The size of each image is 12.8$^{\circ}\times$12.8$^{\circ}$. The
distance between two tickmarks is 42\arcmin.
}
\endfig
 
\begfigwid 0.0cm
\psfig{figure=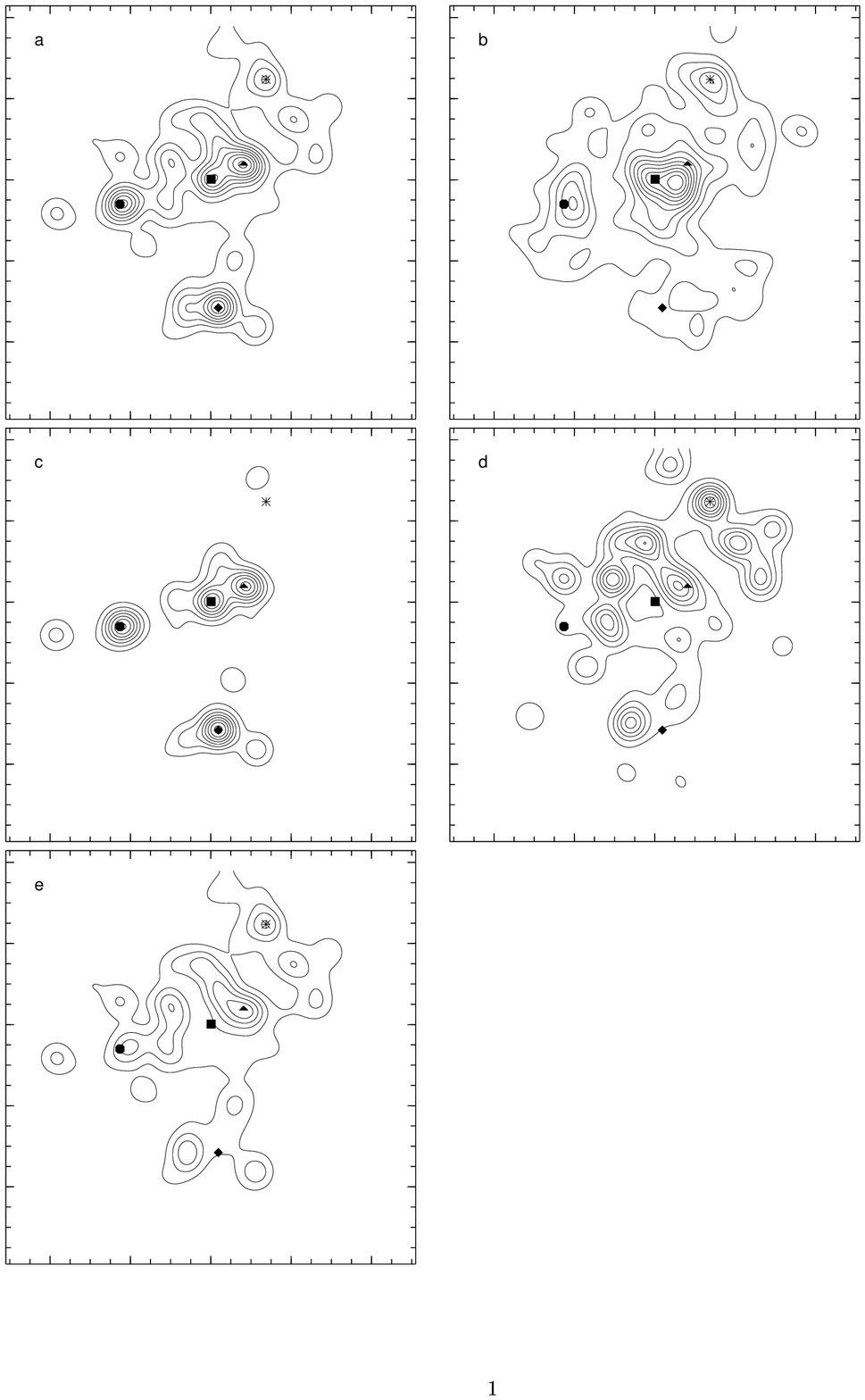,width=14.0cm,clip=} 
\figure{3}{Luminosity-weighted densities of the different galaxy
types. (a) all
galaxy types, (b) dwarf elliptical and dwarf S0 galaxies, (c)
elliptical and S0 galaxies, (d) spiral and irregular galaxies, (e) all
galaxies except for the four elliptical galaxies M87, M49, M86, and M60.
The contours are linear with spacings of 
(a) $7.7\times10^6 \lsol/\sq\arcmin$, 
(b) $4.5\times10^5 \lsol/\sq\arcmin$, 
(c) $6.8\times10^6 \lsol/\sq\arcmin$,
(d) $3.6\times10^6 \lsol/\sq\arcmin$ and 
(e) $7.7\times10^6 \lsol/\sq\arcmin$. 
The level of the first contour line is equal to
the spacing.
Smoothing and symbols as in Fig.2.
}
\endfig

\titlea{A new method for subcluster decomposition}

As the Virgo cluster is not a relaxed, symmetric system
but consists of several subclusters (cf.~above), it is impossible to
treat it as a single entity with spherical
symmetry. Each subcluster has to be 
analysed separately. The next task, then, is to decompose the cluster cleanly 
into the subclusters identified around M87, M86, and M49. 
Two methods to achieve this have been
employed here, each resulting in an independent data set.

In the first, more conventional 
method one simply excludes certain regions around the other subclusters. 
To construct a radial profile of the M87 subcluster we have excluded a
%%%
rectangular 
%%%
area of (70arcmin)$^2$ around M86,
and of (150arcmin)$^2$ around M49. The same procedure can also be
used for the X-ray profile. For profiles centred on
M49 we have excluded all data north of a horizontal line at a distance
of 112\farcm5 from M49. Extracting a profile of the M86 subcluster, however, 
is very difficult because it is so close to, and much smaller and less luminous
than, the M87 subcluster. Contamination by the M87 subcluster is unavoidable. 
Nevertheless, we have made
an attempt to determine a profile of the M86 subcluster by simply subtracting 
the contribution of the M87 subcluster (see Sect.~5).  
In the following we call all of these profiles
constructed by excluding certain regions {\it R-profiles}.

Contaminations from the outer regions of the neighbouring subclusters
are of course a principal shortcoming of the {\it R}-method.
To overcome this difficulty and to test how large the
uncertainties in the profiles are, we have developed a new method of
cluster decomposition. In this
method, each galaxy is assigned to be member of that subcluster to which it
most probably belongs.

Figure 4 shows schematically how the method works.
Suppose the spatial distribution of each subcluster is known
and is taken directly as probability distribution.
The position of a particular galaxy within the cluster 
then determines the probability for this galaxy to belong to
either subcluster: $P_s({\rm M87}), 
P_s({\rm M49})$ and $P_s({\rm M86})$. 
For simplicity we show only one spatial dimension in Fig.~4, but of course 
we use the full two-dimensional information for the subcluster assignment. 
As for all galaxies except the dwarfs velocity information is available
in addition, we apply the same
procedure for (one-dimensional) velocity space, resulting in another 
probability $P_v$. The final membership probability for a galaxy  
is then simply the product $P = P_s \times P_v$, assuming
that $P_s$ and $P_v$ are statistically independent. Finally, a galaxy is
assigned to that subcluster to which it has the highest probability to
belong.

To model the spatial distribution of the subclusters we have used
spherically symmetric $\beta$-models (see Sect.~5.1 below) with
different core radii for the different subclusters and for different
Hubble types. The core radii (for $\beta$ = 1) were taken from BTS87 or,
where not available, estimated from data given in BTS87 and Binggeli et 
al. (1993). The model profiles were centred on the three giant galaxies.
Galaxy types were sorted into  
three groups (as above in Figs.~2 and 3): E+S0, dE+dS0, and Sp+Irr,
with 85, 963 and 244 galaxies, respectively. 
The velocity distributions were modeled by Gaussians, centred on the 
velocities of the central galaxies (M87: $v=1258$ km\,s$^{-1}$, M49:
$v=969$ km\,s$^{-1}$, M86: $v=-227$ km\,s$^{-1}$) and with different
velocity dispersions for
different subclusters and different Hubble types, as given in Binggeli
et al. (1993), or as reckoned for the M86 subcluster from data therein.

All core radii and velocity dispersions adopted are listed in Table~1.  
The distributions were normalized to 
the total numbers of galaxies for the subcluster listed in the
bottom line of Table~1. From these numbers we see that the M87 subcluster 
comprises roughly
three times as many galaxies as the M49 subcluster, which in turn is
about six times the M86 subcluster. These ratios are of course 
different in terms of luminosity or mass (cf.~below).

The final results are not very sensitive to the exact
choice of the input parameters. This becomes evident e.g. in the
$\beta$-value: although for subcluster decomposition always $\beta=1$
was used, there is a large spread in the $\beta$-values in the final results.

\begtabfull
\tabcap{1}{Input parameters for the membership assignments. The first
block shows the core radii in kpc for the various subclusters and
Hubble types. The second block lists the velocity dispersions in
km\,s$^{-1}$. The last line gives the total number of galaxies in each 
subcluster. 
}
\halign{#\quad\hfil&#\quad\hfil&#\quad\quad\hfil&#\quad\quad\hfil&# \cr
\noalign{\hrule\smallskip}\cr
 & & M87 & M49 & M86 \cr
\noalign{\smallskip\hrule\smallskip}\cr
        & E+S0   & 420 & 420 & 260 \cr
$r_c$   & dE+dS0 & 570 & 570 & 340 \cr
        & Sp+Irr &1220 &1220 & 680 \cr
\noalign{\smallskip\hrule\smallskip}\cr
        & E+S0   & 670 & 320 & 120 \cr
$\sigma$& dE+dS0 & 720 & 370 & 120 \cr
        & Sp+Irr & 850 & 580 & 120 \cr
%& & & & & & & &\cr
\noalign{\smallskip\hrule\smallskip}\cr
\multispan2{number of galaxies~~~} & 920 & 317 & 55 \cr
\noalign{\smallskip\hrule}\cr
}
\endtab

\begfig 0.0cm
\psfig{figure=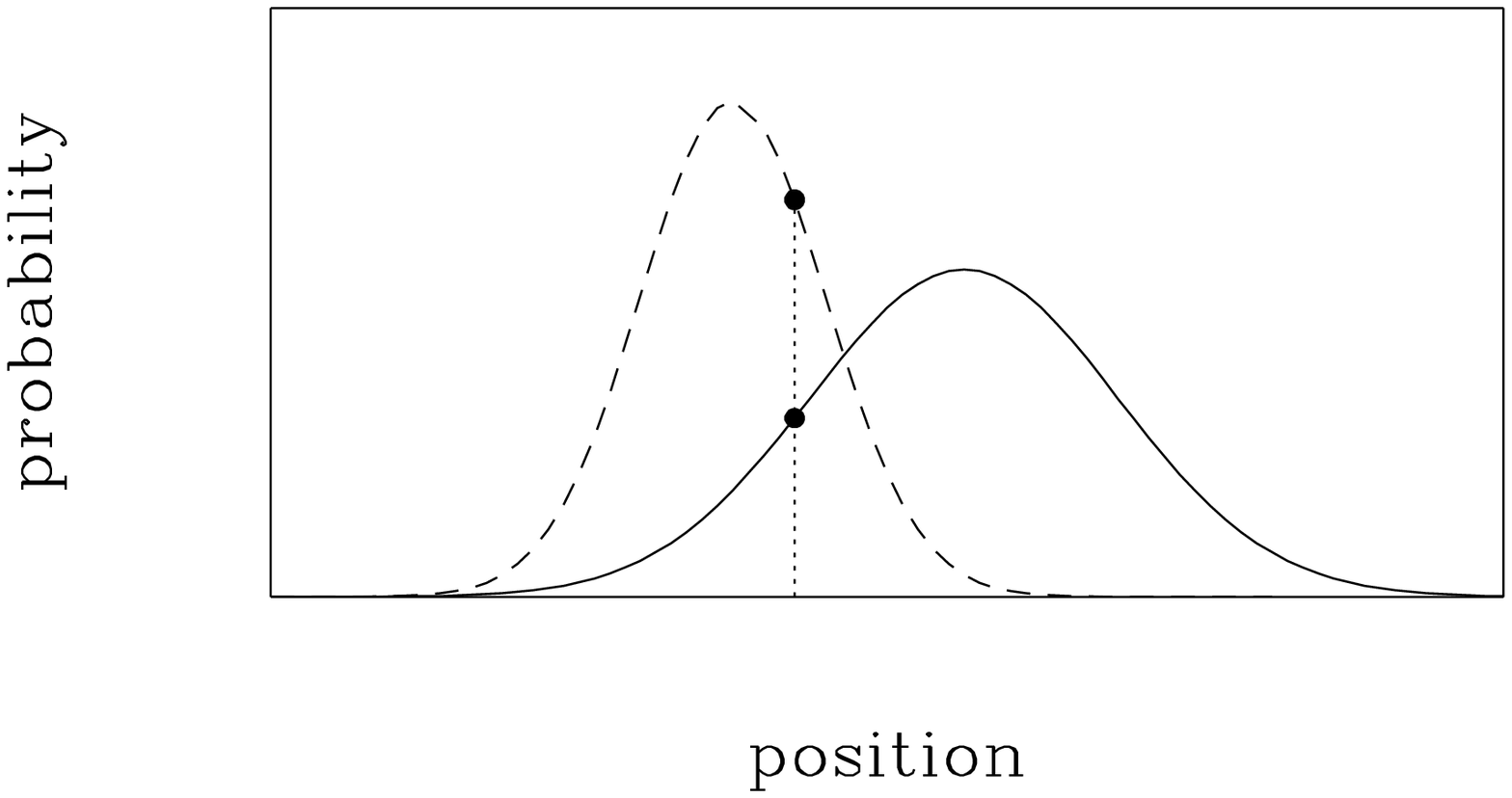,width=8.8cm,clip=}
\figure{4}{Schematic diagram to illustrate the technique of subcluster
assignment. The solid and the dashed lines show the probability
distributions of two subclusters. A galaxy at the position of the
dotted line would have a larger probability to belong to the dashed
subcluster.
}
\endfig

In the following all subcluster profiles constructed
with this probability method are called {\it P-profiles}.
In Fig. 5 we show the three subclusters, separately in three panels, 
as they result from assigning
member galaxies by the new probability method. 
Clearly, this method has its drawbacks, too. 
As the velocities of the central galaxies
M87 and M49 are very similar, and M86 is distinct by its negative velocity,
the decomposition into the M87 and M86
subclusters rests 
mainly on the velocities, while the distinction between the M87
and the M49 subclusters is achieved mainly by the 
positional criterion. This is the reason why the outermost contours of the
M87 and M49 subclusters are artificially flattened in the 
direction towards each other (Figs.~5a and b).

The three subclusters are distinctly different in their appearance. 
In accord with the relative galaxy populations mentioned above,
the M87 subcluster is the largest and most extended one, the M49 subcluster
is second, and the M86 subcluster is very small.
What is more interesting: the smaller the subcluster the more compact
it is, both in the number density distribution and the light distribution.
In particular the M86 subcluster appears very compact. 
This trend is encountered again in 
the profile analysis below. We regard this as a principal result of the 
present paper. 

\begfigwid 0.0cm
\psfig{figure=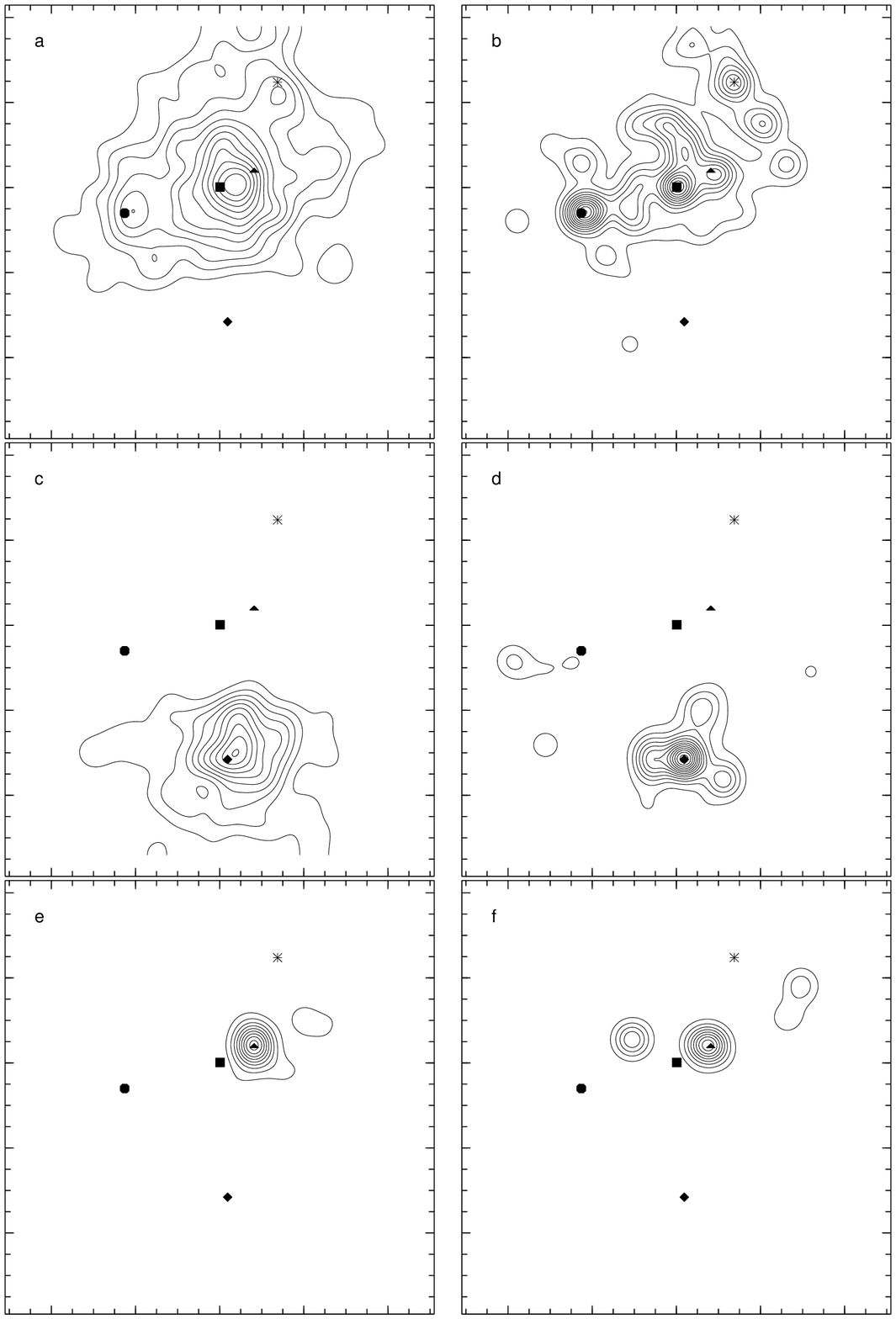,width=15.cm,clip=}
\figure{5}{Galaxy distribution in the subclusters separated by method
$P$. Left panels: number densities, right panels: luminosity-weighted
densities. (a) and (b): M87 subcluster, (c) and (d): M49 subcluster,
(e) and (f) M86
subcluster. All distributions are smoothed with a Gaussian with
$\sigma$ = 24\arcmin. The contours are linear with spacings of 
$1.1\times10^{-3}$ galaxies/\sq\arcmin\ in (a), 
$8.9\times10^{-4}$ galaxies/\sq\arcmin\ in (c), 
$6.7\times10^{-4}$ galaxies/\sq\arcmin\ in (e) and
$4.5\times10^{6} \lsol$/\sq\arcmin\ in (b), (d) and (e). 
The level of the first contour line is equal to 
the spacing. For better
comparison with Fig.~1 the positions of five galaxies are marked: M87
(square), M49 (diamond), M86 (triangle), M60 (octagon), M100
(star). It is obvious that the subclusters are very different in size.
}
\endfig

The luminosity distributions of the three subclusters are shown in
Figs.~5b, d and f. Again we see how these are 
dominated by single giant
galaxies.

\titlea{Quantitative optical -- X-ray comparison of the Virgo cluster}

\titleb{How to compare optical galaxy densities with X-ray surface
brightnesses} 

In different wavelengths different quantities are observed. In the optical one
observes a galaxy density projected in the plane of the sky, while
the X-ray emission measured is proportional to the projection of the 
{\it square}\/ of
the gas density. To make galaxy and gas densities comparable, one
has to derive {\it spatial}\/ densities by deprojecting the observed
quantities. Fortunately, this is
easily done with the $\beta$-model (Cavaliere \& Fusco-Femiano
1976). The optical, projected galaxy density can be modeled by

$$ n = n_0 \left( 1+{r^2 \over r^2_c} \right)^{-{3\over2}\beta +{1\over2}}
   \eqno(1)
$$
where $r_c$ is the core radius, $\beta$ is the slope parameter and 
$n_0$ is the central galaxy density.
Likewise, the surface brightness distribution in X-rays can be modeled as

$$ S = S_0 \left( 1+{r^2 \over r^2_c} \right)^{-{3}\beta +{1\over2}}
   \eqno(2)
$$
with $S_0$ as central surface brightness.
The resulting deprojected density for the optical galaxy distribution, as well
as for the X-ray gas distribution is then

$$ \rho = \rho_0 \left( 1+{r^2 \over r^2_c} \right)^{-{3\over2}\beta}
   \eqno(3)
$$
with the central spatial density $\rho_0$.

This means that we merely have to determine the best-fitting core radii $r_c$
and slope parameters $\beta$ for both the observed optical and X-ray profiles
and we can for each derive a spatial density profile after Eq.~(3) and then
compare them. For convenience, the central spatial densities 
-- galaxy number density, luminosity density, and gas density -- will be 
normalized to 1 for all profiles.

Although the profile of Navarro et al. (1995, NFW profile) might be physically
more meaningful, we do not use it here because it cannot be (de)projected 
in such an elegant way as the $\beta$-profile.

\titleb{Construction of radial profiles: P-profiles and R-profiles}

In the following we describe how the 
radial profiles of the subclusters were constructed. The resulting profiles 
are shown Fig.~6 -- number density profiles in the left panels, luminosity
density profiles in the right panels. The best-fitting
$\beta$-model parameters associated with these profiles, also including
Hubble-type-specific profiles to be discussed in the following subsections,
are listed in Table 2 (number density) and Table 3 (luminosity density). 
It should be remembered that the central spatial density of the profiles is
normalized to 1 throughout.

For the number density profiles we have grouped the galaxies into bins of
equal numbers of galaxies, i.e. depending on the number of galaxies in
the subset we have a different number of bins. This number is also listed in
Table 2. The radius of the bin is defined as the mean value of all
radii of the galaxies in the bin. 

\begfigwid 0.0cm
\psfig{figure=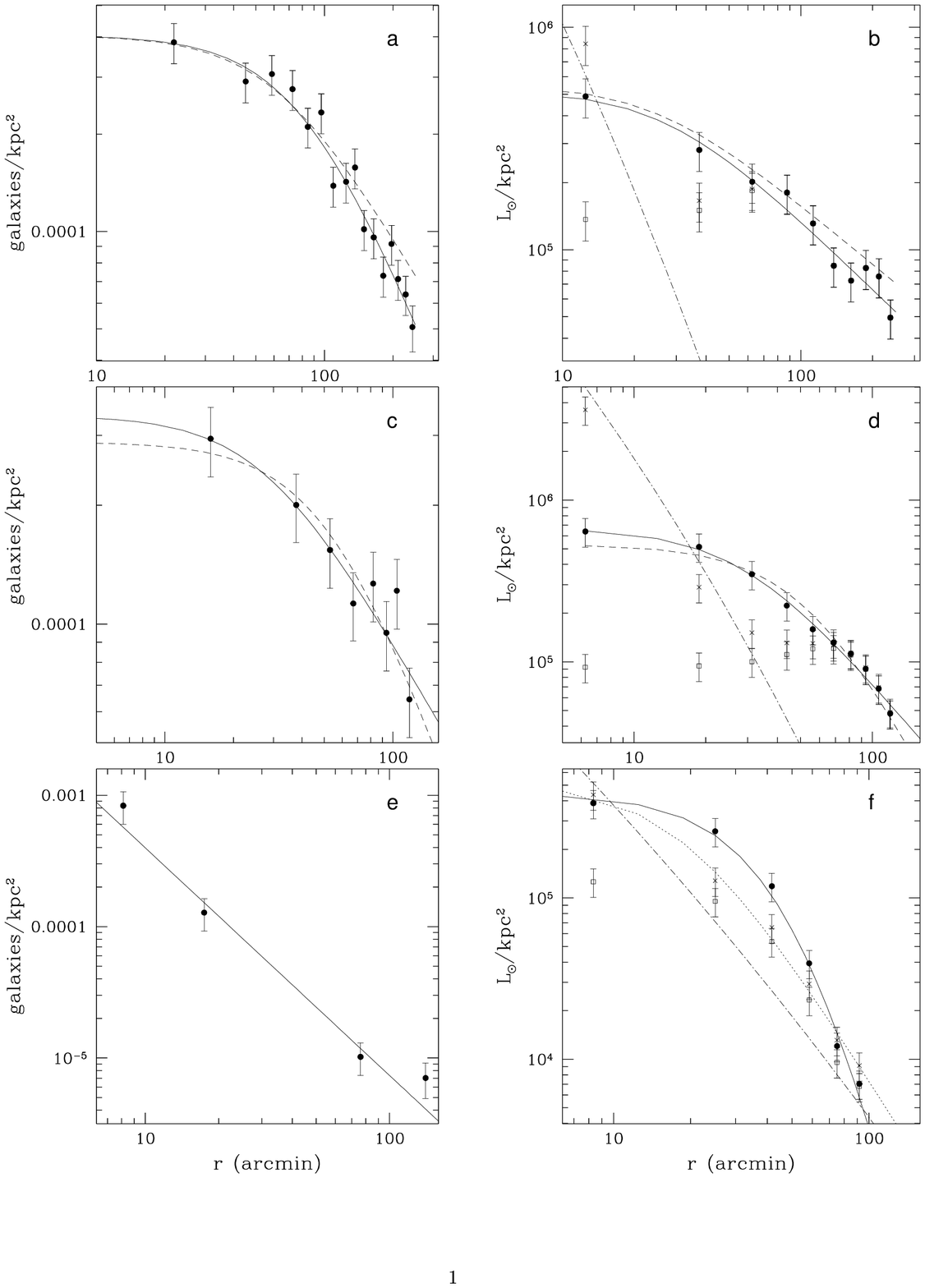,width=16.0cm,clip=} 
\figure{6}{Left panels: profiles of projected number density around M87 (a), 
M49(c), and M86(e) from $P$-selected galaxies (filled
circles). The solid lines represent fits with $\beta$ models. For
comparison, the $\beta$-model fits to
the corresponding $R$-profiles are shown as dashed lines.
The right panels show projected luminosity densities around M87 (b),
M49 (d), and M86 (f) from $P$-selected galaxies. The filled
circles represent the $P$-profiles after applying the Gaussian smoothing
described in the text. Solid and dashed lines are again best-fitting 
$\beta$-models to $P$ and $R$-profiles, respectively.
The open squares are the same $P$-profiles when the
luminosity of the central galaxy is omitted. The profiles marked
with crosses are the sum of the open squares and the true central
galaxy profiles (dash--dotted lines) as observed by Caon et al.~ (1993). 
The only acceptable $\beta$-fit to the crosses for M86 is shown as dotted
line in (f).
}
\endfig

The luminosity profiles pose some difficulties (as noted already by BTS87). The
huge differences between the luminosities of the faint and the bright
galaxies result in a huge scatter in the profile because single bright
galaxies contribute a major fraction to the luminosity of one bin. 
We tried to reduce the scatter by increasing 
the number of galaxies per bin. But even with 100 galaxies per
bin the scatter remains large. 
Also binning the galaxies
into bins with a minimum luminosity does not produce acceptable
profiles. Fits to these profiles would almost give random
results.

The only way to obtain acceptable luminosity profiles is to smooth first the
galaxy distribution -- as it is shown in Fig. 5 -- and to construct
the profiles subsequently from the smoothed image. A Gaussian of
$\sigma = 24\arcmin$ proved to be large enough to obtain good fits.
With smaller smoothing lengths the distribution is
still dominated by single high-luminosity bins. 
With the smoothing applied first we could not do a
binning by equal numbers of galaxies, instead we chose a constant bin
size over the profile. For M87 and M49 we used 10 bins while for
M86 we had to reduce the outer cut-off radius because of the strong
contribution of M90, so we ended up with only 6 bins for this subcluster.
The results of the fits do not depend on the bin size: tests with other 
bin sizes gave very similar results.
%galaxies contribute a lot to single bins, e.g. the galaxy dominating
%the high bins, central,4 and 10 (at radius 11000 pix and 23000pix) are
%M87, M90 and M60.
%results between rc=1.0arcmin beta=5.8 and rc=38arcmin beta=0.78

All profiles have an outer cut-off radius
of $250\arcmin$ or less (see Table 2 and 3.), limited by the coverage of the 
VCC. The error bars shown in Fig. 6 have also been used 
for the fits (see Tables 2, 3 and
5). For the number densities, these are $\sqrt N$ errors, where 
$N$ is the number of galaxies per bin. For the
luminosity profiles this number is not available because of the
smoothing procedure applied before. As here 25 is an average number of
galaxies per bin, we have assumed a corresponding error of 
20\% to determine $\chi^2$ for the
luminosity profile fits. A global change in the error of the data
points would result in different errors of the fit parameters and a
different $\chi^2$, but would not affect the fit parameters themselves.

In Fig.~6, binned profiles are only shown for $P$-selected subclusters.
However, best-fitting $\beta$-models are plotted for both $P$ and $R$-profiles
(the luminosity density $R$-profile for the M86 subcluster, constructed by
subtracting a smoothed image of the M87 subcluster, is very uncertain and is 
not shown in the figure). 
The comparison of $P$ and $R$-profiles
shows that in general they agree quite well, with a tendency of the 
$P$-profiles to be somewhat steeper. 
The flattening in the $R$-profiles is probably
caused by contaminations from the outer (non-excluded) parts of the 
(nominally) excluded
subclusters. The alternative explanation would be that the steeper 
slope of the M87 $P$-profile is due to the
relatively strong gradient in direction to M49 at large radii (see
the outer contours in Fig.5a) which might leave a residual with
the $P$-method. However, this second possibility was rejected by a test,
in which the $P$-fits were repeated with a smaller cutoff radius of
170\arcmin, finding no difference.

\begtabfullwid
\tabcap{2}{Profile parameters and fit results for the number density of
galaxies. The first column gives the galaxy on which the profile
is centred. The second column shows how the galaxies were selected,
either by the probability method ($P$) or by exclusion of rectangular 
regions ($R$)
(see text). Column (3) gives the morphological type selected for the
profile, ``all'' means that no type selection is applied. The binning
(column 4) states how many galaxies are in each bin. Column (5) gives
the outer cut-off radius. The fit parameters: central surface 
number density, core radius and slope
parameter $\beta$ are listed in columns (6), (7) and (8),
respectively, with corresponding 1$\sigma$ errors and $\chi^2$ in
column (9). 
}
\halign{#\quad\hfil&#\quad\hfil&#\quad\hfil&#\quad\hfil&#\quad\hfil&#\quad\hfil
&#\quad\hfil&#\quad\hfil&# \cr
\noalign{\hrule\smallskip}\cr
centre & selection&type & binning& radius & 
$n_0$ & $r_c$ & $\beta$& reduced $\chi^2$\cr
       &          &     &        & (arcmin)     &
       ($10^{-4}$/kpc$^2$)& (arcmin) && \cr
\noalign{\smallskip\hrule\smallskip}\cr
M87 & P&all  & 50& 250&$ 4.0\pm0.5$ &$ 83\pm24$ &$ 0.93\pm0.13$ & 0.85\cr  
M87 & R&all  & 50& 250&$ 4.0\pm0.3$ &$ 65\pm6 $ &$ 0.75\pm0.03$ & 0.44\cr  
M87 & R&E+S0 & 10& 250&$ 0.43\pm0.3$&$ 41\pm65$ &$ 0.79\pm0.33$ & 1.97\cr
M87 & P&dE+dS0&25& 250&$ 4.1\pm 0.6$&$ 84\pm26$ &$ 0.98\pm0.15$ & 0.80\cr
M87 & R&dE+dS0&25& 250&$ 3.5\pm 0.6$&$ 62\pm24$ &$ 0.79\pm0.10$ & 0.88\cr
M87 & R&Sp+Irr&25& 250&$0.25\pm0.01$&$5000\pm20$&$80\pm0.5  $ & 0.28\cr
M49 & P&all  & 25& 125&$ 3.7\pm0.2 $&$ 28\pm2 $ &$ 0.67\pm0.02$ & 0.81\cr
M49 & R&all  & 25& 200&$ 2.9\pm0.6 $&$ 55\pm25$ &$ 0.89\pm0.17$ & 0.86\cr
M49 & R&dE+dS0&25& 200&$ 2.7\pm0.4 $&$ 30\pm5 $ &$ 0.74\pm0.04$ & 0.43\cr
M49 & R&Sp+Irr&16& 200&$0.45\pm0.04$&$3100\pm200$&$ 340\pm40   $ & 0.62\cr
M86 & P&all  & 13& 167&$ 2400\pm300 $&$0.25\pm0.03$&$0.91\pm0.01$ & 4.03\cr
%& & & & & & & &\cr
\noalign{\smallskip\hrule}\cr
}
\endtab

The principal caveat with the smoothed luminosity profiles is of course that
in the inner part the profiles are entirely determined by the smoothed
central giant galaxies. The core radii simply reflect the smoothing scale
length. We have encountered this problem before in Sect.~3 in connection
with Fig.~3. The effect is obvious also in Fig.~5 (right panels).
To see the
exact contribution of the central galaxies we have subtracted their light
from the luminosity profiles and then smoothed the rest as before (open
squares in Fig.~6, right panels). Removing these central giants
renders the luminosity profiles crater-shaped, at least for the M87 and the M49
subclusters where the central galaxies are very massive and dominant.
For the M86 subcluster the effect is less strong.

To get a true luminosity profile for the inner part of a subcluster, one
has to involve the actual light profile of the central galaxy itself 
(assuming, as we do, that the subcluster is exactly centred on it).
Surface brightness profiles, parametrized by the generalized 
de\,Vaucouleurs law $r^{1/n}$, are provided by Caon et al.~(1993) for 
all three giants M87, M49, and M86. Their $B$-band luminosity profiles
based on the Caon et al.~data are shown as dash-dot lines in Fig.~6
(right panels). Clearly, and perhaps not surprisingly, the profiles of 
M87 and M49 are much steeper than the profiles of their embedding subclusters.
It should be noted that the parametrized Caon et al. (1993) profiles are
accurate enough for our purposes; they were also checked to reproduce the 
observed total luminosities when integrated.  

These galaxy profiles can now be added to the decapitated subcluster 
luminosity profiles (i.e.~the profiles constructed without central galaxies). 
The resulting ``true'' luminosity
profiles are shown as crosses in Fig.~6. The profiles of the M87 and M49
subclusters are now no longer monotonic but have a local minimum 
(between $30\arcmin$ and $50\arcmin$) and a local maximum (between $60\arcmin$
and $70\arcmin$). The profile of the M86 subcluster remains monotonic because
M86 itself is rather shallow. This is indeed the only subcluster whose ``true''
luminosity profile can be fitted by a $\beta$-model (the dotted line in 
Fig.~6f). 

The failure of the $\beta$-model for the M87 and M49 subcluster 
luminosity profiles unfortunately 
deprives us of a simple deprojection (by Eq.(3)) and direct
comparison with the gas density profiles. For such a comparison we have 
therefore to rely either on the number density profiles or again on the
smoothed luminosity density profiles; in the latter case, however, it is 
correct only for the outer part, say $r \le 70\arcmin$.

\titleb{Comparison of galaxy number density and luminosity density}

The profiles in Fig.~6 are shown with varying scale, making a comparison
between number and luminosity density difficult. For the M87 subcluster,
the number density and the smoothed luminosity density profiles (for 
both $R$ and $P$) are plotted as deprojected, spatial densities in 
the same Fig.~7. As the profiles are centred
on the brightest galaxy of the subcluster, there
is the trend of the luminosity profile
to be steeper than the number density profile in the inner part -- even with
smoothed luminosities. 
The same trend holds for the other
subclusters as well, which can be read from Tables 2 and 3: 
for the luminosity profiles $\beta$ is larger while the core 
radius $r_c$ is at the same time
smaller by a factor of $\approx$ 2 than for the corresponding 
number density profiles. However, in the outer parts ($r >$ 70\arcmin),
number density and luminosity density are following nearly parallel curves.
The difference is really only due to the central giant galaxies. There is no
general luminosity segregation in the cluster. The galaxies, by number, 
are supposed to trace the gravitational potential of the cluster in a 
non-dissipative way. 
Their profile, if transformed in mass units, is indeed similar to the 
gravitational mass profile (see Sect.~6). In contrast, the central galaxies
themselves (like most individual galaxies) underwent strong dissipation,
leading to a steep luminosity profile -- steeper than the luminosity profile
of the surrounding cluster, as we have seen above (Fig.~6, right panels.).
The deprojected ``true'' luminosity profile (from Fig.~6b) would of course 
look still much steeper than the smoothed one in Fig.~7.

The core radii of the number density fits, 480 and 380 kpc for $P$ and
$R$ - profiles, respectively, are somewhat larger than the standard value
for clusters of galaxies
of 250 kpc (Bahcall 1975), while the core radii of the 
luminosity density fits are
slightly smaller than this (160 and 150 kpc, respectively). 
All our core radii are significantly smaller than the value given by
BTS87 with 650 kpc. One reason for this difference is certainly the fixed 
$\beta$-value in the fit of BTS87. $\beta$ and $r_c$ 
are strongly correlated, i.e. the $\chi^2$-ellipse of the fit is very elongated
and the orientation is such that if one parameter is increased, the other
increases as well. BTS87 used a large
$\beta$ ($\beta=1$) compared to ours, therefore they found a larger
core radius. Another reason might be that in BTS87 the galaxies of the
M86 subcluster were
not excluded, and the centre of the radial bins was positioned between M87
and M86.
Our central luminosity density, with
$5.2\times 10^{11} \lsol/$Mpc$^3$ for $P$-profiles and 
$5.6\times 10^{11} \lsol/$Mpc$^3$ for $R$-profiles, is, however, quite  
similar to the one found by BTS87 ($5\times10^{11} \lsol/$Mpc$^3$).

In Figs. 2, 3 and 5 one sees a second concentration in the M87
subcluster around M60. 
Excluding the region around M60 results in an
increase of $\beta$ by 2-3 \% for the number density and an increase
by at most 10\% for the luminosity density, leaving, however, the
general trends shown in Fig.~7 unchanged. 
For all data given in the tables and
shown in the figures, the M60 emission is included to make the optical
results directly
comparable to the X-ray results.

From Fig.~7 we note again the close agreement between $R$ and $P$-profiles
even for spatial densities.

\begtabfullwid
\tabcap{3}{Profile parameters and fit results for the luminosity 
distribution of galaxies. Columns (1)-(4) same as Table 2. Pc in column (2)
denotes a $P$-profile with the observed luminosity 
profile of the central galaxy incorporated. 
The fit parameters: central surface luminosity density, core radius and slope
parameter $\beta$ are listed in columns (5), (6) and (7),
respectively, with the corresponding 1$\sigma$ errors and $\chi^2$ in
column (8).
}
\halign{#\quad\hfil&#\quad\hfil&#\quad\hfil&#\quad\hfil&#\quad\hfil
&#\quad\hfil&#\quad\hfil&# \cr
\noalign{\hrule\smallskip}\cr
centre & selection&type  &radius & $L_0$ & $r_c$ & $\beta$& reduced $\chi^2$ \cr
       &&&(arcmin)&($10^5\lsol/$kpc$^2$)&(arcmin)&& \cr
\noalign{\smallskip\hrule\smallskip}\cr
M87    & P&all  & 250    &$ 5.2\pm1.4 $&$ 28\pm15$&$ 0.68\pm0.07 $& 0.54\cr  
M87    & R&all  & 250    &$ 5.6\pm0.1 $&$ 25\pm1 $&$ 0.63\pm0.01 $& 0.42\cr
M87    & R&E+S0 & 250    &$ 5.4\pm1.6 $&$ 22\pm9 $&$ 0.81\pm0.06 $& 1.81\cr
M87    & R&dE+dS0&250    &$0.48\pm0.12$&$ 26\pm14$&$ 0.65\pm0.07 $& 0.53\cr
M49    & P&all  & 125    &$ 6.7\pm1.3 $&$ 29\pm1 $&$ 0.92\pm0.01 $& 0.21\cr
M49    & R&all  & 200    &$ 5.3\pm0.3 $&$ 57\pm2 $&$ 1.31\pm0.02 $& 0.71\cr
M86    & P&all  & 100    &$ 4.5\pm0.8 $&$ 56\pm17$ &$ 2.6\pm0.8  $& 0.82\cr   
M86    & R&all  &  66    &$ 5.8\pm0.1 $&$\approx 1000$&$\approx 600$&-\cr
M86    &Pc&all  & 100    &$ 5.3\pm1.5 $&$ 19\pm6 $ &$ 1.2\pm0.1  $& 0.56\cr
\noalign{\smallskip\hrule}\cr
}
\endtab

\begfig 0.0cm
\psfig{figure=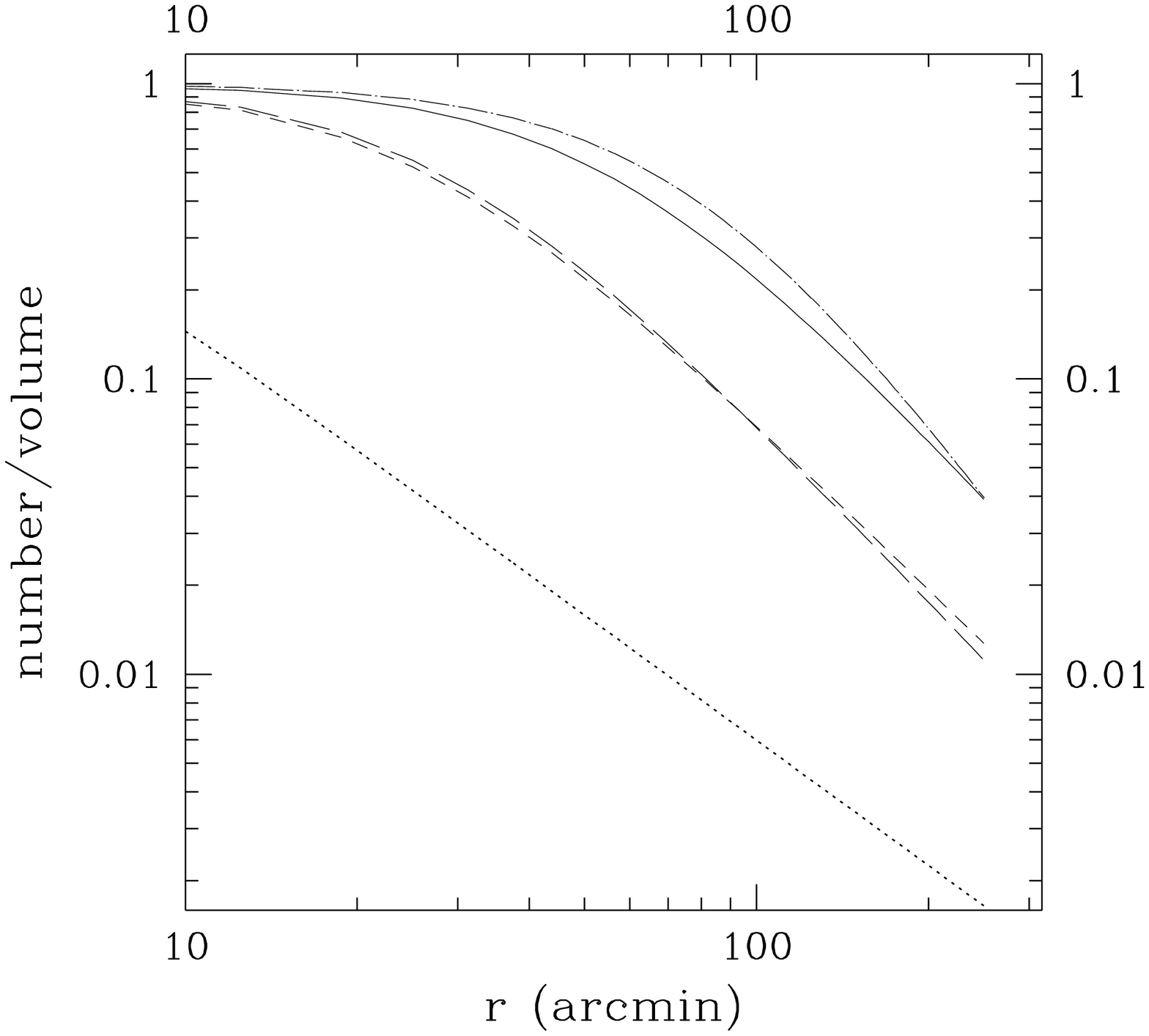,width=8.0cm,clip=} 
\figure{7}{Deprojected (spatial) density profiles for the M87 subcluster. 
All profiles are
normalized to a central value of 1. The two upper lines show the
number density of galaxies (dot-dashed line: $P$-profile, solid line: 
$R$-profile). The two lines in the centre show the luminosity 
density (long dashed line: $P$-profile, short dashed line: $R$-profile). 
The lower straight
line (dotted) is the 3D gas density which is much steeper in the central part
than the galaxy distribution because of a very small core radius.
}
\endfig

\titleb{Gas versus galaxies}

We can compare the gas distribution with the galaxy distribution after
deprojecting both in the way described in Sect. 5.1. The profile parameters
and fit results for the X-ray profiles of the M87 and M49 subclusters
are presented in Table 4. For the M86 subcluster no reliable X-ray profile
could be determined. Note that the X-ray data were not smoothed here
with $\sigma = 24\arcmin$ as they were for Fig.~1; the ROSAT All-Sky Survey
resolution is better than 1\arcmin. The deprojected, spatial
gas density profile for the M87 subcluster is shown in Fig.~7 along with the 
galaxy number and {\it smoothed}\/ luminosity 
densities discussed above. The situation for the M49 subcluster is very
similar (see also Fig.~8). 

Clearly, in the central part (inside 50 to 70\arcmin) the gas density 
falls off much more steeply than either optical profile. Due to the 
normalization of $\rho_0$ = 1, this is seen by the fact that the gas profile 
lies much below the optical profiles. 
The gas profile has a very small core radius (2\farcm7 or
16~kpc, see Table 4), which is an 
order of magnitude smaller than the core radius of the smoothed luminosity
distribution, although the core radius is by far not resolved in the ROSAT
All-Sky Survey: an HRI observation yields a core radius of 10\arcsec\  
%%%
(1~kpc). 
%%%
The artificial flattening due to the Survey point spread
function does of course not affect the X-ray surface profile shown here.  

Is the gas profile also steeper than the ``true'' luminosity profile in the
inner part, i.e.~the luminosity profile of the central galaxy M87 itself?
(Note that only the smoothed 3D luminosity profile is shown in Fig.~7.) 
The answer, also for M49, is no. The profiles of the central giant galaxies
are steeper than the gas density profiles. How can we know without deprojecting
the galaxy profiles? Instead of doing this, which is non-trivial without
$\beta$-model fits, one can invert the problem and treat the spatial gas 
density as a spatial galaxy density
and project it back by putting the X-ray $\beta$-model parameters from
Table 4 into Eq.(1), and then compare this with the actual, observed galaxy
profiles. A steeper slope in projection means also a steeper slope in 3D.
The projected gas density slopes, in the inner part, turn out to be somewhat 
steeper than the projected galaxy density profiles in Fig.~6b 
(M87) and 6d (M49).
But the luminosity profiles of M87 and M49 (even in projection, the dash-dot
lines in those figures) are still very much steeper than the projected gas
densities 

This is plausible. As mentioned before, while the galaxy number density
(and the smoothed luminosity density as well) essentially traces the 
gravitational potential of the cluster, the X-ray gas will have suffered 
a certain (modest) amount of dissipation, thereby steepening the gas profile. 
The principal agent of dissipation here is the cooling flow. The X-ray
component of M87 is indeed known to show a cooling flow with a mass 
accretion rate of about 10 $\msol$/yr within a
cooling flow radius of 100 kpc (Stewart et 
al. 1984; Nulsen \& B\"ohringer 1995). M49 is suspected to have a cooling
flow as well (Irwin \& Sarazin 1996).
But again, the galaxies themselves 
must have suffered much stronger dissipation still 
(by rapid cooling in an early
epoch) and therefore have much steeper profiles than the gas components.

In the outer parts (beyond 50 to 70\arcmin, see Fig.~7) the effect is
reversed. Here the gas profile is flatter 
%%%
than
%%% 
the galaxy density profiles.

\begtabfullwid
\tabcap{4}{X-ray profile parameters and fit results. The first column
gives the galaxy on which the profile is centred. Columns
(2) and (3) give the name of the region and the corresponding angular
selection. Column (4) gives the outer cut-off radius. The fit
parameters: central surface brightness, core radius and slope
parameter $\beta$ are listed in columns (5), (6) and (7), respectively.}
\halign{#\quad\hfil&#\quad\hfil&#\quad\hfil&#\quad\hfil&#\quad\hfil
&#\quad\hfil&#\quad\hfil&# \cr
\noalign{\hrule\smallskip}\cr
centre & region &sector& outer radius & $S_0$ & $r_c$ &
       $\beta$ \cr
       &        &(N over E)& (arcmin
%%%
[Mpc]
%%%
)&(counts/arcmin$^2$/s)& (arcmin) & &\cr
\noalign{\smallskip\hrule\smallskip}\cr
M87    & all    &$~~~0^\circ-360^\circ$  & 250~[1.5] & 0.21 & 2.7 & 0.47  \cr
M87    & SW     &$180^\circ-270^\circ$    & 250~[1.5] & 0.17 & 4.4 & 0.52  \cr
M87    & N      &$~~~0^\circ-45^\circ$   & 250~[1.5] & 0.25 & 2.4 & 0.46  \cr
M87    & S      &$135^\circ-225^\circ$    & 250~[1.5] & 0.25 & 2.4 & 0.47  \cr
M49    & all    &$~~~0^\circ-360^\circ$  & 30~[0.17]  & 0.029& 3.0 & 0.76  \cr
%M49 (outer part)& all    &$~~~~0^\circ-360^\circ$  &125 &$3\times10^{-4}$& 19.2 
& 0.51 \cr
%& & & & & & & \cr
\noalign{\smallskip\hrule}\cr
}
\endtab

\titleb{The subclusters in comparison}

Already from a glance at Fig.~5 (left panels) we have noted
a systematic trend in the appearance of the three subclusters (Sect.~4),
in the sense that the poorer (less populous) a subcluster the more compact
it appears. This can now be quantified by a comparison of the density 
profiles derived for the three subclusters. 

In Fig.~8 the deprojected, normalized profiles of the subclusters are
shown. Consider first the optical profiles. Up to a radius of about 
10 - 15\arcmin\ the galaxy number density profiles
are similar, but in the outer parts the fits confirm the tendency
already noted by eye: the
poorer the subcluster the steeper the radial profile. 
For clarity only the $P$-number density profiles are shown in
Fig.~8, but the $R$-profiles, as well as the luminosity profiles 
show the same systematic trend.

To compare also the X-ray components of different subclusters,
we tried to fit also the X-ray
emission around M49 with a $\beta$-model, although there are clearly
asymmetries in the X-ray halo (see Irwin \& Sarazin 1996). The fit
parameters are listed in Table 4. The core radius of the gas profile
of the M49 subcluster, with 3\farcm0 or 17 kpc, is almost as small as the 
one of M87, which is a signature that M49 has a cooling flow as well.

Furthermore, there is a faint extended X-ray
component at larger radii around M49 with the following parameters:
$S_0=3\times10^{-4}$counts/\sq\arcmin/s, $r_c=19\arcmin$, and
$\beta$=0.51 for a fit between 30 and 125$\arcmin$. However, that far out
the X-ray emission is already less than 30\% of the background level, and
we therefore neglect it for the following analysis.

The X-ray halo around M86 is even more asymmetric with its plume in
the North-West (Rangarajan et al. 1995). As
the $\beta$-model assumes spherical symmetry we do not present a
$\beta$-fit of the X-ray emission around M86 here.

The tendency seen in the optical -- a steeper profile in
poorer subclusters  -- is clearly 
reflected in the X-ray profiles of M87 and M49 (see Fig.~8). 
The luminosity profiles ($R$ and $P$) and X-ray profiles of the M49 subcluster
all have much larger
$\beta$ values than the corresponding profiles of the M87 subcluster,
demonstrating its stronger concentration (see Tables 2, 3 and 4). 

The compactness of the M86 subcluster is even more extreme 
(see Fig.~8), although 
the fit to the number
density profile has to be regarded with some caution as the profile
consists only of 4 bins. The luminosity $P$-profile had to be restricted
to a radius of $100\arcmin$ because of a second bright galaxy at a
distance of $150\arcmin$ from M86 (see Fig.~5d). 
Excluding that galaxy the fit
gave plausible results (Table 3). For the luminosity $R$-profile,
where the influence by the M87 subcluster was subtracted, the data could be
used out to a radius of merely 66\arcmin, resulting again in 4 bins.
However, the compactness of the M86 subcluster is unlikely to be an
artifact of the selection procedure but is probably real, as also some 
galaxies far away (out to 175\arcmin)
from M86 were selected to belong to it.

%In Fig.~8 one sees again (compare Fig.~7) the difference of the optical
%and X-ray profiles. The X-ray profiles are more concentrated towards
%the centre,
%because the X-ray core radii are smaller. In the outer parts the slope
%of the optical distribution is again slightly steeper.

\begfig 0.0cm
\psfig{figure=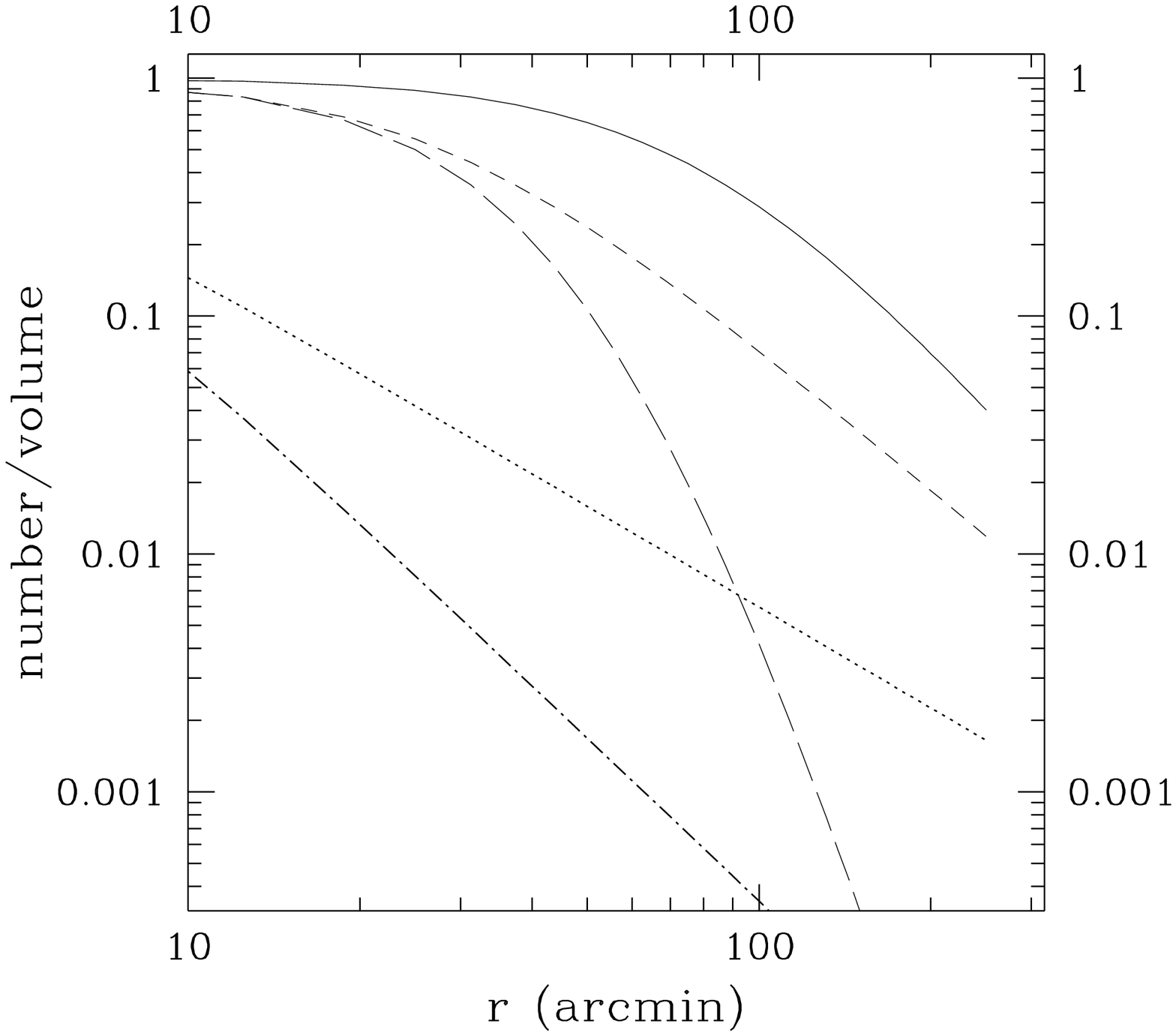,width=8.0cm,clip=} 
\figure{8}{Deprojected density profiles ($P$-profiles) 
of the various subclusters. All profiles are
normalized to a central value of 1. 
The three upper lines show the
number density of galaxies (solid line: M87 subcluster, short dashed
line: M49 subcluster, long dashed line: M86 subcluster). The two lower
lines show the 3D gas density derived from X-rays
of the M87 subcluster (dotted line) and
the M49 subcluster (dot-dashed line). Obviously, the poorer the
subcluster the steeper is its profile, i.e. the more 
concentrated it is. 
}
\endfig

\titleb{Distribution of the Hubble types}

From the different distributions of the various Hubble types in
Fig.~2 (number densities) and Fig.~3 (luminosities) one expects 
also different profiles. $\beta$-model parameters for different types in 
the M87 and M49 subclusters, for cases where such fitting was feasible 
at all, are given
in Tables 2 and 3. The profiles for number densities of the
different Hubble types around M87 are shown in Fig.~9. 
The trends, in accord with the morphology-density relation (Dressler 1980,
also BTS87), are evident. While the
profile for dwarf ellipticals is very similar to the all-type
profile, the profile for elliptical plus S0s is somewhat steeper than both,
because these types are most concentrated in the cluster.
However, they are still much less 
concentrated than the gas in the inner part. Beyond $r \approx 50\arcmin$
this turns around and the gas component becomes shallower than these types.
However, the gas component remains everywhere much steeper than 
spirals and irregulars, 
whose distribution
is most dispersed. The luminosity profiles (not plotted) show the same
trends. 

\begfig 0.0cm
\psfig{figure=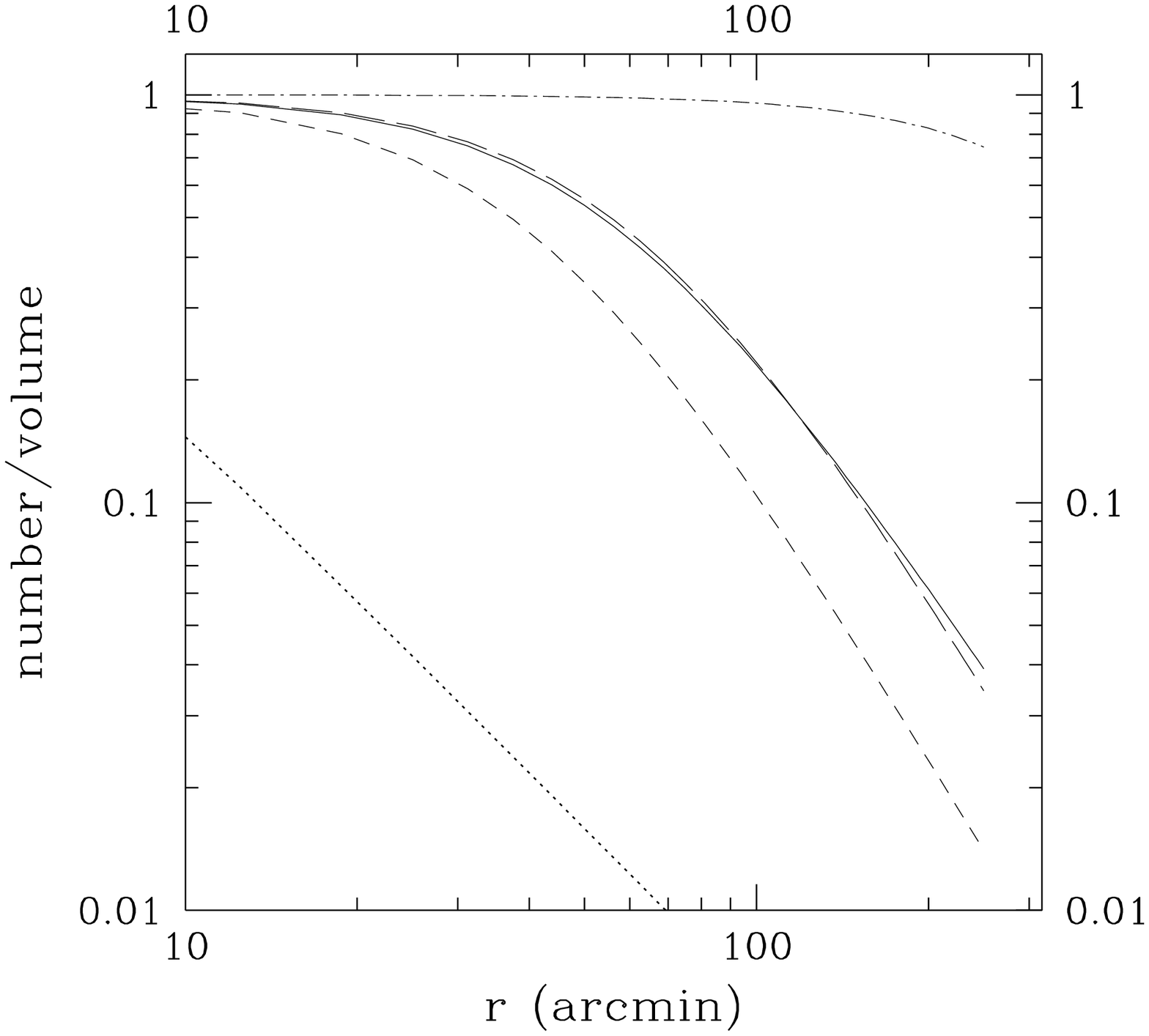,width=8.0cm,clip=} 
\figure{9}{Deprojected number density profiles ($R$-profiles) 
of the various morphological types of galaxies. The profiles are
centred on M87 and are normalized to a central density of 1. 
The profile for dwarf elliptical galaxies (long dashed line) is
very similar to the combined profile for all types (solid line). The
profile for elliptical and S0 galaxies (short dashed line) is somewhat
steeper, reflecting the compact distribution. On
the other hand, the
spiral and irregular galaxies (dot-dashed line) show a very flat profile.
For comparison also the 3D gas density of the M87 subcluster 
is shown (dotted line).
}
\endfig

\titleb{Different regions around M87}

A closer look at the X-ray emission around M87 (Figs.~1 and 2e) 
shows that it is not
exactly spherically symmetric, but shows a bridge in the direction to M49,
an elongation towards North, and a steep cut-off in the
South-West (SW). To see the effect of these asymmetries, we have fitted 
X-ray profiles for different sectors of the M87 subcluster. The results
are given in Table 4 and shown in Fig.~10. 
The region in the SW turns out to have a larger
core radius and a larger $\beta$ than the other sectors. 
The profiles of the other regions do not look very
different from the overall profile.

For comparison, the same regions around M87 were extracted from the
optical data. The corresponding optical profile parameters and fit results 
are listed in
Table~5, and the profiles are again shown in Fig.~10. 
Obviously, the fits of the Northern and the Southern sector
do not give significant results: the errors are exceedingly large 
due to the low number of galaxies.
Only for the SW sector could an acceptable fit be achieved -- 
again (as for the X-rays) with a large core
radius and a large $\beta$. In Fig.~10 this shows up by a SW profile which
is slightly steeper then the overall profile at large radii (more pronounced
in the luminosity profile). This is the steep drop in density to the SW
visible in Figs.~2a and 2e.

\begfig 0.0cm
\psfig{figure=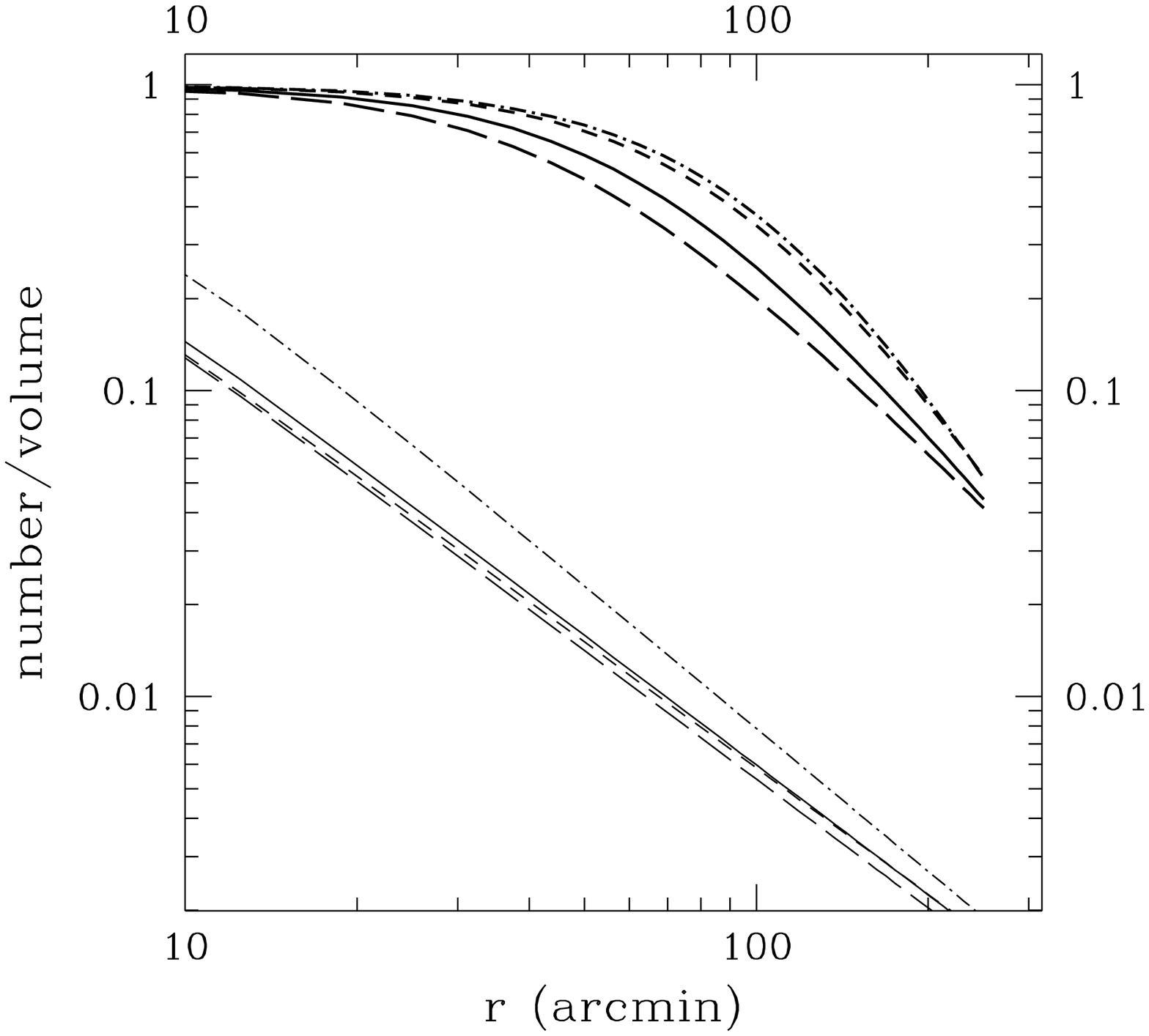,width=8.0cm,clip=} 
\figure{10}{Deprojected profiles 
of various regions (sectors) around M87. All profiles are
normalized to a central density of 1. 
The four upper profiles (bold lines) show the number density
of the galaxies ($R$-profiles) while the four 
lower profiles (thin lines) show the
gas density: Total cluster (solid line), North (short dashed
line), South (long dashed line) and South-West (dot-dashed line). For the
definitions of the sectors see Tables~4 and 5. 
}
\endfig

\begtabfullwid
\tabcap{5}{Profile parameters and fit results for different regions
around M87. All numbers listed are for $R$-profiles with an outer cutoff
radius of 250\arcmin. Column (1) indicates whether the projected
number density or the projected luminosity density was fitted. Columns
(2) and (3) give the name of the region and the corresponding angular
selection. The binning (Column 4) is the number of galaxies per bin
for the number densities and the constant bin size of $25\arcmin$
for the luminosity densities. Column (5) is the central 
projected number density
and column (6) the central projected luminosity density. The core radius and
$\beta$ are listed in columns (7) and (8). In column (9) the reduced
$\chi^2$ is given.
}
\halign{#\quad\hfil&#\quad\hfil&#\quad\hfil&#\quad\hfil&#\quad\hfil&#\quad\hfil
&#\quad\hfil&#\quad\hfil&# \cr
\noalign{\hrule\smallskip}\cr
density &region&sector& binning & $n_0$&$L_0$ & $r_c$ &$\beta$& reduced 
$\chi^2$\cr
        &      &(N over E)&   &($10^{-4}/$kpc$^{2}$)&($10^5\lsol/$kpc$^2$)
&(arcmin)& \cr
\noalign{\smallskip\hrule\smallskip}\cr
number    &all&$~~~0^\circ-360^\circ$& 50  &$ 4.0\pm0.3$ &&$ 65\pm6$ & 
$0.75\pm0.03$ & 0.44\cr
luminosity&all&$~~~0^\circ-360^\circ$&const&&$ 5.6\pm0.1$ &$ 25\pm1$ & 
$0.63\pm0.01$ & 0.42\cr
number    &SW &$180^\circ-270^\circ$& 25  & $4.1\pm0.9$ &&$ 110\pm70$& 
$1.1\pm0.4 $ & 0.27\cr
luminosity&SW &$180^\circ-270^\circ$&const&& $3.6\pm0.6$ &$ 120\pm40$& 
$1.5\pm0.4 $ & 2.0 \cr
number    &N  &$~~~0^\circ-45^\circ$ & 25  & $3.6\pm1.7$ &&$ 95\pm115$ & 
$0.95\pm0.57$ & 0.37\cr
luminosity&N  &$~~~0^\circ-45^\circ $&const&& $3.1\pm0.3$ &$ 1100\pm600$& 
$40\pm40 $  & 2.5 \cr 
number    &S  &$135^\circ-225^\circ$& 25  & $3.8\pm1.5$ &&$ 47\pm52$ & 
$0.63\pm0.16 $& 1.5 \cr
luminosity&S  &$135^\circ-225^\circ$&const&& $8.3\pm8.2$ &$ 8.2\pm12$& 
$0.62\pm0.05 $& 0.60\cr
%& & & & & & & \cr
\noalign{\smallskip\hrule}\cr
}
\endtab

\titlea{Mass profiles and mass-to-light ratios}

From the galaxy luminosity densities we can easily construct radial mass
density and integrated mass profiles by assigning every galaxy a 
mass-to-light ratio. For simplicity, and because we are mainly interested
in the overall trends, we have adopted a constant mass-to-light
ratio of 20 for every galaxy type. This number is of course arbitrary
but seems realistic as an average (e.g.~Binney \& Tremaine 1987). 
%%%
Although there are hints that the mass-to-light ratio varies with the Hubble
type (e.g., Burstein et al. 1995; Gavazzi et al. 1996), it should be 
emphasized that the mass determinations involved there refer to the integrated
galaxy mass out to a given radius. The {\it total}\/ masses of galaxies, which 
we would need here, are essentially unknown. It seems therefore safest 
at present to assume a constant mass-to-light ratio.
Nevertheless, we should keep in mind that this simplifying assumption might 
introduce an additional uncertainty in the galaxy mass profiles because of the
morphological segregation.
%%%

The resulting profiles 
are shown in Fig.~11 (M87 subcluster) and Fig.~12 (M49 subcluster),
where differential mass density profiles are given in the lower panels (b) and
integrated (cumulative) mass profiles in the upper panels (a). 
A differentiation with respect to $P$ and $R$-selected profiles shows 
little difference. 

These mass profiles were constructed from the {\it smoothed}\/ luminosity
profiles, which strictly make sense only for the outer part, $r \ge$ 
50 - 70\arcmin (remember Sect.~5.2). Further in one would have to treat
the luminous mass of the central giant galaxies M87 and M49, which is beyond
our scope. The galaxy profiles shown in Figs.~11 and 12 therefore start only
from $r = 60\arcmin$, corresponding to 0.35 Mpc with the adopted distance scale.

\begfig 0.0cm
\psfig{figure=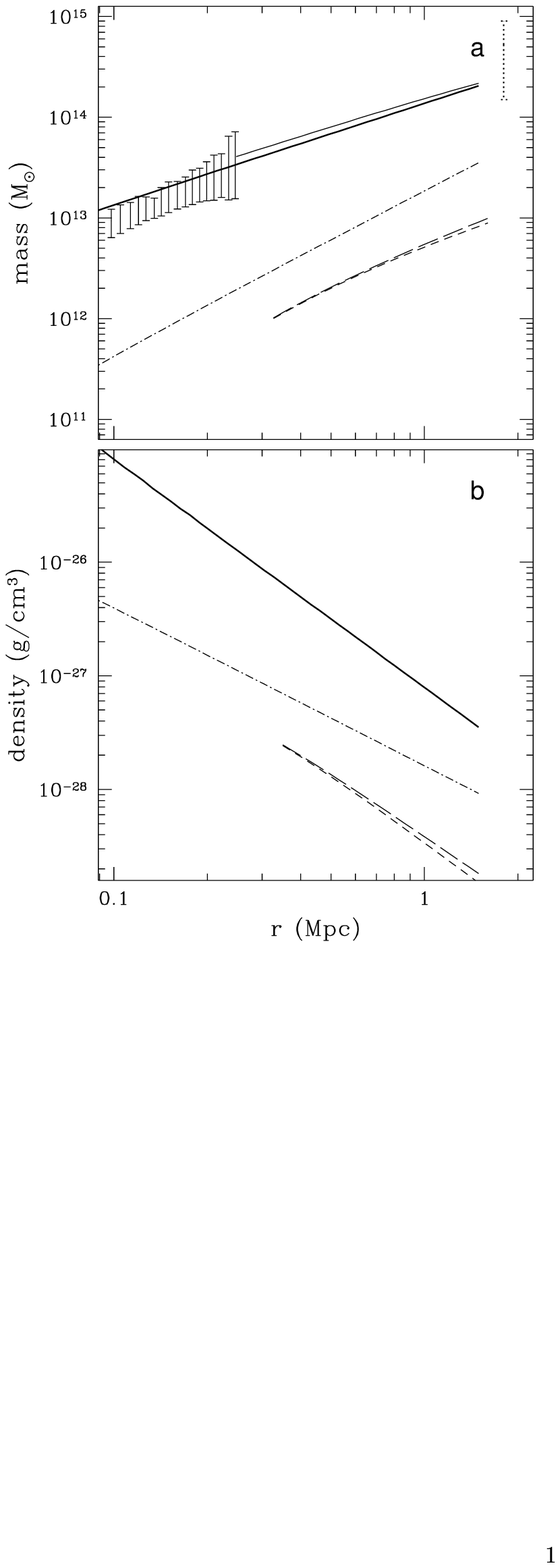,width=8.0cm,clip=} 
\figure{11}{Integrated mass (a) and differential mass density (b)
profiles around M87. Shown are the gas mass (dash-dotted lines),
the galaxy mass from a $P$-profile (short dashed line) and a $R$-profile
(long dashed line), as well as the total (gravitating) mass assuming 
isothermal cluster gas
(bold solid line). In panel (a) the thin solid line gives the 
integrated mass when
a temperature gradient is assumed (see text). The solid error bars are from
Nulsen \& B\"ohringer (1995), and the dotted error bar indicates the total
mass of the M87 subcluster estimated by B\"ohringer et al.~(1994).
}
\endfig

\begtabfullwid
\tabcap{6}{Masses of the subclusters centred on M87, M49 and M86
(integrated values). Galaxy masses are based on $P$-profiles with a
constant $\msol/\lsol$ of 20. The total masses
of the M87 and M49 subclusters were calculated on the assumption of an
isothermal gas. The total mass of 
the M86 subcluster is taken from B\"ohringer et al. (1994). The 1$\sigma$ 
error for the galaxy masses is $\approx$ 20\%, that for the total masses
is 30-50\% (except for M86).
}
\halign{#\quad\hfil&#\quad\hfil&#\quad\hfil&#\quad\hfil&#\quad\hfil&# \cr
\noalign{\hrule\smallskip}\cr
&\multispan2\hfil M87 \hfil&\multispan2\hfil M49 \hfil& M86        \cr
& ($r$=400kpc) & ($r$=1Mpc) & ($r$=400kpc)&($r$=750kpc) & ($r$=240kpc) \cr
\noalign{\smallskip\hrule\smallskip}\cr
$M_{tot} [10^{13}\msol]$ & 5.5 & 14   & 4.7 & 8.7  &  $1-3$ \cr
$M_{gas} [10^{13}\msol]$ & 0.42& 1.9  &0.026&0.044 &  -     \cr
$M_{gal} [10^{13}\msol]$ &0.14& 0.51 & 0.17&0.34  & 0.06   \cr
$M_{gas}/M_{tot}$        & 8\% & 14\% &0.6\%&0.5\% &  -     \cr
$M_{gal}/M_{tot}$        & 3\% &  4\% & 4\% & 4\%  & $2-6$\%\cr
$M_{gal}/M_{gas}$        &0.34 & 0.28 & 6.6 & 7.7  &  -     \cr
$M/L [\msol/\lsol]$      &$\approx 500$ &$\approx 500$&$\approx 500$
&$\approx 600$  & $300-1000$\cr
\noalign{\smallskip\hrule}\cr
}
\endtab

From the X-ray data we can derive mass profiles for the intra-cluster
gas components of the two subclusters, which are 
also shown in Figs.~11 and 12. 
In the M87 subcluster the gas mass is dominating over
the galaxy mass. 
At a radius of 1 Mpc the ratio $M_{gal}/M_{gas}$ is 0.28 and 0.29
for the $P$- and the $R$-profile, respectively (see Table 6 for 
various mass ratios). 
These numbers are uncertain by $\approx$ 20\%, which stems entirely from
the uncertainty in the fitted luminosity profiles. The errors in the
gas mass profiles are very small. The case is reversed for the M49 subcluster:
here the gas mass is only a fraction of the mass in galaxies; at 750 kpc
we have $M_{gal}/M_{gas} \approx$ 8 (Table 6).

With the additional assumption of hydrostatic equilibrium we can use the
X-ray gas as tracer of the total, gravitating mass profiles of the 
subclusters and derive these quantities by simply plugging our
$\beta$-models with the parameters given in Table 4 into the
well-known formula
$$
M_{tot}(r) = {-kr\over \mu m_p G} T \left({ d \ln \rho \over d \ln r }+
                                    { d \ln T    \over d \ln r }\right),
   \eqno(4)
$$
where $\rho$ and $T$ are the density and the temperature of the
intra-cluster gas, and $r$, $k$, $\mu$, $m_p$, and $G$ are the  
radius, the Boltzmann constant,
the molecular weight, the proton mass, and
the gravitational constant, respectively. 

In Fig.~11 we show an integrated
mass profile of the M87 subcluster (bold solid line) 
assuming an isothermal cluster of 2.7 keV. According
to B\"ohringer et al.~(1994), this temperature is
acceptable for the radius range considered here (the 
cooling flow region is not included in this analysis).
We show also a mass profile for a
constant temperature gradient taken from the ROSAT data, 
which is a good approximation only for
$r >$ 240 kpc (the thin solid line). This temperature gradient is roughly in
agreement with ASCA results of observations in North-West direction of
M87 (Matsumoto 1998; Ohashi et al. 1998).
Obviously, both profiles are very similar. 
We can compare these profiles with other X-ray derived masses (see Fig.~11a): 
the estimate at a large radius by B\"ohringer et al. (1994), also 
from the ROSAT All-Sky Survey data, and the detailed mass estimate by
Nulsen \& B\"ohringer (1995) derived from a deep pointed ROSAT/PSPC
observation with 30000sec. Within the errors, both mass estimates agree with
our total mass profile.
The error bars
give an idea of the uncertainty in the total mass derived in this
work. The largest uncertainty, with 30\%, is probably in the
temperature (B\"ohringer et al. 1994), so that we
estimate an uncertainty in the total mass of 30-50\%.
Because of this large error, all
results from this mass analysis must be taken with caution, showing
tendencies rather than exact values.
The masses for two different radii are listed in Table~6.

The total, gravitating mass is overwhelmingly larger than either the gas
or galaxy mass, which is of course due to the presence of dark matter.
The total mass profile is also slightly flatter than the 
gas mass profile in the M87 subcluster (see Fig.~11a). 
Within 400 kpc the gas mass accounts 
%%%
for
%%% 
8\% of the total mass, at 1 Mpc
this fraction becomes 14\% (Table 6), i.e.~the gas-to-total mass ratio
is increasing outwards, which is obvious from Fig.~11. Due to the
large uncertainty in the total mass the increase between these two
radii is not very significant. But when compared to the ratio at
100~kpc (3\%) there is certainly a trend visible.
The galaxy-to-total mass
ratio, on the other hand, is only slightly increasing with radius:
at 400 kpc, $M_{galaxies}/M_{tot}$= 3\%, while at 1 Mpc, 
$M_{galaxies}/M_{tot}$ = 4\%. 
%These are relatively low
%$M_{gas}/M_{tot}$ ratios while the $M_{galaxies}/M_{tot}$ ratios are
%in the expected range. 

The total mass profile of the M49 subcluster (Fig.~12) was
calculated on the assumption of an isothermal gas with temperature 1.4
keV. This temperature was found by Irwin \& Sarazin (1996) in the
region around M49 (excluding the central part of M49). 

\begfig 0.0cm
\psfig{figure=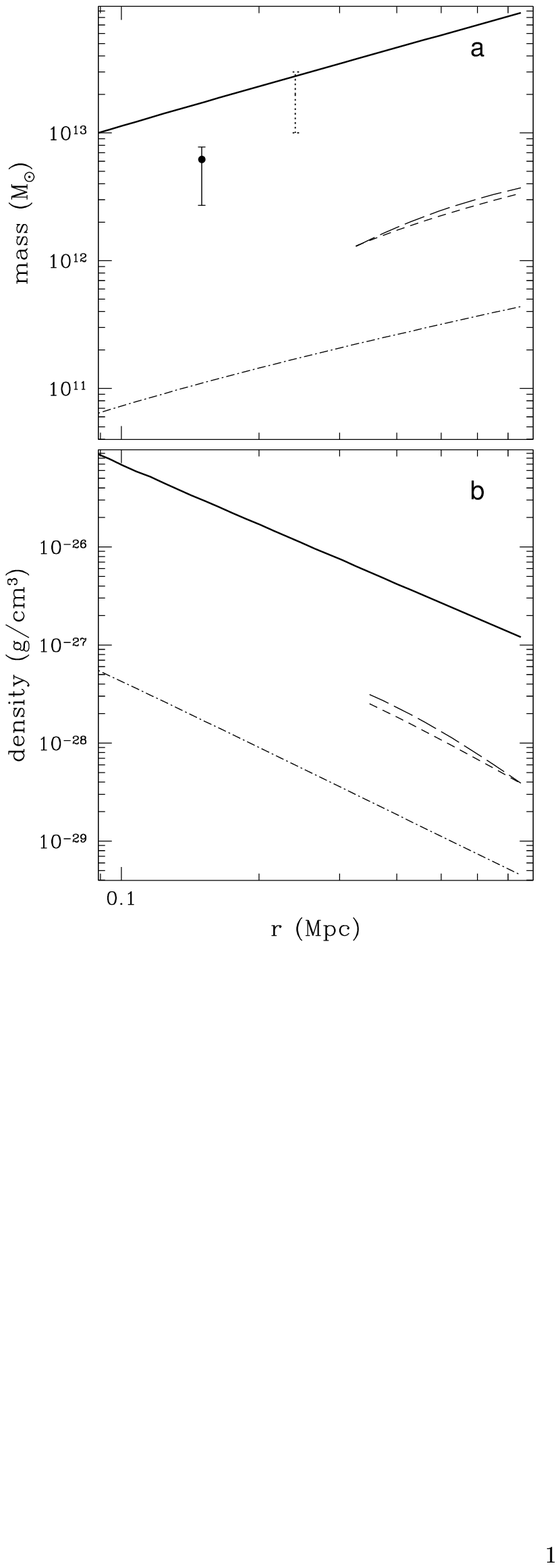,width=8.0cm,clip=} 
\figure{12}{Integrated mass (a) and differential mass density (b)
profiles around M49. Symbols are the same as in Fig.~11.
The solid error bar and data point in panel (a) 
shows the outermost mass estimate 
by Irwin \& Sarazin (1996). The dotted error bar 
indicates the total mass given in B\"ohringer et al. (1994).
}
\endfig

Figure 12a shows also a comparison with 
other X-ray derived masses. The mass
estimate by B\"ohringer et al. (1994) was again derived
from the ROSAT All-Sky Survey data and is consistent with our estimate
for the total mass (the dotted error bar).
A mass estimate by Irwin
\& Sarazin (1996) based on ROSAT/PSPC and HRI pointed observations, after
a scaling to the distance adopted here, 
disagrees with our mass profile by a factor of three (the data point with a
solid error bar). We have no explanation for this discrepancy.

Compared to the M87 subcluster, the total mass profile of the M49 subcluster
is systematically shifted to smaller values (see also Table~6). 
To compare the total, overall masses of the two subclusters, one has to take
into account 
%%%
the fact
%%%
that the M87 subcluster is more
extended than the M49 subcluster. If one assumes a radius of 1.5 Mpc for
the M87 subcluster and 0.75 Mpc for the M49 subcluster, one finds masses 
of $M_{tot,M87}=2.1\times10^{14}\msol$ and
$M_{tot,M49}=8.7\times10^{13}\msol$, rendering the M87
subcluster about
2.4 times more massive than the M49 subcluster.

As noted above, the gas mass of the M49 subcluster is rather small, and
even smaller than the galaxy mass.
The gas mass profile in the M49 subcluster is also somewhat flatter than 
in the M87 subcluster (Fig.~12a), 
which comes from a rather steep density gradient (see Fig.~12b; note the
different radius scale when comparing between Fig.~11 and Fig.~12). 
The gas mass in the M49 subcluster is
always less than 1\% of the total mass, which is much less than in the M87 
subcluster (8 - 14\%) and
in other clusters. In contrast, the galaxy-to-total mass fraction is similar
to the M87 subcluster (around 4\%, see Table 6).

The galaxy masses are slightly different here for
the $P$ and the $R$-profiles. This difference is an indication of how large
the errors are, which is hard to quantify otherwise. The difference
in the integrated galaxy masses ($P - R$)/$P$ is at maximum 36\%.

For comparison, a mass estimate for the M86 subcluster, taken from
B\"ohringer et al. (1994) is also 
listed in Table~6. The total masses and galaxy 
masses are of course smaller than for the other two subclusters, but the
ratio is very similar.
%velocity dispersions A 760 km/s
%                     B 390 km/s

From the total mass profile and the (galaxy) luminosity profile we can 
calculate how the
mass-to-light ratio changes with radius. This is shown in Fig.~13.
Global mass-to-light ratios for the three subclusters are also given in 
Table 6.
For the luminosities, $P$-profiles were used. We had again to restrict
the fitted luminosity profiles of
the M87 and M49 subclusters to $r >$ 350 kpc, to avoid M87 and M49 themselves.
Note that these $M/L$ profiles are differential, i.e.~are based on the 3D
density
profiles shown in Figs. 11b and 12b. Note also that this is independent
of the mass-to-light ratios of single galaxies; here we deal essentially
with the dark matter {\it between}\/ the galaxies.

From Fig.~13a we can see that the mass-to-light ratio of the M87 subcluster 
is around 500 $\msol/\lsol$ with
a tendency to decrease with radius. 
The ratio for the M49 subcluster is again around 500, but follows the
inverse trend with radius: between 350 and 750 kpc it increases from 400 to 
600 $\msol/\lsol$.
The mass-to-light ratio of the M86 subcluster is shown as large error bar,
which is from the mass error bar given by B\"ohringer et al. (1994).
The error in the mass-to-light ratio is dominated by the uncertainty
in the mass (30-50\%). 

For comparison we show also, in Fig.~13b, the mass-to-light ratio 
projected onto the plane of the sky, i.e.~calculated from the observed
luminosity, and the projected mass profiles. This way of presenting the 
ratio has two advantages.
(1) The optical data can be used directly without fitting them with a
model, so that we can use also the inner parts of the luminosity
profiles, i.e. the ones with the correct central galaxy profile.
(2) It is a quantity which can in principle be derived directly from
observations, e.g. by a weak lensing analysis. However, the disadvantage is
that we have to extrapolate the X-ray profiles to quite large radii in
order to include all the mass. For the M87 subcluster, an extrapolation 
radius of 5~Mpc was necessary, while for the M49 subcluster a radius of 
1.5~Mpc was found sufficient. 

\begfig 0.0cm
\psfig{figure=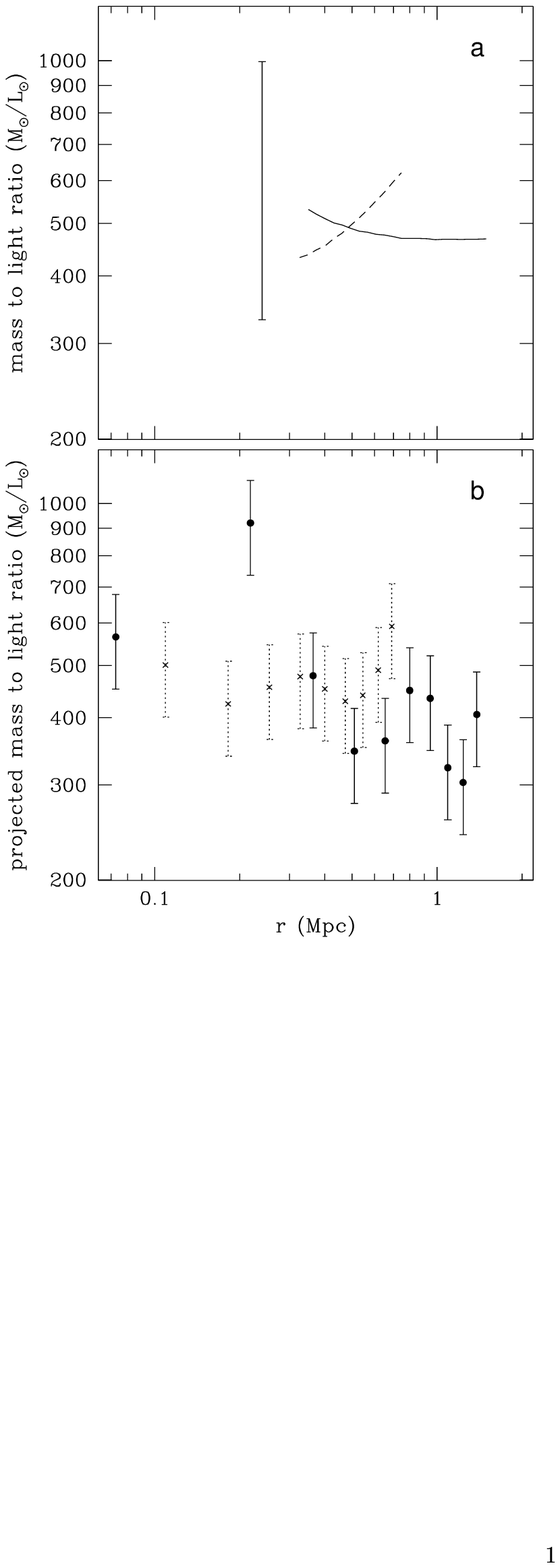,width=8.0cm,clip=} 
\figure{13}{a) 3D mass-to-light ratio profile for the M87 subcluster
(solid line) and the M49 subcluster (dashed line). 
The range of the mass-to-light ratio for the M86 subcluster is shown
as an error bar. b) 2D mass-to-light ratio (i.e.~projected onto the plane of
the sky) for the M87 subcluster (filled circles and solid error bars) and 
the M49 subcluster (crosses
and dotted error bars). The errors plotted are only the errors in the
galaxy luminosity profiles. The total uncertainties of all mass-to-light 
ratios (except for the M86 subcluster) are 30-50\%. 
}
\endfig

\titlea{Summary and Discussion}

The Virgo cluster, as the nearest cluster of galaxies, is ideally suited
for a detailed comparison of the distribution of two
components -- intra-cluster gas and galaxies. 
For this purpose we have used the X-ray data
from the ROSAT All-Sky Survey and the optical information from the VCC.
Following the morphological studies of Binggeli et al.~(1987,
1993) for galaxies, and B\"ohringer et al.~(1994) for the X-ray gas, we 
have made first a qualitative, global comparison, and then concentrated
on the three major subclusters of Virgo centred on M87, M49, and M86.
For both components we have fitted the observed subcluster 
profiles with isothermal $\beta$-models, which allows easy deprojection to 
three-dimensional densities.

Apart from the new comparison with the distribution of the X-ray gas, as well 
as the construction of mass density and mass-to-light ratio profiles which 
can be inferred from the X-ray gas,
we have extended the analysis by BTS87 in four respects: 
\noindent
(1) While BTS87 used a 
constant exponent of $-1$ for the projected galaxy density
(corresponding to a $\beta$ = 1), we have left
this parameter free for the fitting and -- as can be seen in the tables 
-- it indeed varies
considerably. 
\noindent
(2) We have developed a new technique (probability method $P$) 
for separating, i.e.~decomposing the subclusters, 
based on both spatial and velocity
information.
\noindent
(3) We have determined radial profiles not only around M87, but also
for the M49 and M86 subclusters. 
\noindent
(4) We have explored also profiles in different
sectors around M87.

The comparison of the X-ray and optical density profiles of the Virgo 
subclusters has led to the following results:

-- The X-ray profile is steeper than the optical profile in the central part
($r <$ 70\arcmin) but with a slight inverse trend at larger radii -- 
for both the M87 and M49 subclusters.
This is reflected in the much smaller X-ray core radii and the smaller
X-ray $\beta$-values than the corresponding optical values. The reasons
for the steep X-ray profiles in the central parts are probably the
cooling flows in M87 and M49.

A comparison with a model of isothermal gas in hydrostatic equilibrium in a
cluster where the dark matter is distributed with a NFW profile
(Navarro et al. 1995) shows
that the distributions of gas and dark matter
have the same slope at radii larger than the core radius, while the
gas distribution is flatter in the central part (Navarro et
al. 1996). Obviously, the galaxies are not distributed like dark
matter, but have a much flatter profile in the centre, even flatter than
the gas distribution (however, this excludes the central galaxy
itself whose luminosity profile is much steeper than the gas profile, 
and whose {\it mass}\/ profile may indeed be similar to 
the gas profile, cf.~Sect.~5.4). 
This effect can also
be seen from the mass-to-light ratio profile which is not
constant but shows a tendency of increasing mass for small radii.
The same effect is reported by 
Carlberg et al. (1997) who also found an increasing mass-to-light
ratio in the centre from their sample of 14 clusters. This effect
cannot be an artifact of the total-mass determination because
different methods were used. Carlberg et al. applied a virial approach for the
determination of the total mass, while our total
masses are based on X-ray observations.  

-- The more massive the subcluster the less compact is its structure - both in
optical and in X-rays, i.e.~poorer subclusters have a steeper radial profile.
This behaviour is in agreement with the results
of N-body simulations by Navarro et al. (1997) who also found  
steeper profiles for
low-mass haloes than for high-mass haloes.
The same systematic effect is well-known to hold also for normal
elliptical galaxies (Kormendy 1977, 1985), i.e. low-luminosity ellipticals
(like M32, but unlike ``dwarf'' ellipticals) appear much more compact
than giant ellipticals. However, it is not clear 
%%%
why 
%%%
dissipation-less
(sub)clusters and dissipational galaxies have so much in common, physically.

-- Different Hubble types show different slopes in the subcluster
   profiles. We confirm the results by BTS87 that
   early-type galaxies are much more strongly clustered than late-type
   galaxies, reflecting of course the well-known 
   morphology-density relation (Dressler 1980).

-- There is a region South-West of M87 which shows a steeper gradient
   than the rest of the M87 subcluster - both in the optical and in
   X-rays. Steep gradients in the X-ray emission can be caused by
   shock waves emerging after the collision and merging of subclusters
   (Schindler \& M\"uller 1993; Kulp \& Schindler 1998). However, in
   such cases one finds usually a bulge or elongation of the
   collisionless particles in this direction, i.e. contrary to what
   is found in the galaxy distribution here. One can find steep
   gradients in the same directions only for non-central collisions
   where everything is twisted so that at a certain radius the bulge of
   the galaxy distribution and the shocks in the gas have an offset of
   $90^\circ$. In this case, however, one would expect to see other
   twists in the gas distribution. Therefore, the shock scenario is not a
   viable explanation here. Shortly before the collision the gas
   and the galaxy distributions have the same orientation, but this is
   also an unlikely scenario for the subcluster M87, because there are
   no other indications (like, e.g., two very close concentrations,
   much closer than M87 to M86)
   of a forthcoming merger in the near future. It
   is possible though that a merger occurred a long time ago, at least
   3 Gigayears ago, and what one sees is still a slight elongation in the
   direction of the original collision axis (which in this
   case would be SE-NW). From simulations one
   finds that this small elongation has the same direction for both
   components and 
   remains stable for several Gigayears if there is no other event
   disturbing it. The reason why there is no steep slope on the other
   side of the assumed former collision axis (i.e. in the NE),
   might simply be due to an infall or passage of a small group on this side
   destroying the symmetry.

-- Differential and integrated profiles of the galaxy mass, gas mass, 
   and total gravitating mass density are presented
   for both the M87 subcluster and the M49 subcluster. For the galaxy
   mass density we have excluded
   the central 60\arcmin = 0.35 Mpc, because for that innermost region one 
   would have to consider the radial profile of the central galaxy (i.e.~
   M87 or M49) itself. The gas
   mass fraction in M87, with 8\% and 14\% at 400~kpc and 
   1~Mpc, respectively, is slightly on the low side for clusters, but
%%%
   is still in the normal range, cf. e.g. 
   B\"ohringer (1994, 5 clusters: 12-45\%);
   White \& Fabian (1995, 19 clusters: 10-22\%); 
   Buote \& Canizares (1996, 5 clusters: 8-22\%);
   Briel et al. (1992, Coma: 30$\pm$14\%);
   White et al. (1994, A478: 28\%);
   Elbaz et al. (1995, A2163: $\approx 30\%$);
   Mohr et al. (1996, A576: 8\%);
   Feretti et al. (1997, A2319: 20-35\%);
   Neumann \& B\"ohringer (1997, Cl0016+16: 14-32\%);
   Schindler \& Prieto (1997, A2634: 12$^{+10}_{-5}$\%);
   Sarazin et al. (1998, A2029: 26$\pm$14\%);
   Schindler et al. (1998, Cl0939+4713: 4-9\%) 
   (all scaled to $H_0 = 50$ km s$^{-1}$ Mpc$^{-1}$).
   One finds
   the usual behaviour that the gas distribution is somewhat flatter than the
   distribution of the total mass, i.e. the gas distribution is more
   extended. This tendency was found in many clusters e.g. 
   David et al. (1995, 10 clusters and groups); %2-30\%
   Allen et al. (1996, PKS0745-191); %18-23\%
   B\"ohringer et al. (1996, A3627); %8-58\%
   Sarazin \& McNamara (1997, A2597); %21\%
   Schindler et al. (1997, RXJ1347.5-1145); %34-52\%
   B\"ohringer et al. (1998, A2390). %14-40\%
   The integrated gas mass is about 3 times the galaxy mass. 
   This is about the same factor that was found in other clusters,
   e.g. 
   Henry et al.  (1993, A2256); 
   B\"ohringer (1995, Perseus);
   Bardelli et al. (1996, A3558).
   A comparison of the gravitating mass of the M87 subcluster
   ($M_{tot,M87}=2.1\times10^{14}\msol$) with the masses of other 
   clusters shows that they are in general more massive than the M87
   subcluster, e.g. the Coma and the Perseus cluster
   have almost a factor of ten more gravitating mass (B\"ohringer 1994).
%%%
   The galaxy mass
   density is getting flatter towards the centre, which is the reason
   why the mass-to-light ratio also 
   tends to increase with decreasing radius (but again excluding M87 itself,
   i.e.~for $r <$ 60\arcmin). 
   We find
   projected mass-to-light ratios between 300 and 500 $\msol/\lsol$ at radii
%%%
   larger than 200~kpc which are relatively large mass-to-light ratios
   compared to the
   values of 90-250 $h_{50} \msol/\lso$ generally found in other clusters 
   (cf. e.g. Hughes et al. 1989; Carlberg et al. 1996; Smail et al. 1997;
   Squires et al. 1997). But on the other hand some much higher values
   were found, e.g. 3300 $h_{50} \msol/\lso$ in AXJ2019+1127 (Hattori
   et al. 1997). In particular for the Virgo cluster a high value of
   800~$\msol/\lso$ (scaled to a distance of 20Mpc) was found by 
   modeling velocity fields (Tully \& Shaya 1998).
%%%

   The mass distributions for the M49 subcluster are similar. 
   The only quantity which differs substantially between the M87 and M49 
   subclusters, however, is the
   gas mass fraction. For the M49 subcluster we find a very small gas
   mass fraction of less than 1\% of the total mass. 
   Even with the low total mass value
   of Irwin and Sarazin (1996), this fraction remains below 2\%.
%%% 
   Such a small gas fraction is very unusual for clusters. Even for
   galaxy groups the fraction is quite low, cf. e.g. 
   Mulchaey et al. (1993, NGC2300: 2\%);
   David et al. (1994, NGC5044: 6\%);
   Pildis et al. (1995, 13 groups: maximum 2.8\%;
   Neumann \& B\"ohringer (1995, AWM7: 11-27\%)
   (all scaled to $H_0 = 50$ km s$^{-1}$ Mpc$^{-1}$).
%%%
   We have not 
   found any plausible scenario which could explain this.

\acknow{It is a pleasure to thank Doris Neumann for important discussions
and Carlo Izzo for his most helpful EXSAS support.
S.S. and H.B. thank the
Verbundforschung, and B.B. the Swiss National Science Foundation 
for financial support. }

\begref{References}

\ref Allen S.W., Fabian A.C., Kneib J.-P., 1996, MNRAS 279, 615
\ref Bahcall N.A., 1975, AJ 198, 249
\ref Bardelli S., Zucca E., Malizia A., Zamorani G., Scaramella R.,
       Vettolani G., 1996, A\&A 305, 435
\ref Binggeli B., Sandage A., Tammann G.A., 1985, AJ 90, 1681
\ref Binggeli B., Tammann G.A., Sandage A., 1987, AJ 94, 251 (= BTS87)
\ref Binggeli B., Popescu C.C., Tammann G.A., 1993, A\&AS 98, 275
\ref Binggeli B., 1998, in: Proceeding of the Ringberg workshop ``M87'', 
     R\"oser H.-J., Meisenheimer K. (eds.), Springer, Berlin, in press
\ref Binney J., Tremaine S., 1987, Galactic Dynamics, Princeton University
     Press, Princeton, p. 589ff.
\ref B\"ohringer H., 1994, in: Cosmological Aspects of X-ray Clusters
     of Galaxies, Seitter, W.C. (ed.). NATO ASI Series C 441, Kluwer,
     Dordrecht, p. 123
\ref B\"ohringer H., 1995, in: Relativistic Astrophysics and Cosmology, 
     Proceedings of the 17th Texas Symposium, B\"ohringer, H. et
     al. (eds.). The New York Academy of Sciences, New York, p. 67
\ref B\"ohringer H., Briel U.G., Schwarz R.A., Voges W., Hartner G.,
     Tr\"umper J., 1994, Nat. 368, 828
\ref B\"ohringer H., Neumann D.M, Schindler S., Kraan-Korteweg R.C.,
     1996, ApJ 467, 168
\ref B\"ohringer H., Tanaka Y., Mushotzky R.F., Ikebe Y., Hattori M.,
       1998, A\&A 334, 789 
\ref Briel U.G., Henry J.P., Schwarz R.A., B\"ohringer H., Ebeling H.,
     Edge A.C., Hartner G., Schindler S., Tr\"umper J., Voges W., 1991,
     A\&A 246, L10
\ref Briel U.G., Henry J.P., B\"ohringer H., 1992, A\&A 259, L31
\ref Buote D.A., Canizares C.R., 1996, ApJ 457, 565
\ref Burns J.O., Roettiger K., Pinkney J., Perley R., Owen F.N., 
     Voges W., 1995, ApJ 446, 583
\ref Burstein D., Bender R., Faber S., Nolthenius R., 1995,
       Ap. Lett. \& Comm. 31, 95
\ref Caon N., Capaccioli M., D'Onofrio M., 1993, MNRAS 265, 1013
\ref Carlberg R.G., Yee H.K.C., Ellingson E., Abraham R., Gravel P.,
       Morris S., Pritchet C.J., 1996, ApJ 462, 32
\ref Carlberg R.G., Yee H.K.C., Ellingson E., 1997, ApJ 478, 462
\ref Cavaliere A., Fusco-Femiano R., 1976, A\&A 49, 137
\ref David L.P., Jones C., Forman W., Daines S., 1994, ApJ 428, 544 
\ref David L.P., Jones C., Forman W., 1995, ApJ 445, 578
\ref Dressler A., 1980, ApJ 236, 351
\ref Drinkwater M.J., Currie M.J., Young C.K., Hardy E., Yearsley
     J.M., 1996, MNRAS 279, 595 
\ref Elbaz D., Arnaud M., B\"ohringer H., 1995, A\&A 293, 337
\ref Federspiel M., Tammann G.A., Sandage A., 1998, ApJ 495, 115
\ref Feretti L., Giovannini G., B\"ohringer H., 1997, New Astron., 2, 501
\ref Fitchett M.J., 1988, in: The Minnesota Lectures on Clusters of Galaxies,
      ASP Conf.~Ser.~5, Dickey, J.M. (ed.). Astronomical Society of the Pacific,
      San Francisco, p. 143 
\ref Forman W., Schwarz J., Jones C., Liller W., Fabian A.C., 1979,
     ApJ 234, L27
\ref Gavazzi G., Pierini D., Boselli A., 1996, A\&A 312, 397
\ref Girardi M., Escalera E., Fadda D., Giuricin G., Mardirossian F., 
     Mezzetti M., 1997, ApJ 482, 41
\ref Hattori M., Ikebe Y., Asaoka I., Takeshima T., B\"ohringer H.,
       Mihara T., Neumann D.M., Schindler S., Tsuru T., Tamura T., 1997,
       Nat. 388, 146 
\ref Henry J.P., Briel U.G., Nulsen P.E.J, 1993, A\&A 271, 413
\ref Hughes J.P., 1989, ApJ 337, 21 
\ref Irwin J.A., Sarazin C.L., 1996, ApJ 471, 683
%\ref Jacoby G.H., Branch D., Ciardullo R., Davies R.L., Harris W.E.,
%     Pierce M.J., Pritchet C.J., Tonry J.L., Welch D.L., 1992, PASP
%     104, 599
\ref Kormendy J., 1977, ApJ 218, 333
\ref Kormendy J., 1985, ApJ 295, 73
\ref Kulp K., Schindler S., 1998, in preparation
\ref McMillan S.L.W., Kowalski M.P., Ulmer M.P., 1989, ApJS 70, 723
\ref Matsumoto H., 1998, Ph.D thesis, in preparation
\ref Mohr J.J., Geller M.J., Fabricant D.G., Wegner G., Thorstensen
       J., Richstone D.O., 1996, ApJ 470, 724 
\ref Morris P.W., Shanks T., 1998, MNRAS 298, 451
\ref Mulchaey J.S., Davis D.S., Mushotzky R.F., Burstein D., 1993, ApJ
       404, L9
\ref Navarro J.F., Frenk C.S., White S.D.M., 1995, MNRAS 275, 720
\ref Navarro J.F., Frenk C.S., White S.D.M., 1996, ApJ 462, 563
\ref Navarro J.F., Frenk C.S., White S.D.M., 1997, ApJ 490, 493
\ref Neumann D.M., B\"ohringer H., 1995, A\&A 301, 865
\ref Neumann D.M., B\"ohringer H., 1997, MNRAS 289, 123
\ref Nulsen P.E.J., B\"ohringer H., 1995, MNRAS 274, 1093
\ref Ohashi T., Kikuchi K., and the ASCA Virgo project team, 1997, 
       in: Proceedings of the International Astronomical
        Union Symposium no. 188, The Hot Universe, Kyoto, p. 148 
\ref Pildis R.A., Bregman J.N., Evrard A.E., 1995, ApJ 443, 514
\ref Rangarajan F.V.N., White D.A., Ebeling H., Fabian A.C., 1995, MNRAS
      277, 1047
%\ref Sarazin C.L., 1996, Rev. of Mod. Phys. 58, 1
\ref Sarazin C.L., McNamara B.R., 1997, ApJ 480, 203
\ref Sarazin C.L., Wise M.W., Markevitch M.L., 1998, ApJ 498, 606 
\ref Schindler S., M\"uller E., 1993, A\&A 272, 137 
\ref Schindler S., Prieto M.A., 1997, A\&A 327, 37
\ref Schindler S., Hattori M., Neumann D.M., B\"ohringer H., 1997,
      A\&A 317, 646
\ref Schindler S., Belloni P., Ikebe Y., Hattori M., Wambsganss J.,
       Tanaka Y., 1998, A\&A 338, 843
\ref Schr\"oder A., 1995, Ph.D. thesis, Universit\"at Basel.
%\ref Shapley H., Ames A. 1926, Harvard Circ. No. 294
\ref Smail I., Ellis R.S., Dressler A., Couch W.J., Oemler A.Jr.,
       Sharples R.M., Butcher H., 1997, ApJ 479, 70
\ref Squires G., Neumann D.M., Kaiser N., Arnaud M., Babul A.,
       B\"ohringer H., Fahlman G., Woods D., 1997, ApJ 482, 648 
\ref Stewart G.C., Canizares C.R., Fabian A.C., Nulsen P.E.J., 1984,
     ApJ 278, 536
\ref Takano S., Awaki H., Koyama K., Kunieda H., Tawara Y., Yamauchi
     S., Makishima K., Ohashi T., 1989, Nat. 340, 289
\ref Tammann G.A., Federspiel M., 1996, in: The Extragalactic Distance
     Scale, Proceedings of the ST ScI May
     Symposium,  Livio, M. et al. (eds.). Cambridge University Press, p. 137.
\ref Tonry J.L., Ajhar E.A., Luppino G.A., 1990, AJ 100, 1416 
\ref Tr\"umper J., 1993, Science 260, 1769  
\ref Tully R.B., Shaya E.J., 1998, in: Proceedings of the MPA/ESO
       Cosmology Conference: Evolution of Large Scale Structure held
       in Garching, in press (astro-ph/9810298)
\ref Voges W., Boller T., Dennerl K., Englhauser J., Gruber R.,
        Haberl F., Paul J., Pietsch W., Tr\"umper J.E., Zimmermann H.U., 1996,
        in: Proceedings of the workshop R\"ontgenstrahlung from the
        universe, held in W\"urzburg, Germany, Zimmermann H.U.,
        Tr\"umper J.E., Yorke H. (eds.). MPE Report 263, p. 637
\ref West M.J., 1994, in: Clusters of Galaxies, Proceedings of the XIVth
      Moriond Astrophysics Meeting, Durret, F. et al. (eds.). 
      Editions Frontieres,
      Gif-sur-Yvette, p. 23
\ref White D.A., Fabian A.C., 1995, MNRAS 273, 72
\ref White D.A., Fabian A.C., Allen S.W., Edge A.C., Crawford C.S.,
      Johnstone R.M., Stewart G.C., Voges W., 1994, MNRAS 269, 589
\ref White S.D.M., Briel U.G., Henry J.P., 1993, MNRAS 261, L8
\ref Zabludoff A.I., Zaritsky D., 1995, ApJ 447, L21
\endref
\bye